\documentclass[pre,aps,a4,12pt]{revtex4}

\usepackage{graphicx}
\usepackage{subfig}
\usepackage{caption}
\usepackage[usenames,dvipsnames]{color}
\usepackage[sort&compress]{natbib}
\usepackage{amsmath}

\makeatletter
\def\gsim{\compoundrel>\over\sim}
\def\lsim{\compoundrel<\over\sim}
\def\compoundrel#1\over#2{\mathpalette\compoundreL{{#1}\over{#2}}}
\def\compoundreL#1#2{\compoundREL#1#2}
\def\compoundREL#1#2\over#3{\mathrel
      {\vcenter{\hbox{$\m@th\buildrel{#1#2}\over{#1#3}$}}}}
\makeatother

\newcommand{\paDir}[2]{\frac{\partial #1}{\partial #2}}
\newcommand{\paDirH}[3]{\frac{\partial^{#3} #1}{\partial #2^{#3}}}
\newcommand{\fnDir}[2]{\frac{\delta #1}{\delta #2}}
\newcommand{\vx}{{\bf x}}
\newcommand{\delPhi}{\delta \hspace{-0.8mm} \phi}
\newcommand{\chiArray}{\begin{pmatrix} 1 \\ \chi \end{pmatrix}}
\newcommand{\mat}[4]{\begin{pmatrix}#1 & #2 \\ #3 & #4 \end{pmatrix}}

\newcommand{\barA}{\bar{\phi_a}}
\newcommand{\barB}{\bar{\phi_b}}
\newcommand{\mx}{\mathrm{max}}

\newcommand{\mn}{\mathrm{min}}

\usepackage{ulem}

\begin{document}

\title{Modelling fluids and crystals using a two-component modified phase field crystal model}
\author{M.J.~Robbins$^\ast$, A.J.~Archer$^\ast$, U.~Thiele$^\ast$ and E.~Knobloch$^\#$}
\affiliation{$^\ast$Department of Mathematical Sciences, Loughborough University, Leicestershire 
		LE11 3TU, UK\\
		$^\#$Department of Physics, University of California, Berkeley, California 94720, USA}

\begin{abstract}
\vspace{1mm}
\begin{center}
\begin{minipage}{0.8\linewidth}
	A modified phase field crystal model in which the free energy may be minimised
	by an order parameter profile having isolated bumps is investigated.  The phase diagram 
	is calculated in one and two dimensions and we locate the regions where modulated and 
	uniform phases are formed and also regions where localised states are formed.  We investigate 
	the effectiveness of the phase field crystal model for describing fluids and crystals with defects.  
	We further consider a two component model and elucidate how the structure transforms from 
	hexagonal crystalline ordering to square ordering as the concentration changes.  Our conclusion
	contains a discussion of possible interpretations of the order parameter field. 
\end{minipage} 
\end{center}
\end{abstract}

\maketitle

\section{Introduction}
Modelling materials at the atomic scale is a task which, for example, may be performed using Molecular
Dynamics simulations.  This involves solving coupled equations of motion to calculate the position of 
each particle at every time step.  The resulting calculations can be very computationally expensive, 
especially when one seeks to consider phenomena which involve a large number of particles.  Only short 
atomic time scales can be feasibly accessed with this or other such approaches.  However, there are 
some instances where it is important to consider materials on the atomic length scale for much longer 
diffusive time scales, e.g., when investigating freezing or glass transitions.  One approach to such problems 
that may be adopted is to develop a phase field model capable of describing the structure of materials on the 
scale of the individual particles.  In contrast to traditional phase field models, the recently developed phase 
field crystal (PFC) models are capable of just such a description and are now widely used in the literature to model 
crystalline structures \cite{EKHG02,ElGr04,TGTD11,MKP08,WPV10}.  The PFC model consists of a Swift-Hohenberg-like
equation \cite{SwHo77}, but with conserved dynamics rather than the non-conserved dynamics of the regular 
Swift-Hohenberg equation.  Similar models with conserved dynamics arise in quite different contexts as well 
\cite{Knob89, MaCo00}.  The regular PFC model is governed by the following equation:  
\begin{equation}
	\paDir{\phi(\vx,t)}{t} = \alpha \nabla^2 \fnDir{F[\phi]}{\phi(\vx,t)},
	\label{eqPFCDyn} 
\end{equation}
where the free energy functional
\begin{equation}
	F[\phi] = \int d\vx \hspace{2mm} f(\phi), 
	\label{eqPFCFree}
\end{equation}
with
\begin{equation}
	 f(\phi) = \frac{\phi}{2}[r + (q^2 + \nabla^2)^2] \phi + \frac{\phi^4}{4},
	\label{eqPFCOne}
\end{equation}
where $\alpha$ is the mobility coefficient, $r$ is the undercooling parameter that decreases with decreasing 
temperature, $q$ is a constant which determines the typical microscopic length scale in the system and 
$\phi(\vx, t)$ is the order parameter.  For certain parameter values this free energy functional is minimised by 
an order parameter profile consisting of a periodic array of bumps which somewhat resembles the density 
distribution of particles in a crystalline material.  This interpretation is bolstered by the fact that it has been 
shown that the PFC model (Eqs.~\eqref{eqPFCDyn} and \eqref{eqPFCFree}) may be derived from the 
density functional theory of freezing \cite{EPBS07} and the dynamical density functional theory for colloidal 
particles \cite{TBVL09, HEP10} with certain approximations.  The free energy is minimised by either 
periodic structures or by a homogeneous flat profile, depending on the values of $q$, $r$ and $\bar{\phi} = 
\frac{1}{L^d} \int d\vx \hspace{1mm} \phi(\vx, t)$, where $L^d$ is the size of the system.  In two 
dimensions ($d = 2$), one observes a homogeneous phase, two hexagonal phases (hexagonally ordered 
bumps/holes) and a stripe phase \cite{EKHG02,ElGr04,OhSh08,TGTD11}.  The literature largely focuses 
on the region of the 2d phase diagram which contains the hexagonally arranged bumps and their 
transition to the homogeneous state \cite{EKHG02,ElGr04,EPBS07,OhSh08,MKP08,TBVL09,WPV10,
TGTD11}.  The uniform profile $\phi(\vx, t) = \bar{\phi}$ represents the order parameter in a uniform liquid 
and the hexagonal phase is treated as a crystal.  The model is  then used to consider a number of 
problems including melting and freezing \cite{TBVL09,TGTD11,EPBS07} and grain boundary effects 
\cite{EKHG02, ElGr04, MKP08}.

In the `standard' PFC model (Eqs.~\eqref{eqPFCDyn} and \eqref{eqPFCFree}) the hexagonally arranged 
bumps are considered to be particles/colloids in a crystalline structure.  The interpretation of the striped and
hexagonally ordered hole structures is unclear and as such these phases are commonly ignored.  The 
conjecture that the ordered bumps represent crystalline particle structures can be extended by including a 
`vacancy term' in the free energy \cite{CGD09,BeGr11} which strongly breaks the hole-bump ($\phi \to -\phi$) 
symmetry of Eq.~\eqref{eqPFCOne}:  
\begin{equation}
	F[\phi] = \int d\vx \hspace{2mm} \bigg[ f(\phi) + f_{vac}(\phi) \bigg]. 
	\label{eqVPFCOne}
\end{equation}
Using this free energy \eqref{eqVPFCOne}, it is possible to obtain structures which contain a mixture of 
bumps and vacant areas (areas where the order parameter is approximately uniform around the value 
$\phi \approx 0$), which in the interpretation of Ref.~\cite{CGD09} resemble snapshots of fluid configurations 
or crystalline structures with defects. We will return to the issue of the precise interpretation of the nature of the 
order parameter field in the conclusion.  In this paper we investigate the thermodynamics and the structures 
formed in this augmented conserved Swift-Hohenberg model, or `vacancy phase field crystal' (VPFC) model, 
and also in a two component generalisation of this model.  The vacancy term takes the following form 
\cite{CGD09,BeGr11}:
\begin{equation}
	f_{vac}(\phi) = H\phi^2 (|\phi| - \phi),
	\label{eqPFCVacTerm}
\end{equation}
where $H$ is a constant. We use the value $H = 1500$, as in Refs.~\cite{CGD09,BeGr11}. This acts as a 
piecewise function which is zero for positive values of $\phi$ and takes an increasingly large value when 
$\phi < 0$. Hence, this term penalises negative values of $\phi$. This leads to the VPFC model forming 
periodic structures which are somewhat different from those of the regular PFC. In addition, the VPFC model
has a large region of parameter space at small $\bar{\phi}$, where {\it spatially localised} structures form.  
The time evolution of the order parameter $\phi$ is governed by the conserved dynamics used in the standard 
PFC model \eqref{eqPFCDyn}.

We begin in Sec.~\ref{secOneComp} by considering the phase behaviour of the model, investigating 
the transition between periodic and localised states.  We focus on understanding the bifurcation diagrams
connecting the various uniform, periodic and localised states exhibited by the model.  We then go on to 
consider how individual localised states or particles interact with one another.  In Sec.~\ref{secTwoComp} we 
extend the model to consider a two component system, and we determine how the particles in the binary 
mixture interact with one another.  We find a transition between hexagonal and square ordering of the 
particles as the concentration changes.  Our conclusions follow in Sec.~\ref{secConc}, and include
a discussion of the proper interpretation of the order parameter field $\phi$.

\section{One-Component System}
\label{secOneComp}

\subsection{Linear stability of a homogeneous profile}
We begin by considering the phase behaviour of the VPFC model (Eqs.~\eqref{eqPFCDyn} and 
\eqref{eqVPFCOne}).  We calculate the limit of linear stability for a homogeneous flat state using a linear 
stability analysis.  In the context of colloidal suspensions exhibiting  microphase separation and fluids of 
charged particles, this limit of linear stability is referred to as a `$\lambda$-line' 
\cite{stel95, CGE03, Ale04, APER07}. Since $f_{vac}$ is non differentiable at $\phi = 0$ we treat 
it in a piecewise manner, by treating the two cases $\bar{\phi} > 0$ and $\bar{\phi} < 0$ separately 
(in fact, if $\phi(\vx)$ takes the form of Eq.~\eqref{eqLinStabPhi} and $\bar{\phi} > |\xi|$, then $f_{vac} = 0$ 
everywhere and the thermodynamics of the VPFC model reduces to that of the regular PFC model).  
We assume that the order parameter $\phi$ takes the form of a flat profile plus an additional small 
amplitude harmonic modulation:
\begin{equation}
	\phi = \bar{\phi} + \delPhi = \bar{\phi} + \xi e^{i{\bf k \cdot x}} e^{\beta t},
	\label{eqLinStabPhi}
\end{equation}
where $\bar{\phi}$ is the average value of the order parameter and the amplitude $|\xi| \ll 1$.  Substituting this 
expression into the functional derivative of the free energy \eqref{eqVPFCOne} we obtain:
\begin{equation}
	\fnDir{F}{\phi} = (r + q^4) \bar{\phi} + 3H \bar{\phi} (|\bar{\phi}| - \bar{\phi}) + \bar{\phi}^3 
	+ \big[(k^2 - q^2)^2 + \Delta\big] \delPhi + O(\delPhi^2),
	\label{eqPFCFuncDer}
\end{equation}
where
\begin{equation}	
	\Delta = r + 6H (|\bar{\phi}| - \bar{\phi}) + 3\bar{\phi}^2.
	\label{eqDelta}
\end{equation}
Inserting this expression for the functional derivative \eqref{eqPFCFuncDer} into the dynamical equation 
\eqref{eqPFCDyn} and then linearising we arrive at the following dispersion relation:

\begin{equation}
	\beta = -k^2 \alpha[(k^2 - q^2)^2 + \Delta]. 
	\label{eqPFCDispRel} 
\end{equation}
When the growth rate $\beta(k) > 0$, any small amplitude
modulation with wave number $k = |{\bf k}|$ will grow over time. There
is a local maximum in $\beta$ (which becomes the global maximum when
the uniform state is unstable) at the wave number:
\begin{equation}
	k_m = \frac{1}{3}\sqrt{6q^2 + 3 \sqrt{q^4 - 3\Delta}}.
	\label{eqTypWaveNum}
\end{equation}
Thus, if one takes an initially almost flat profile $\phi(\vx, t=0) = \bar{\phi} + \mathcal{X}(\vx)$,
where $\mathcal{X}(\vx)$ is composed of a sum of a large number of small-amplitude harmonic modulations [cf. 
Eq.~\eqref{eqLinStabPhi}] with different wave numbers $k$ (in practice $\mathcal{X}(\vx)$ is
generated by adding a small random number to the discretised initial profile), then as the system
evolves in time $\phi(\vx,t)$ will develop spatial modulation on the length scale $\frac{2\pi}{k_m}$, 
since this scale corresponds to the maximum growth rate $\beta_m\equiv\beta(k_m)$.  This length 
scale has an inverse dependence on the value of $q$, i.e., increasing the value of $q$ reduces the 
length scale of the structures which are formed.

\begin{figure}[t]
	\includegraphics[width = 0.7\linewidth]{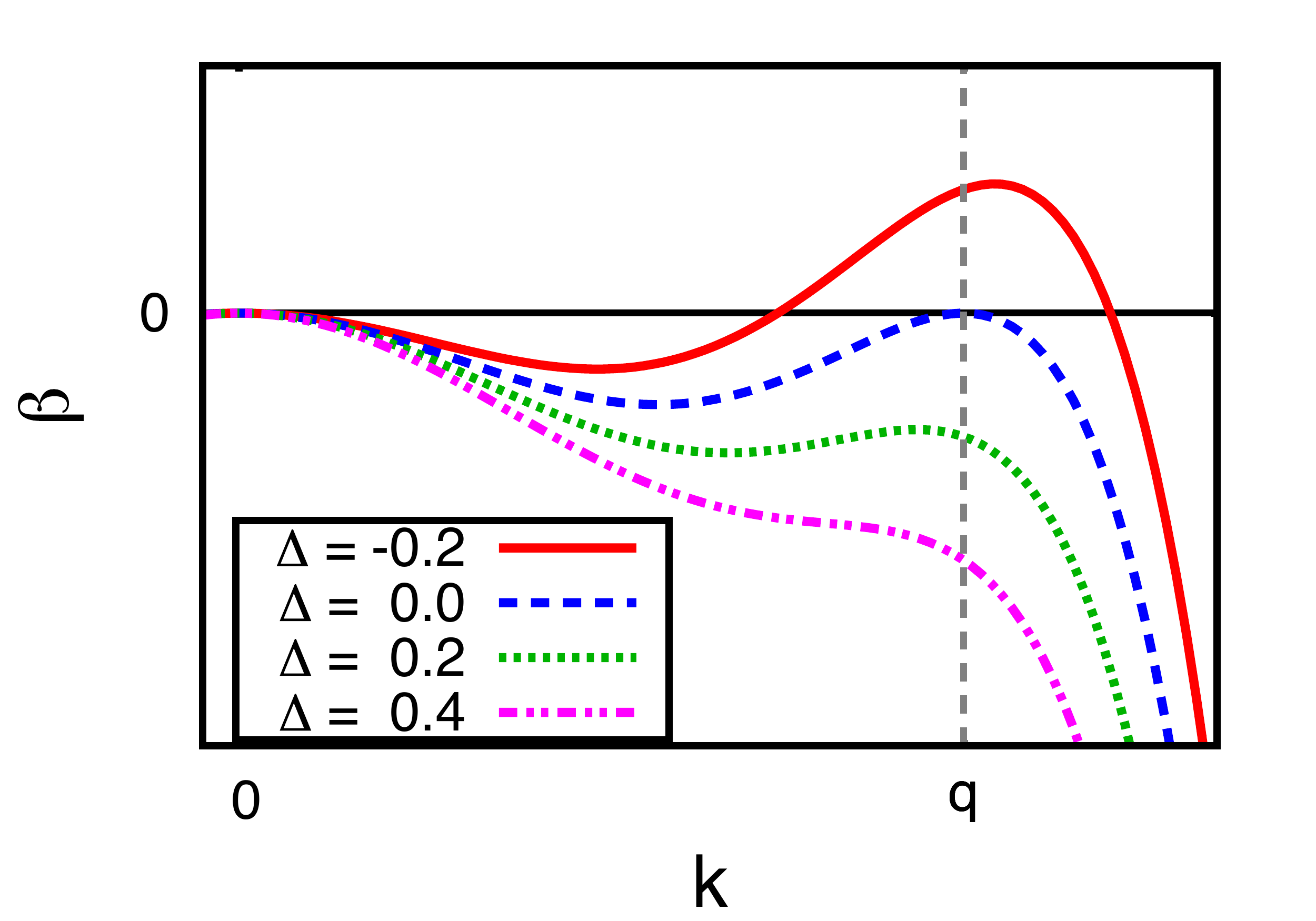}
	\caption{Dispersion relation curves for the VPFC model (Eqs.~\eqref{eqPFCDyn}
and \eqref{eqVPFCOne}) when $q = 1$.  Four cases are shown, with (i) $\beta(k_m) > 0$ (red solid line), 
(ii) $\beta(k_m) = 0$ (blue dashed line), (iii) $\beta(k_m) < 0$ but $k_m>0$ (green dotted line) and 
(iv) $\beta \le 0$ and $k_m=0$ (magenta dash-dotted line).}
	\label{figDispRel}
\end{figure} 

The limit of linear stability is defined as the locus of points in parameter space where the maximum 
in the dispersion relation \eqref{eqPFCDispRel} is at zero, i.e.,~$\beta_m = 0$.  The conditions 
$\beta = \paDir{\beta}{k} = 0$, subject to the requirement that $k_m\ne0$ yield $\Delta = 0$, $k_m = \pm q$.
Thus $\Delta$ in Eq.~\eqref{eqDelta} can be considered as a measure of stability: when $\Delta < 0$ the 
system is linearly unstable and when $\Delta > 0$ the system is linearly stable. The magnitude 
of $\Delta$ indicates how `far' we are from the limit of stability.  Figure~\ref{figDispRel} shows the 
dispersion relations $\beta(k)$ for various values of $\Delta$.  In accordance with Eq.~\eqref{eqTypWaveNum} 
the maximum at $k_m \approx q$ disappears when $\Delta > \frac{q^4}{3}$; in this case only the maximum at 
$k=0$ remains. It is important to note that these results are identical to those of the regular PFC 
model (Eqs.~\eqref{eqPFCDyn} and \eqref{eqPFCFree}) for positive values of the order parameter $\bar{\phi} > 0$. 

\subsection{One-dimensional model}

In order to develop a better understanding of the effect of the `vacancy term' \eqref{eqPFCVacTerm} 
we initially consider the phase diagram for the system in one spatial dimension.  The regular PFC model 
(Eqs.~\eqref{eqPFCDyn} and \eqref{eqPFCFree}) in one dimension exhibits two distinct phases \cite{ElGr04}: 
a non-uniform state in which the order parameter profile resembles a sinusoid and a uniform state in which 
the order parameter is a constant.  The phase diagram of the regular PFC model is symmetric around 
$\bar{\phi} = 0$ owing to the symmetry of the free energy \eqref{eqPFCFree} with respect to $\phi \to - \phi$. 
This is no longer the case when the vacancy term \eqref{eqPFCVacTerm} is added.

\begin{figure}[t!]
	\includegraphics[width=0.6\linewidth]{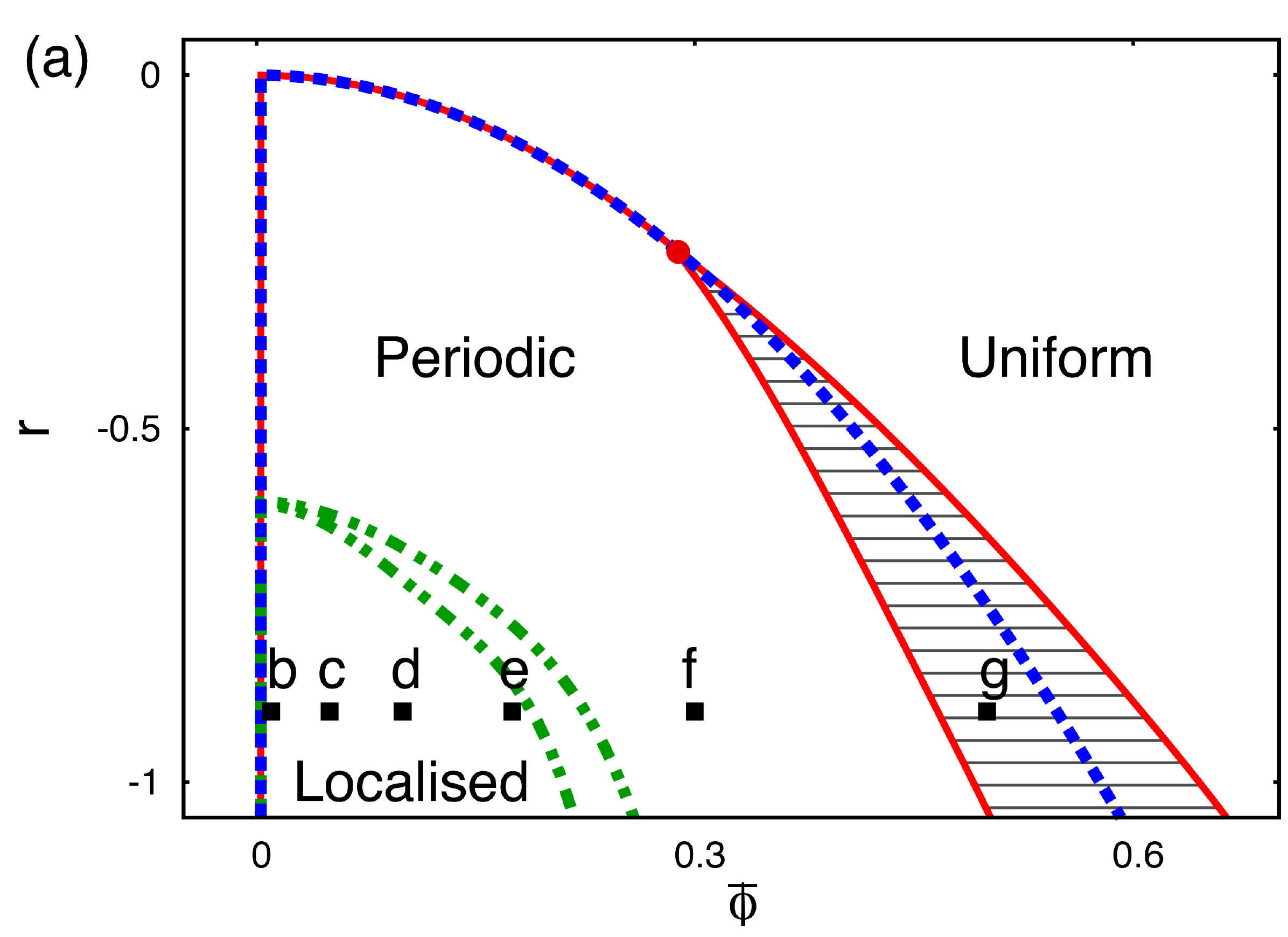} \\
	\includegraphics[width=0.3\linewidth]{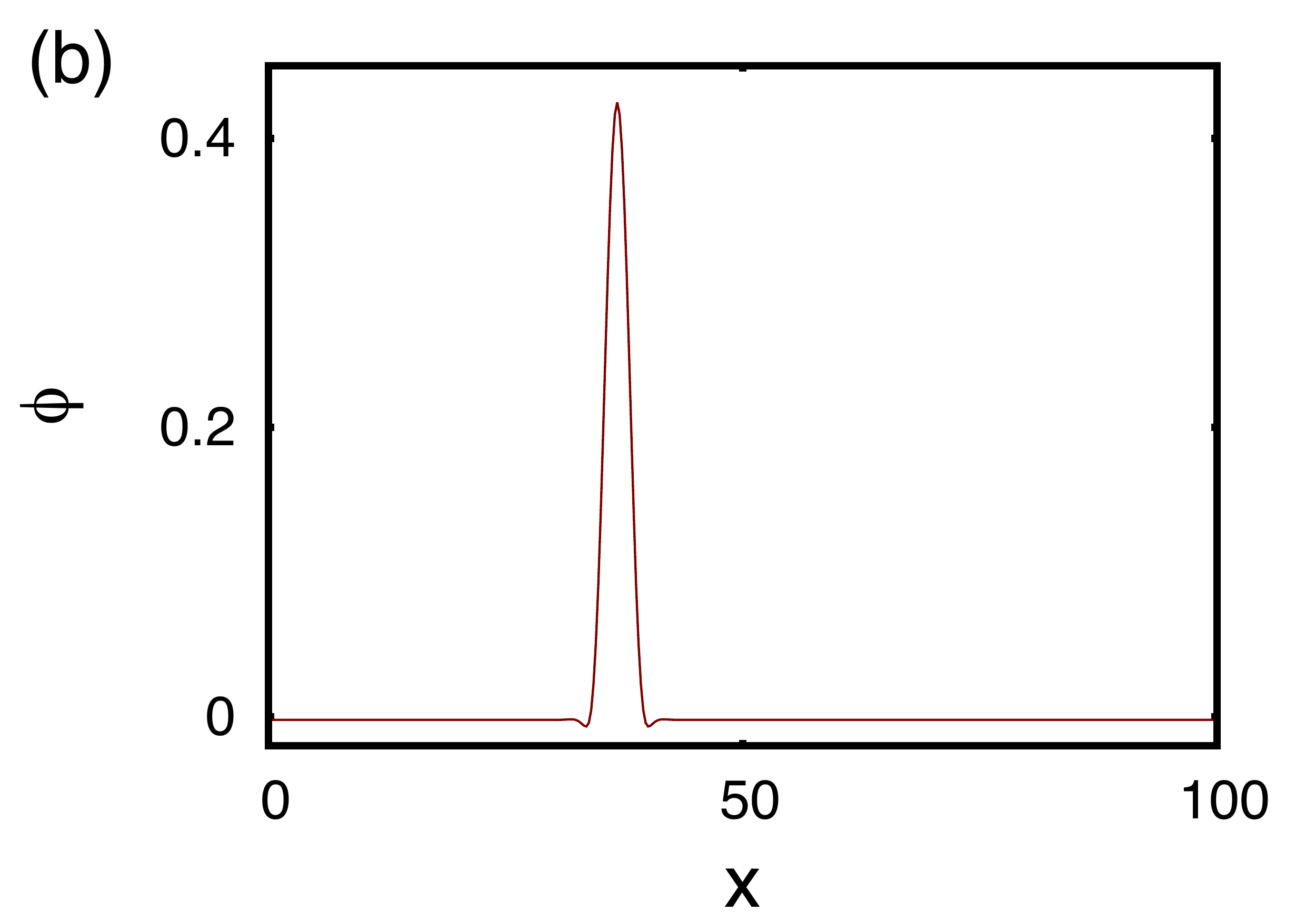}
	\includegraphics[width=0.3\linewidth]{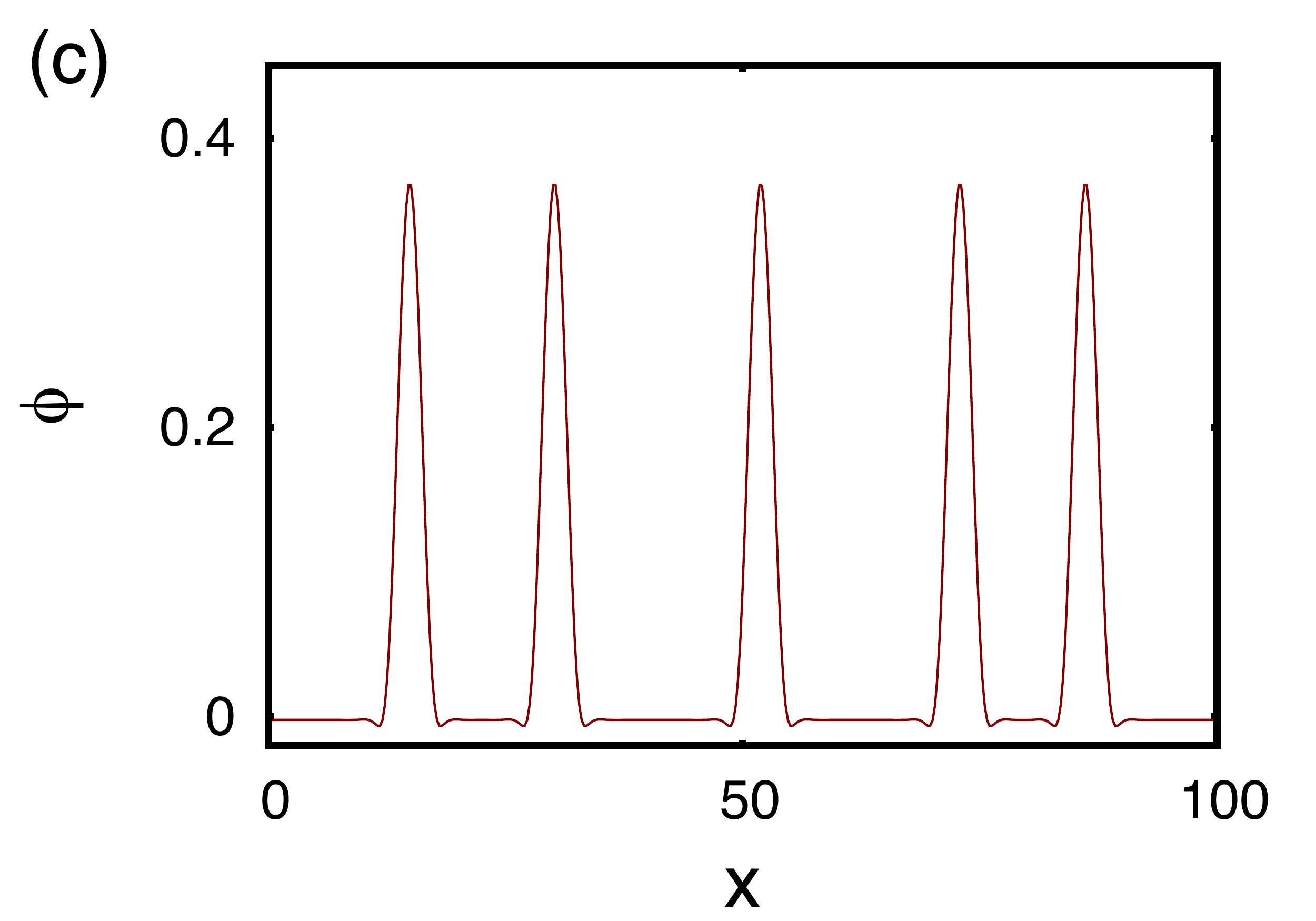}
	\includegraphics[width=0.3\linewidth]{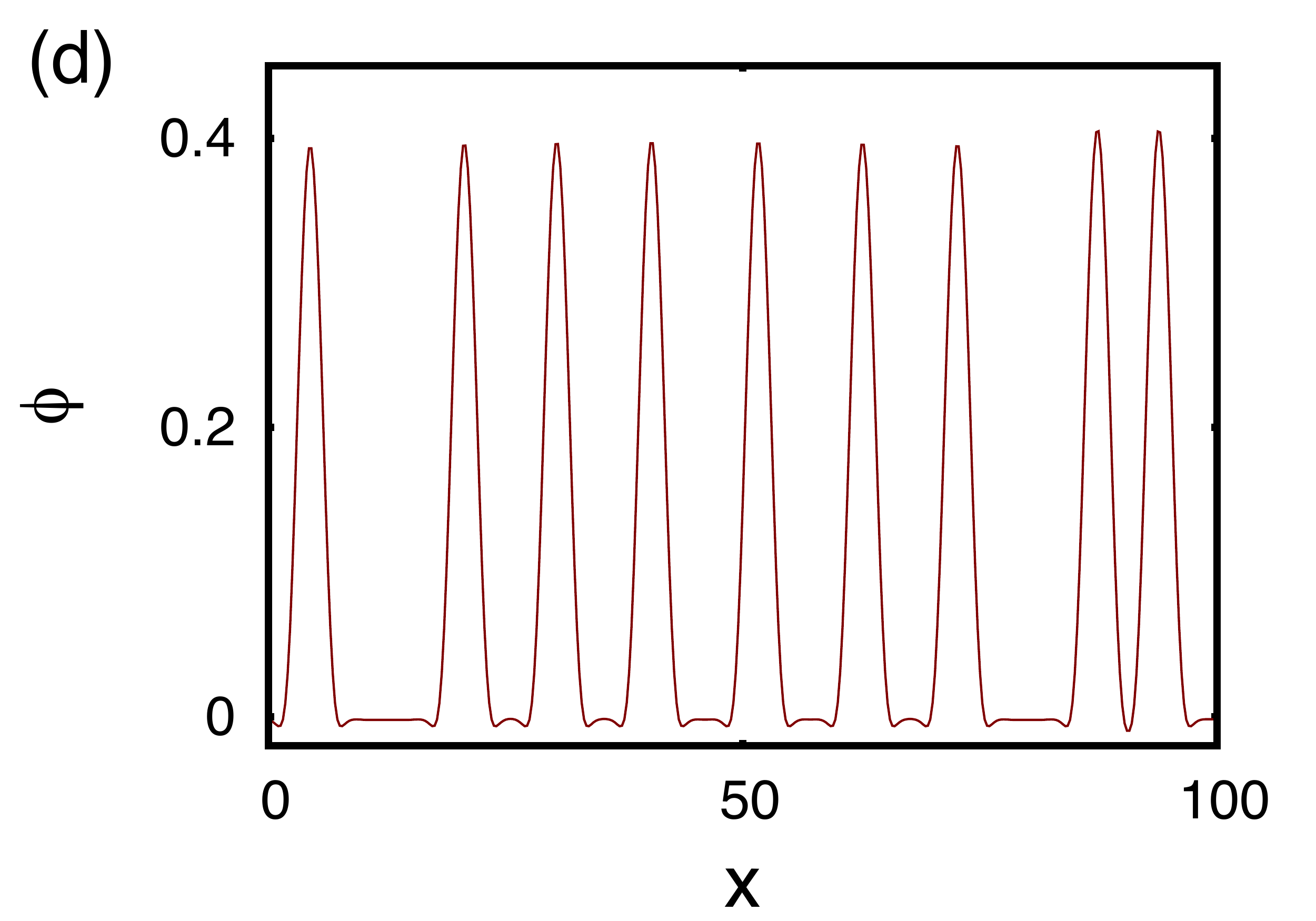}\\
	\includegraphics[width=0.3\linewidth]{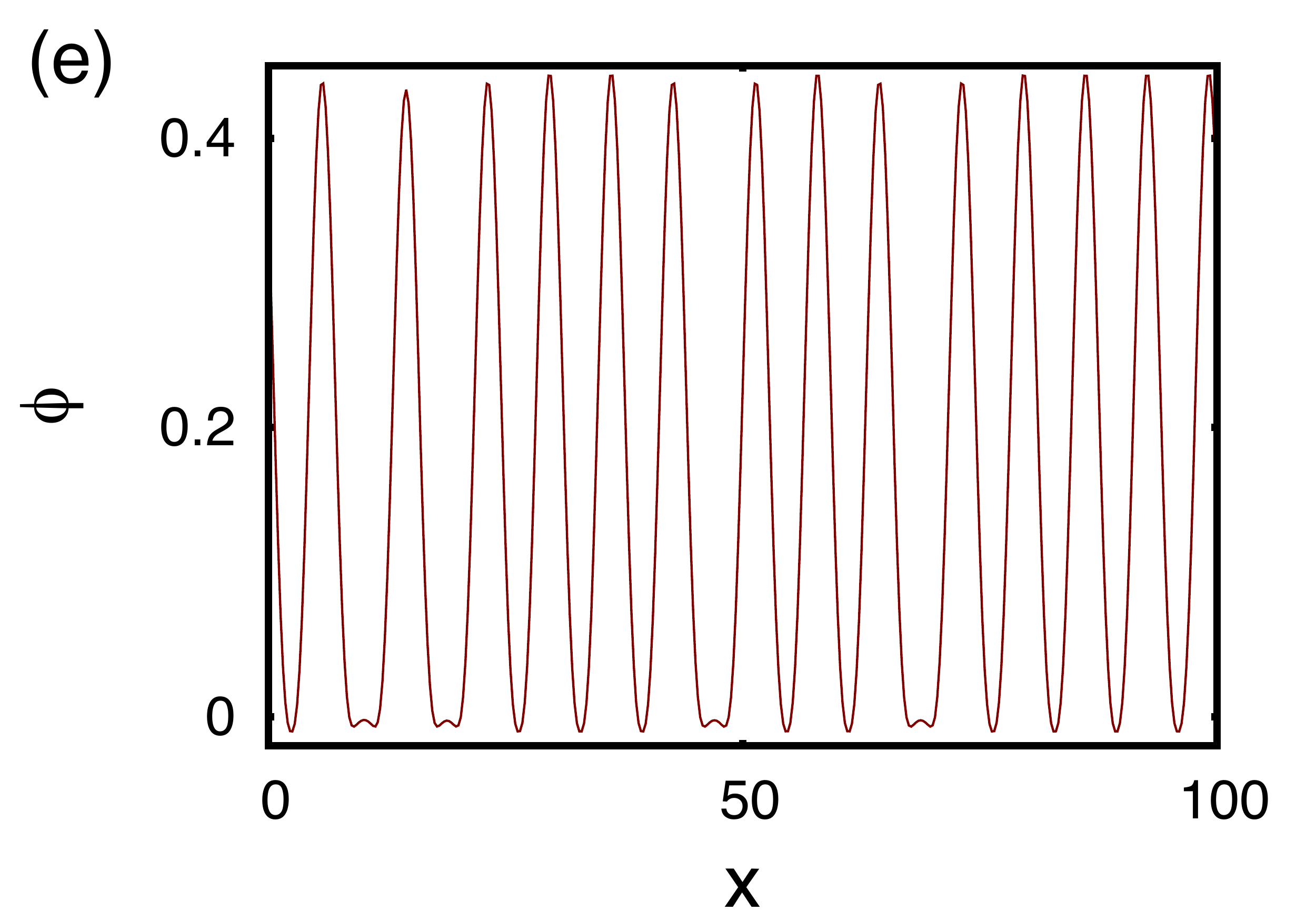}
	\includegraphics[width=0.3\linewidth]{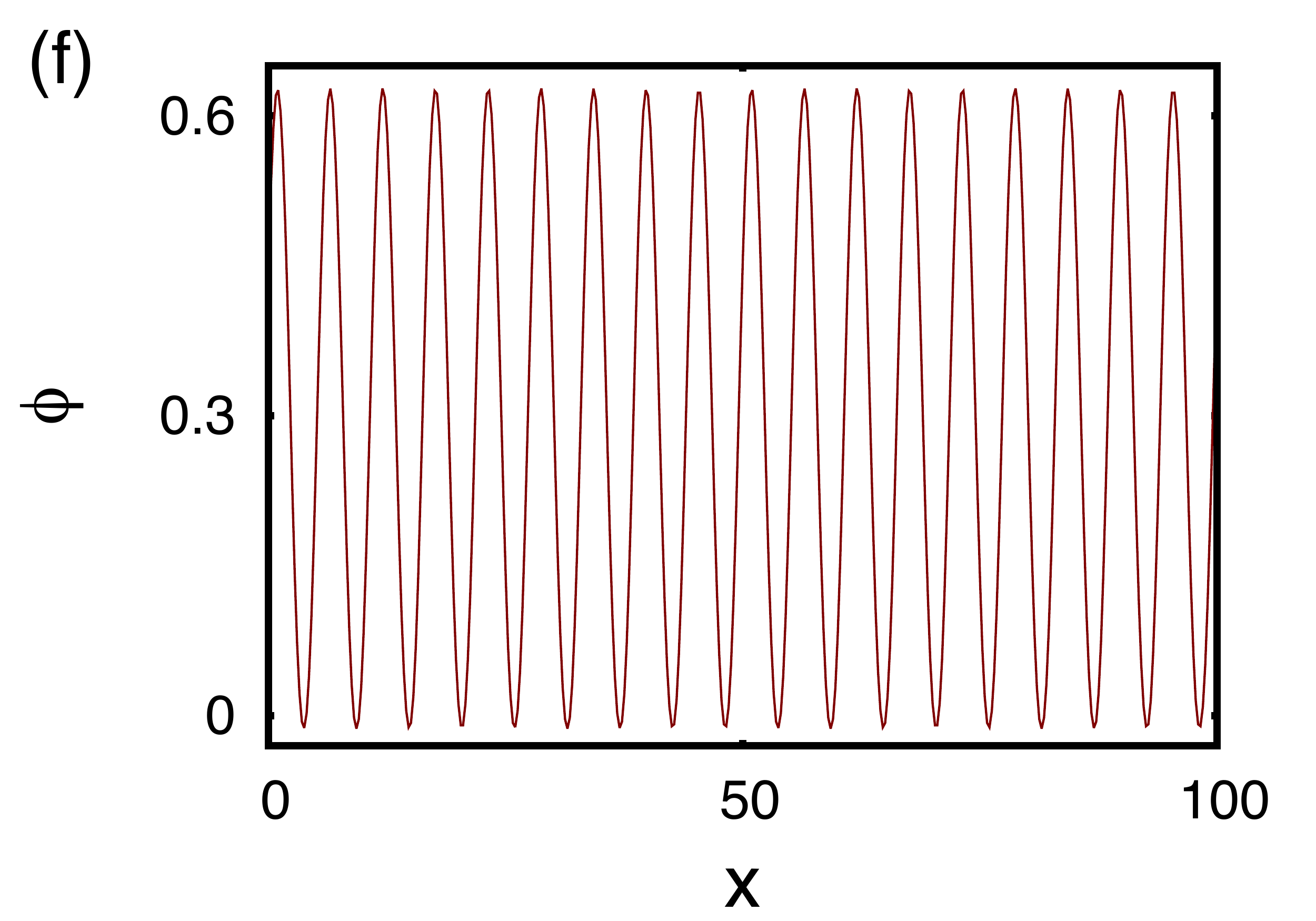}
	\includegraphics[width=0.3\linewidth]{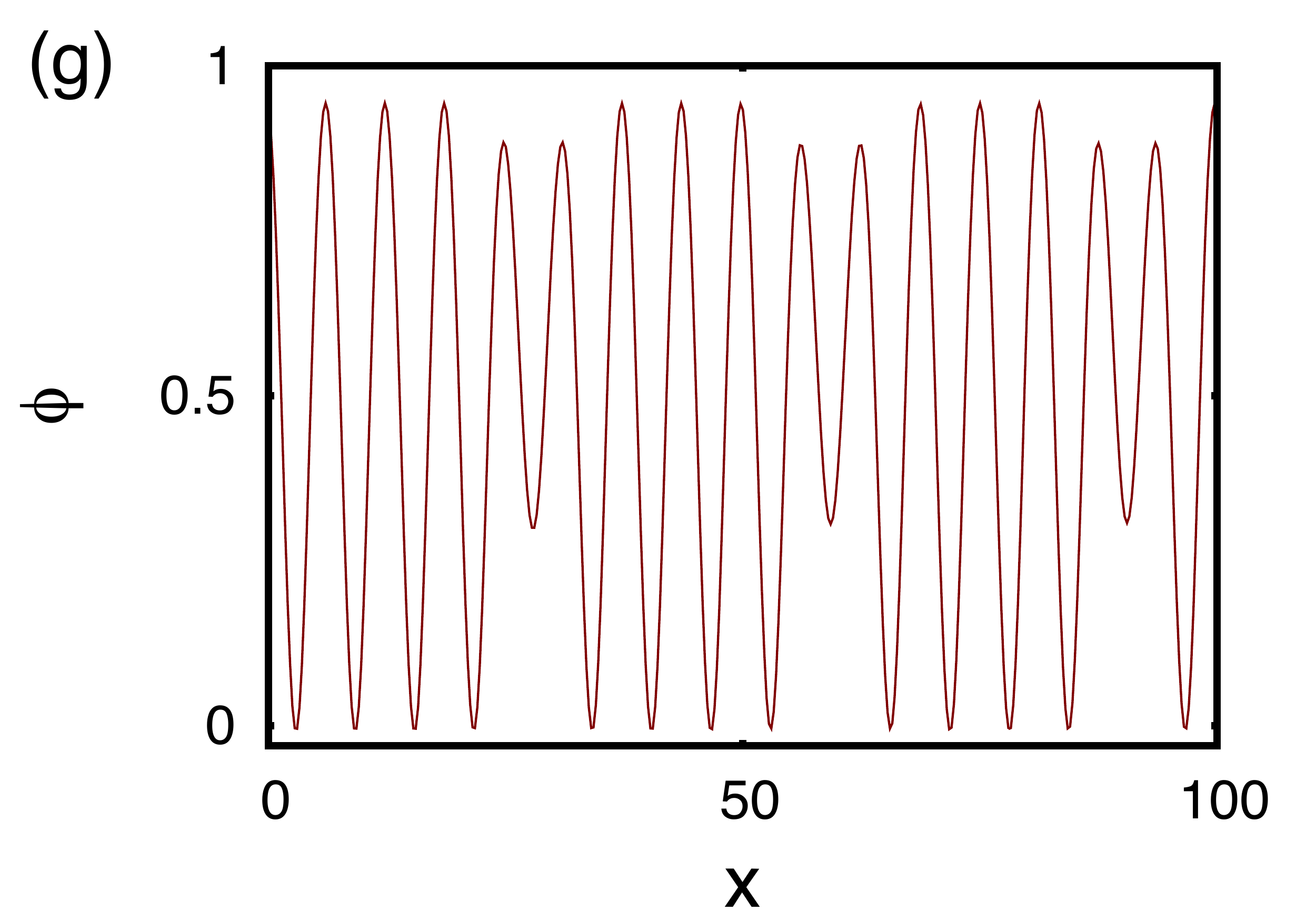}
	\caption{The phase diagram for the 1D VPFC model (Eqs.~\eqref{eqPFCDyn} 
	and \eqref{eqVPFCOne}) is displayed in (a) for the case $q = 1$.  The red solid lines are the 
	coexistence curves between the periodic and uniform phases; the red circle is the tricritical point.  
	The blue dashed line is the locus $\Delta = 0$, which is the limit of linear stability for uniform profiles.  
	The green dot-dashed lines are a guide showing the parameter space where local and periodic 
	structures are formed. (b) - (g) show examples of order parameter profiles from numerical simulations 
	corresponding to (local) minima of the free energy, for the values of $\bar{\phi}$ and $r$ indicated in (a).  
	The parameter values are: $q = 1$, $r = -0.9$, $\alpha = 1$ and (b) $\bar{\phi} = 0.01$, (c) 
	$\bar{\phi} = 0.05$, (d) $\bar{\phi} = 0.1$, (e) $\bar{\phi} = 0.175$, (f) $\bar{\phi} = 0.3$ and
	(g) $\bar{\phi} = 0.5$.}
	\label{figPhaseD1D}
\end{figure} 

The phase diagram for the 1D VPFC model is shown in Fig.~\ref{figPhaseD1D}(a) and is very different from
that of the regular PFC \cite{ElGr04}. As with the regular PFC model, modulated profiles are present below the limit of 
linear stability $\Delta = 0$ (blue dashed line) provided $\bar{\phi}<\sqrt{3/2}q^2$. However, with the added 
vacancy term \eqref{eqPFCVacTerm} the lower limit for the presence of the modulated phase is at $\bar{\phi} \gsim 0$
($H\gg1$). The tricritical point with $\bar{\phi}>0$ (red dot) familiar from the PFC model remains. Above this
point the phase transition between the periodic and homogeneous phases is of second order. Below this 
point a periodic phase with $\bar{\phi} = \bar{\phi}_p$ coexists with a homogeneous phase with 
$\bar{\phi} = \bar{\phi}_h$ and the phase transition between these phases is of first order.  Figure~\ref{figPhaseD1D}(a)
shows the coexisting phases using fixed temperature (horizontal) tie-lines connecting $\bar{\phi}_p$ and $\bar{\phi}_h$
(solid red lines).

We can calculate the location of the tricritical point as follows: Since the wavenumber near $\Delta=0$ is $k_m\approx q$
we assume that the order parameter profile takes the form 
\begin{equation}
	\phi = \bar{\phi} _p+ A \cos qx + B \cos 2qx + \dots
	\label{eqOneModeApp}
\end{equation}
and compute the free energy $F$. When $\bar{\phi}_p>0$ the vacancy term drops out and we obtain the following expression for the free energy per unit length $f_p=F/L$ of the periodically modulated phase:
\begin{equation}
f_p=\frac{1}{2}(r+q^4)\bar{\phi}_p^2+\frac{1}{4}\bar{\phi}_p^4
+\frac{1}{4}rA^2+\frac{1}{4}(r+9q^4)B^2+\frac{3}{4}\bar{\phi}_p^2(A^2+B^2)
+\frac{3}{4}\bar{\phi}_pA^2B+\frac{3}{32}(A^4+4A^2B^2+B^4).
\label{eq:F_modulated}
\end{equation}
We refer to the Ansatz (\ref{eqOneModeApp}) as the two-mode approximation. This approximation is reliable around and above the tricritical point since the amplitude of the modulation in $\phi$ is small when $|\Delta|\ll1$. Moreover the two-mode approximation appears to be exact at the tricritical point, since the location of the tricritical point is unaffected by the inclusion of $\cos 3qx$ and other higher order modes. In contrast, the mode $B\cos 2qx$ must be retained in order to obtain the correct value of the amplitude $A$ in the vicinity of the tricritical point (see below). 

To demonstrate this we minimise $f_p$ in Eq.~\eqref{eq:F_modulated} with respect to the amplitudes $A$ and $B$, obtaining the following two conditions
\begin{eqnarray}
r+3\bar{\phi}_p^2 + 3\bar{\phi}_pB + \frac{3}{4}A^2 + \frac{3}{2}B^2 = 0,\label{eqA}	\\
(r+9q^4)B + 3\bar{\phi}_p^2B+\frac{3}{2}\bar{\phi}_pA^2+\frac{3}{2}A^2B+\frac{3}{4}B^3=0.\label{eqB}
\end{eqnarray}
Solving these for the amplitudes $A$ and $B$, and substituting into Eq.\ \eqref{eq:F_modulated}, we obtain an approximation for the free energy density of the periodic phase $f_p$. Linearising Eq.\ \eqref{eqB} in $B$, we find that
\begin{equation}
B=-\frac{\bar{\phi}_pA^2}{6q^4}+O(\Delta_p A^2,A^4)
\end{equation}
where $\Delta_p\equiv r+3\bar{\phi}_p^2$ and hence, from Eq.~\eqref{eqA}, that
\begin{equation}
	A = 2\sqrt{-{\frac{\Delta_p}{3}}}\biggl(1-\frac{2\bar{\phi}_p^2}{3q^4}\biggr)^{-1/2}+O(\Delta_p).
	\label{eqOneModeAppAmp1}
\end{equation}
The free energy density of the homogeneous phase $f_h$, having $\phi(x)={\bar \phi}_h$, is obtained simply setting $A=B=0$ in Eq.\ \eqref{eq:F_modulated} to obtain $f_h=\frac{1}{2}(r+q^4)\bar{\phi}_h^2+\frac{1}{4}\bar{\phi}_h^4$. The chemical potential in the homogeneous phase is $\mu_h=\partial f_h/\partial \bar{\phi}_h$, and in the periodic phase $\mu_p=\partial f_p/\partial \bar{\phi}_p$.

To calculate the location of the tricritical point we recall that at coexistence
between the periodic state and the homegeneous state we must have $\mu_{p}=\mu_{h}$. We write the average value of $\phi$ in the periodic state $\bar{\phi_p}=\bar{\phi}_h+C$, where $C$ is the difference between the average value of the order parameter in the two coexisting phases, implying that the coexistence condition is 
\begin{equation}
\mu_{p}(\bar{\phi}_h+C)-\mu_h(\bar{\phi}_h)=0, \label{eqcoexistence}
\end{equation}
or equivalently:
\begin{equation}
(r+q^4)C + 3\bar{\phi}_h^2C + \frac{3}{2}\bar{\phi}_h A^2 +  O(A^4,CA^2,C^2) = 0.	
\label{eqC}
\end{equation}
Since the amplitude $A$ of the modulated phase at coexistence is small when $|\Delta_h|\ll1$, where 
$\Delta_h\equiv r+3\bar{\phi}_h^2$, Eqs.~(\ref{eqB}) and (\ref{eqC}) yield, for $|\Delta_h|\ll1$, the expressions
\begin{equation}
B=-\frac{\bar{\phi}_hA^2}{6q^4}+O(\Delta_h A^2,A^4),\qquad C=-\frac{3\bar{\phi}_hA^2}{2q^4}+O(\Delta_h A^2,A^4).
\end{equation}
Equation~(\ref{eqA}) then yields
\begin{equation}
	A = 2\sqrt{-{\frac{\Delta_h}{3}}}\biggl(1-\frac{38\bar{\phi}_h^2}{3q^4}\biggr)^{-1/2}+O(\Delta_h).
	\label{eqOneModeAppAmp}
\end{equation}
Taking the limit $C\to 0$ now takes us to the tricritical point.
At the tricritical point $\Delta_h=\Delta_p=0$ and the chemical 
potentials $\mu_{p}(\bar{\phi}_h)$ and $\mu_h(\bar{\phi}_h)$ are identical. Thus from
Eq.\ \eqref{eqOneModeAppAmp} we see that the tricritical 
point occurs at $\bar{\phi}=\sqrt{3/38}q^2\approx 0.281q^2$, $r=-(9/38)q^4\approx -0.237q^2$.
These coordinates agree precisely with the result obtained from the common tangent construction 
between the free energy of the periodic phase and the free energy of the homogeneous phase 
to determine phase coexistence, and also with our numerical simulations of the VPFC model 
performed with $q=1$. Thus when $q=1$ the phase transition is of second order for $r>-9/38$ 
and of first order for $r<-9/38$. As already mentioned this result is {\it exact} in the sense 
that it is unchanged if more modes are included in the Ansatz (\ref{eqOneModeApp})
and improves on the prediction $r=-1/4$ obtained for the PFC model using a {\it one-mode} 
approximation \cite{ElGr04}. Note that the vacancy term \eqref{eqPFCVacTerm} does not affect 
the transition because $\phi$ is positive everywhere.

As discussed above the transition between the periodic and the uniform phases 
is of first order below the tricritical point. As $r$ decreases, this region of 
the phase diagram is increasingly affected by the vacancy term \eqref{eqPFCVacTerm}, 
as the amplitude of the structures becomes large enough to reach negative $\phi$ 
values. We observe that including the vacancy term \eqref{eqPFCVacTerm} decreases 
the distance between the coexistence curves.  This is because the vacancy term 
increases the free energy of the profiles in the periodic phase, which decreases 
the difference between the free energy of the periodic structures and the homogeneous 
state and hence a common tangent construction between the two yields values which 
are closer to the linear stability line $\Delta=0$. We calculate the coexistence 
values below the tricritical point by numerically solving for the order parameter 
profiles, because the two-mode approximation becomes inaccurate when $r < -0.3$,
where the order parameter profile develops regions where $\phi<0$ and so the vacancy
term makes a contribution to the free energy. Note that for the regular PFC model
(i.e.,when $H=0$) the two mode approximation for the free energy works very well,
agreeing to two significant figures or more with the exact free energy for $r\geq-0.9$.

To calculate the coexisting phases numerically we select a value of $r$ and determine the order parameter
profile along this line for different values of $\bar{\phi}$.  The free energy at each of these points is 
minimised with respect to the domain size, which effectively gives us the minimum free energy for the infinite 
system.  A polynomial is then fitted to these values to produce a continuous curve which gives the free energy 
of the periodic phase for the chosen value of $r$.  We then make the common tangent construction between 
the free energy of the periodic phase and the uniform phase to calculate the two coexisting $\bar{\phi}$ 
values at the chosen value of $r$.  This process is then repeated for different values of $r$.  The resulting 
coexistence curves are plotted as the solid red lines in Fig.~\ref{figPhaseD1D}(a).

The periodic structures which are formed by the VPFC model [Fig.~\ref{figPhaseD1D}(f)] are qualitatively 
very similar to the structures which can be found in the regular PFC model.  However, the amplitude of the
modulations is restricted by the large penalty in the free energy accumulated when $\phi < 0$.  Inside the 
coexistence region between the periodic and uniform states, we observe interesting structures where the 
amplitude of the peaks does not remain constant and a second length scale is visible in the structures 
[Fig.~\ref{figPhaseD1D}(g)]. This is also an effect which is present in the regular PFC model and will be 
discussed in detail in future work.  What is most intriguing, and is perhaps the most appealing aspect of the 
VPFC model, is the appearance of localised states for small positive values of $\bar{\phi}$ when the 
magnitude of $r$ is sufficiently large ($r \lsim -0.6$).  We obtain order parameter profiles by numerically integrating 
forward in time Eqs.~\eqref{eqPFCDyn} and \eqref{eqVPFCOne} until a stationary solution is 
reached, starting from the initial profile $\phi(\vx,t=0) = \bar{\phi} + \mathcal{X}(\vx)$, where $\mathcal{X}$ is a 
small amplitude random noise profile with zero mean. A rich variety of different patterns is observed, 
including periodic structures mixed with almost flat regions [Fig.~\ref{figPhaseD1D}(d)] and individual 
isolated peaks [Figs.~\ref{figPhaseD1D}(b) and (c)].  In Fig.~\ref{figPhaseD1D}(a) the green dot-dashed 
curves indicate the boundary of the region where one observes regular periodic structures and where the 
localised structures are formed.  Note that these are guidelines only and are not thermodynamic coexistence 
curves.  The lower-left dot-dash curve roughly denotes the linear stability limit of the regular periodic 
structures, such as that in Fig.~\ref{figPhaseD1D}(f).  This is determined numerically. We begin with a periodic 
profile and reduce the value of $\bar{\phi}$ gradually, minimising the free energy at each step, while keeping 
$r$ constant.  The limit point is then defined as the value of $\bar{\phi}$ where the periodic profile becomes 
linearly unstable and a vacancy is introduced.  In a similar way, we determine the upper-right dot-dash line, 
which is the limit of linear stability of the structures with defects.  This is found by starting with a profile 
containing a single vacancy and increasing $\bar{\phi}$ until the vacancy disappears.  These two points are 
calculated for different values of $r$ and then a best fit to this data is shown in Fig.~\ref{figPhaseD1D}(a).  
There is some hysteresis in the region between these two curves, with the type of profile produced depending 
heavily upon the initial conditions.

Within the localised state region of the phase diagram it is possible to obtain order parameter profiles with a 
varying number of peaks for a given system of length $L$.  Keeping $r \lsim -0.6$ constant and varying 
$\bar{\phi}$ allows us to control the number (density) of bumps as shown in Figs.~\ref{figPhaseD1D}(b)--(e).  
Beginning with $\bar{\phi} = 0$ we find isolated peaks in large vacant areas (where $\phi$ is approximately 
uniform with $\phi \lsim 0$).  As $\bar{\phi}$ is increased the number of peaks increases until we return to the 
familiar regular periodic structures.  The assumption of Ref.~\cite{CGD09} is that unlike in the regular PFC, 
where the uniform phase is associated with the liquid and the modulated phase with the crystal, in the VPFC 
model one may associate each bump in $\phi(\vx)$ as corresponding to a particle and so the model can 
describe fluids [Figs.~\ref{figPhaseD1D}(c) and (d)], crystals with vacancies and defects 
[Fig.~\ref{figPhaseD1D}(e)] and regular crystals [Fig.~\ref{figPhaseD1D}(f)].

\begin{figure}
	\textbf{(a)}\includegraphics[width=0.45\hsize]{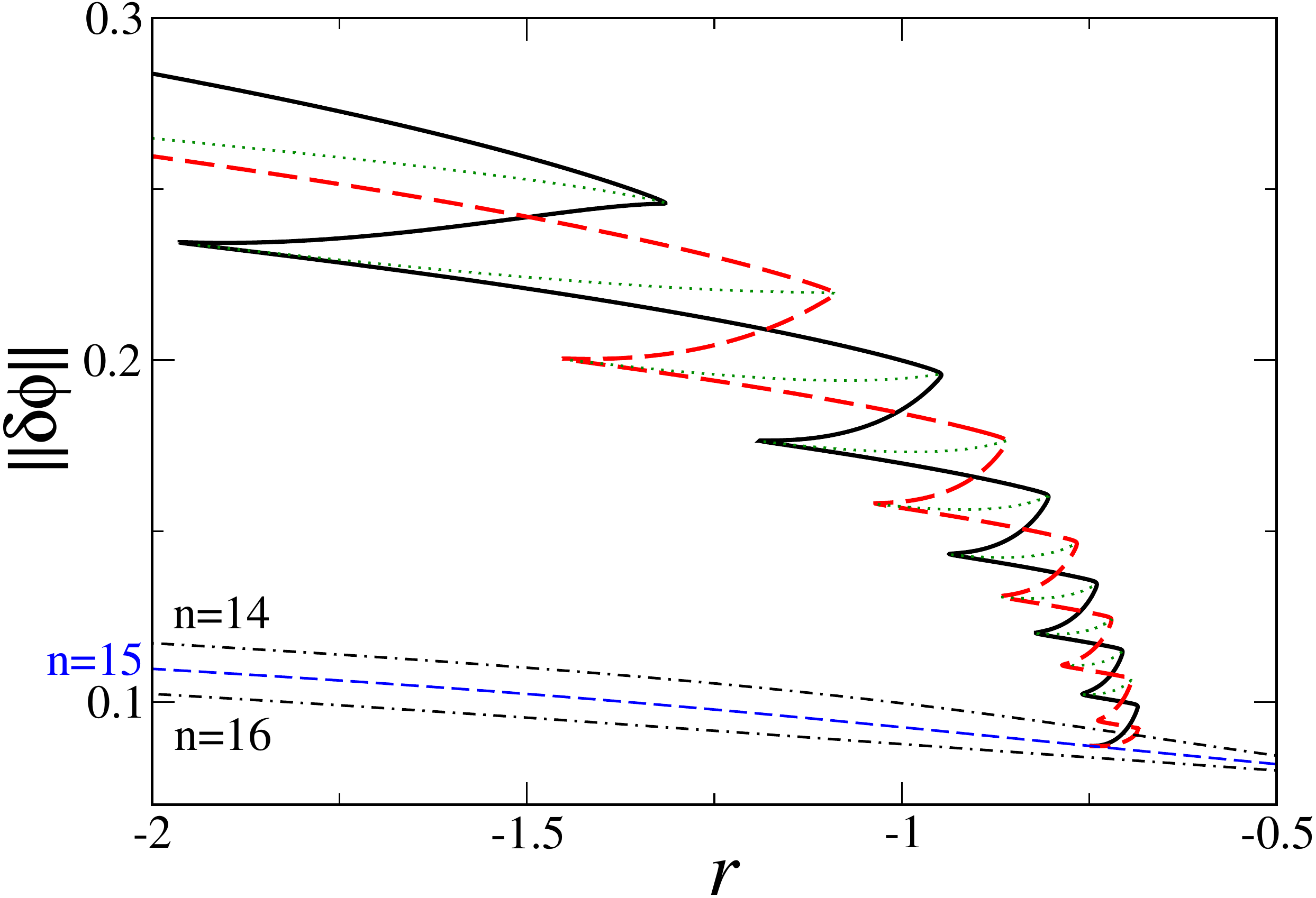}
	\includegraphics[width=0.45\hsize]{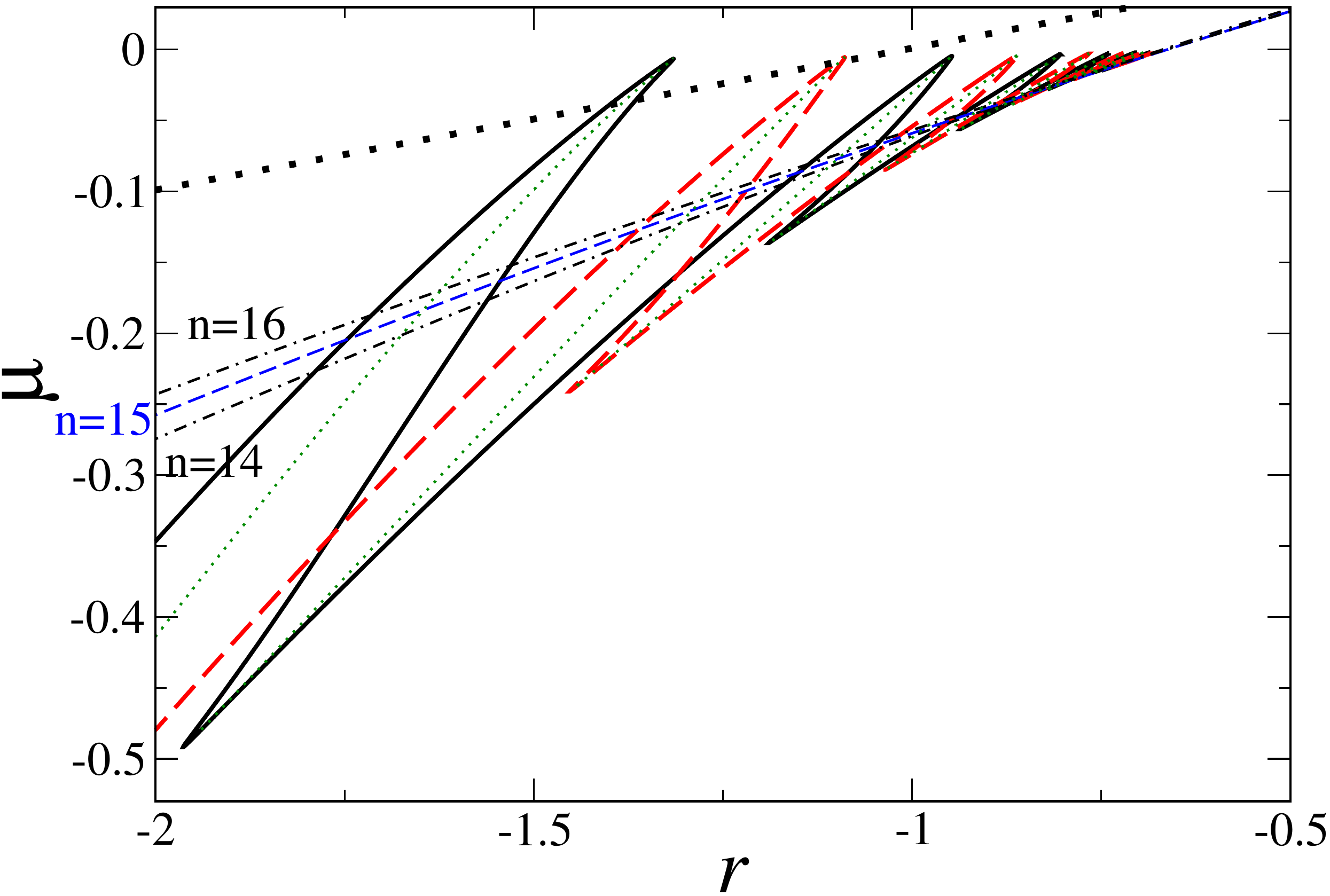}\textbf{(b)}\\
	\textbf{(c)}\includegraphics[width=0.45\hsize]{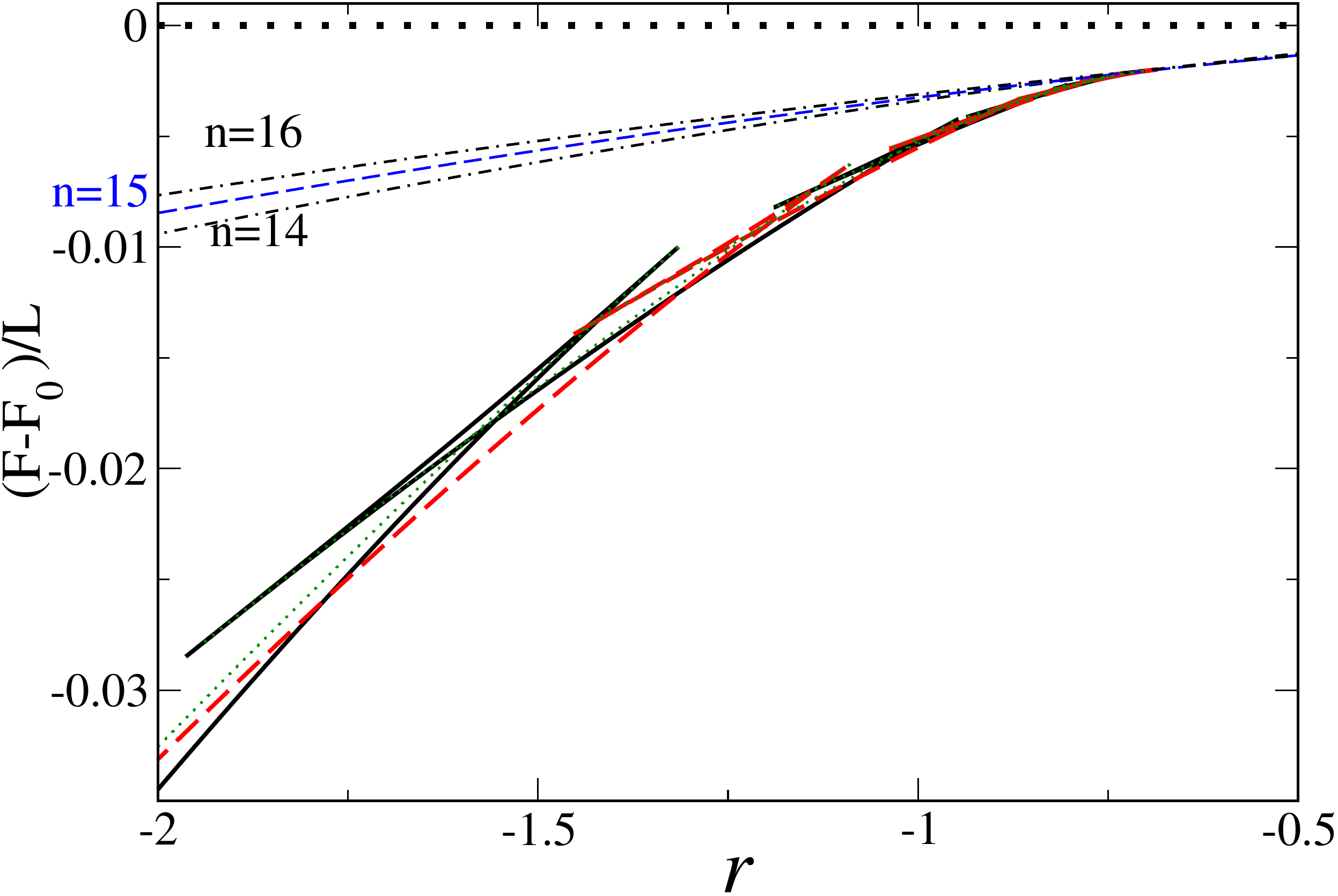}
	\includegraphics[width=0.45\hsize]{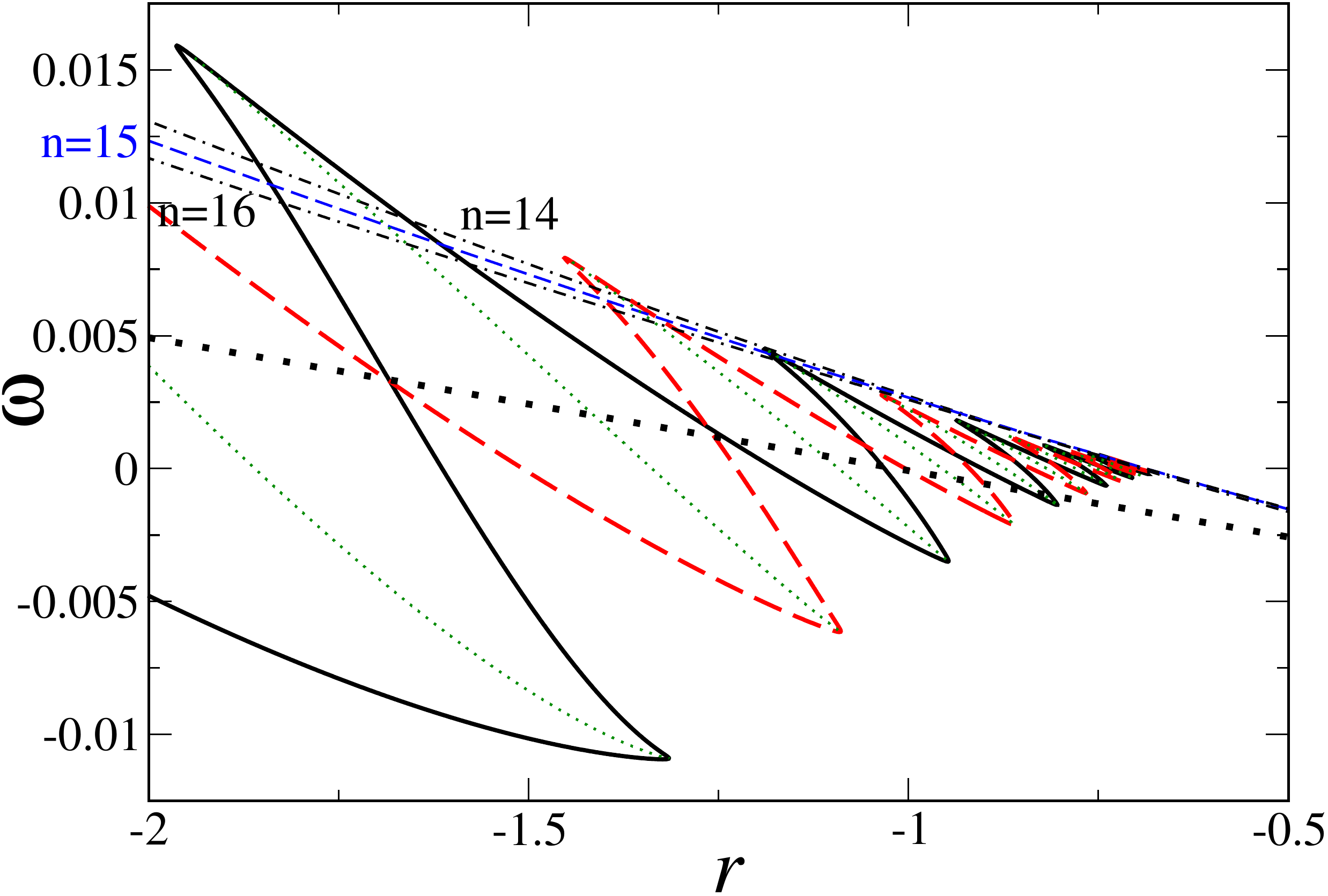}\textbf{(d)}\\
	\caption{(color online) Bifurcation diagram showing localized solutions of the VPFC (an augmented 
		      conserved Swift-Hohenberg equation) [Eqs.~(\ref{eqPFCDyn}) and (\ref{eqPFCVacTerm})] 
		      with $H=1500$, as a function of the parameter $r$, for the mean order parameter 
		      $\bar{\phi}=0.1$ and a fixed domain size of $L=100$. The various solution profiles are 
		      characterised by their (a) $L^2$ norm, (b) chemical potential $\mu$, (c) mean free energy 
		      density $(F-F_0)/L$, and (d) mean grand potential density $\omega \equiv F/L-\bar{\phi}\mu$.
		     The heavy black dash-dotted line corresponds to the homogeneous solution $\phi(x)=\bar{\phi}$. 
		     Periodic solutions with $n=15$ bumps are shown as a thin blue dashed line, whereas the 
		     nearby thin black dotted lines represent the $n=14$ and $n=16$ solutions as indicated in 
		     the plot.  The heavy solid black and dashed red lines that bifurcate from the $n=15$ periodic 
		     solution represent symmetric localized states with a maximum (odd states) and a minimum 
		     (even states) at the center, respectively.  The green dotted lines that connect the two branches 
		     of symmetric localized states correspond to asymmetric localized states. Together the branches 
		     of localized states form a slanted snakes-and-ladders structure.
	}
	\label{fig:loc-fam-phi01}
\end{figure}

\begin{figure}
	\textbf{(a)}\includegraphics[width=0.45\hsize]{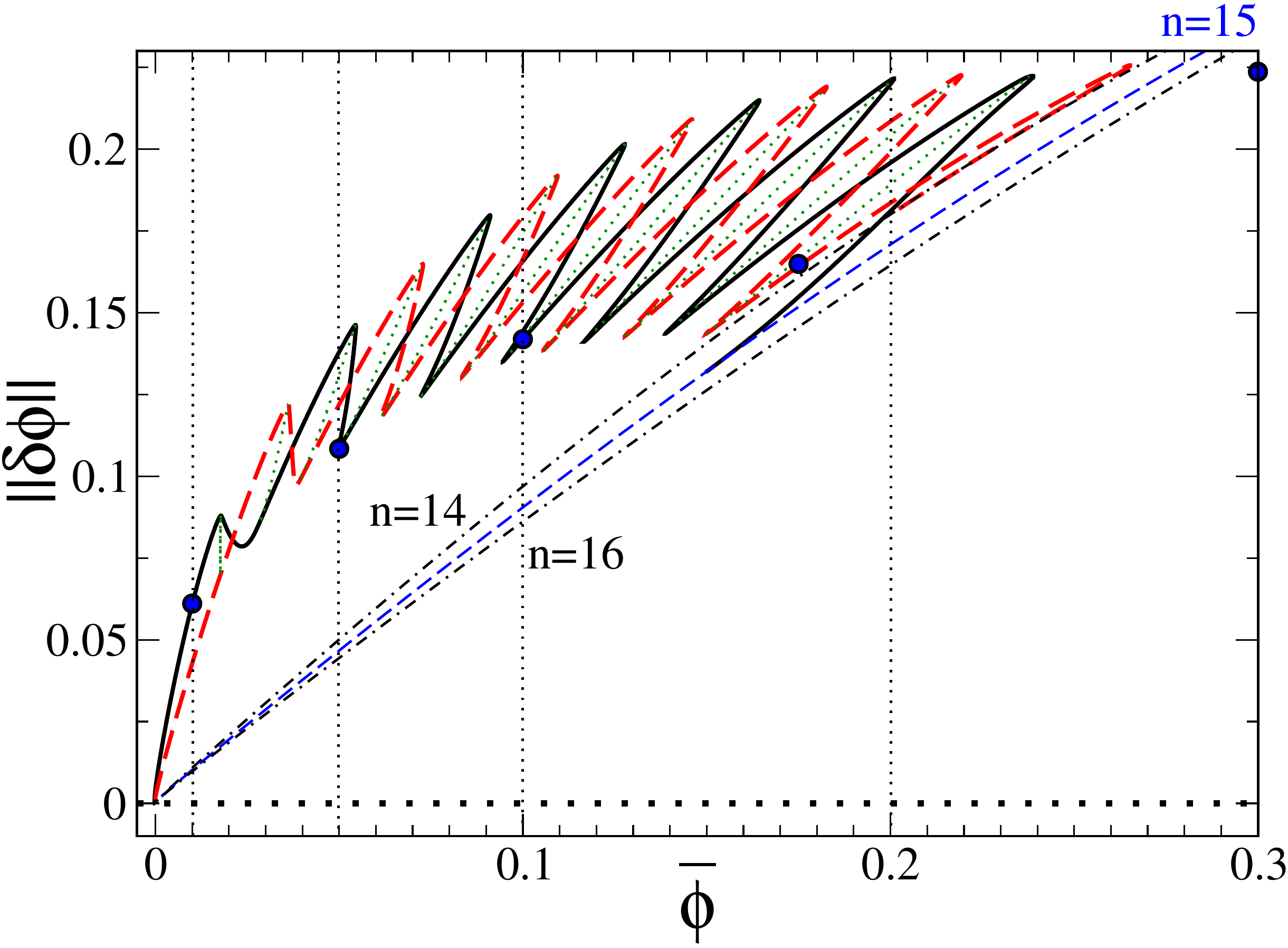}
	\includegraphics[width=0.45\hsize]{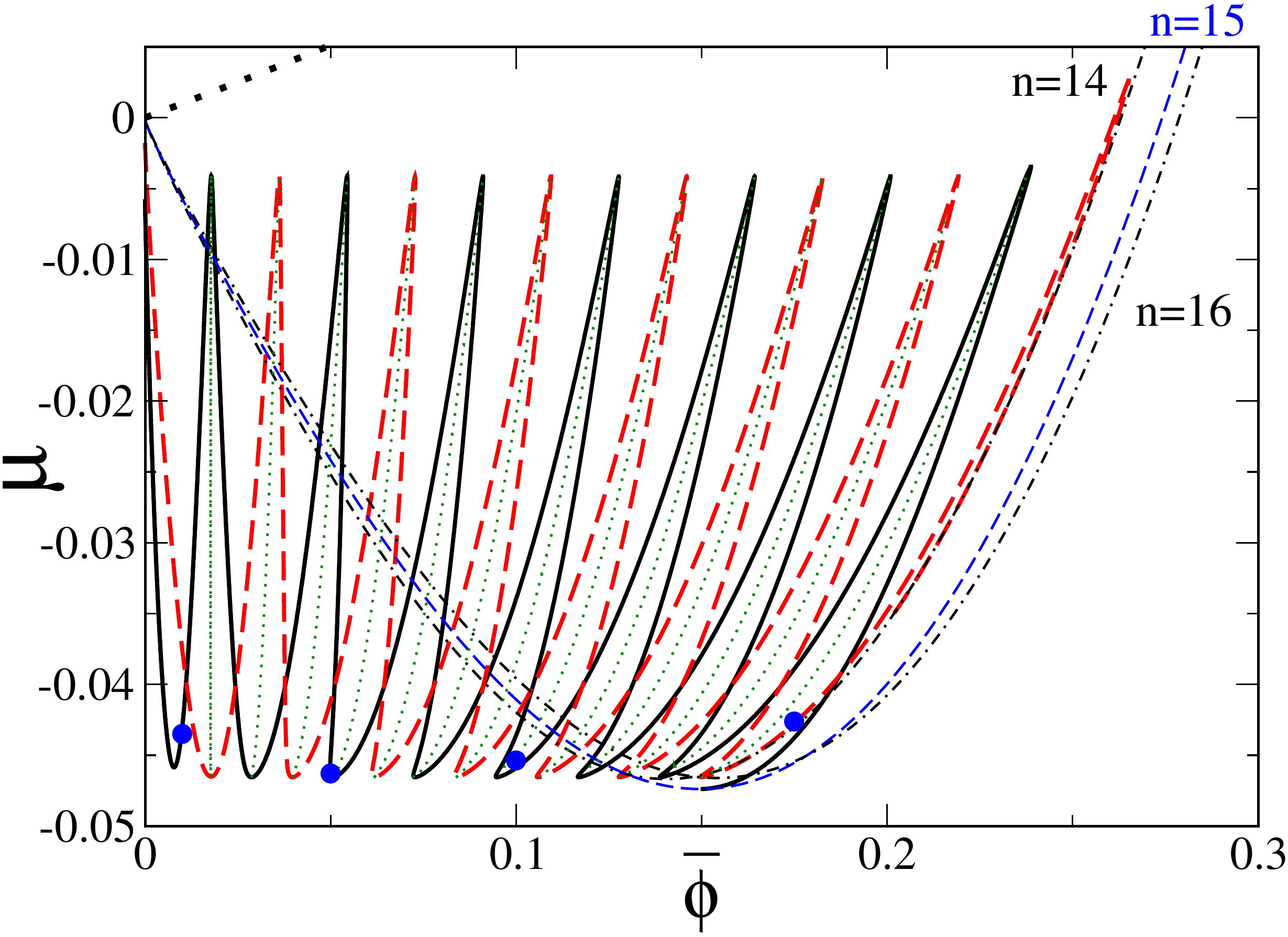}\textbf{(b)}\\
	\textbf{(c)}\includegraphics[width=0.45\hsize]{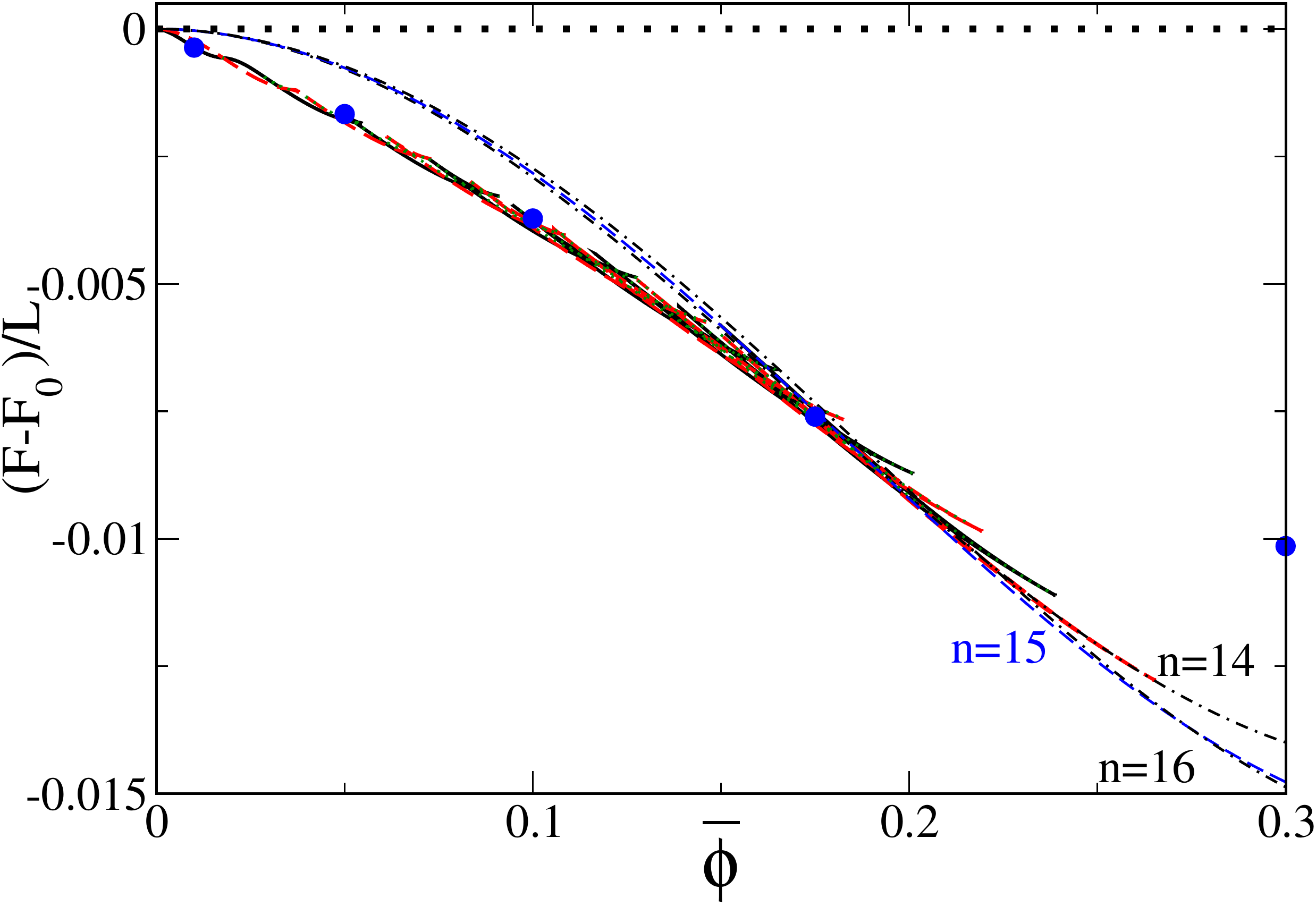}
	\includegraphics[width=0.45\hsize]{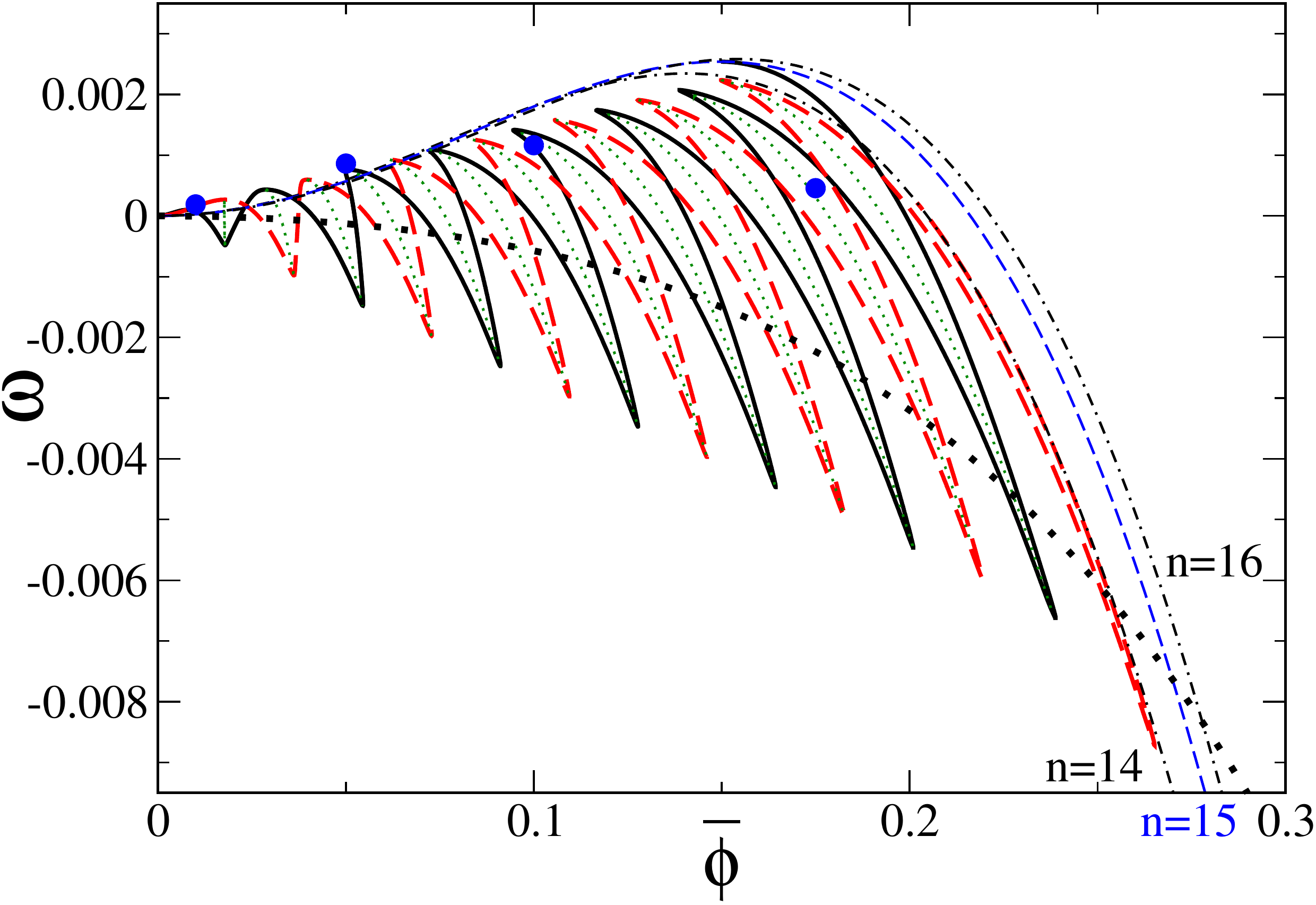}\textbf{(d)}\\
	\caption{(color online) Bifurcation diagram showing localized solutions of the VPFC as a function 
		      of the mean order parameter $\bar{\phi}$, for $r=-0.9$ and a fixed domain size of $L=100$.  The 
		      various solution profiles are characterised by their (a) $L^2$ norm, (b) chemical potential $\mu$, 
		      (c) mean free energy density $(F-F_0)/L$, and (d) mean grand potential density 
		      $\omega \equiv F/L-\bar{\phi}\mu$.
		     The line styles are as in Fig.~\ref{fig:loc-fam-phi01}. Here, however, the heavy solid 
		     black and dashed red lines bifurcate at large $\bar{\phi}$ from the $n=15$ and $n=14$ 
		     periodic solutions, respectively. Typical profiles for all the branches of localized states 
		     are given in Fig.~\ref{fig:loc-prof-rm09}.  The vertical dotted lines in (a) correspond to values 
		     of $\bar{\phi}$ for results in Fig.~\ref{fig:loc-prof-rm09}.  The blue dots correspond to the five
		     time simulation profiles shown in Fig.~\ref{figPhaseD1D}(b) - (f).} 
	\label{fig:loc-fam-rm09}
\end{figure}

The findings presented in Fig.~\ref{figPhaseD1D} indicate the existence of a hysteretic transition 
between periodic and localised states, and are a consequence of \textit{homoclinic snaking} 
\cite{BuKn06,BuKn07,BuKn07b,Knob08} in the present system. In the standard homoclinic scenario such
localised states are present within a part of the coexistence region called the pinning region.
The localised states in the lower left part of the parameter plane $(\bar{\phi},r)$ in Fig.~\ref{figPhaseD1D}(a)
correspond to the {\it global} energy minimum or to other deep but local energy minima. Families of such 
steady state solutions can be obtained for the VPFC model that we study here [Eq.~(\ref{eqPFCDyn}) with 
Eqs.~(\ref{eqPFCOne}), (\ref{eqVPFCOne}) and (\ref{eqPFCVacTerm})] by employing the path continuation 
techniques bundled in the package AUTO07p \cite{DPC01}. As an example, in Figs.~\ref{fig:loc-fam-phi01} 
and \ref{fig:loc-fam-rm09} we show the characteristics of localised solutions along cuts through the plane 
$(\bar{\phi},r)$. In particular, Figs.~\ref{fig:loc-fam-phi01} and~\ref{fig:loc-fam-rm09} give results for 
changing $r$ (at constant $\bar{\phi}=0.1$) and changing $\bar{\phi}$ (at constant $r=-0.9$), respectively. 
All solutions are characterised by their $L^2$ norm 
$||\delta \phi||\equiv\sqrt{(1/L)\int_0^{L}(\phi(x)-\bar{\phi})^2dx}$, chemical potential $\mu=\delta F/\delta\phi$, 
mean free energy density difference $(F[\phi(x)]-F_0)/L$, where $F_0=F[\bar{\phi}]$ and mean grand potential 
density $\omega \equiv F[\phi(x)]/L -\bar{\phi}\mu$, and satisfy periodic boundary conditions on the domain $0\le x\le L$.

It turns out that there exist three types of localised steady states: (i) the heavy solid 
black line consists of $x\to-x$ symmetric localized states that have a maximum at the 
centre, i.e., the overall number of bumps within the structure is odd.  (ii) The dashed 
red line also represents $x\to-x$ symmetric localized states but this time with a hole 
(minimum) at the centre. (iii) The localized solutions of the third type are not symmetric 
under $x\to-x$ and are called ``asymmetric states''.  These reside on branches that connect 
(via pitchfork bifurcations) the two branches of symmetric localized states.  These
branches are included in the bifurcation diagrams as dotted green lines. 

\begin{figure}
	\includegraphics[width=0.9\hsize]{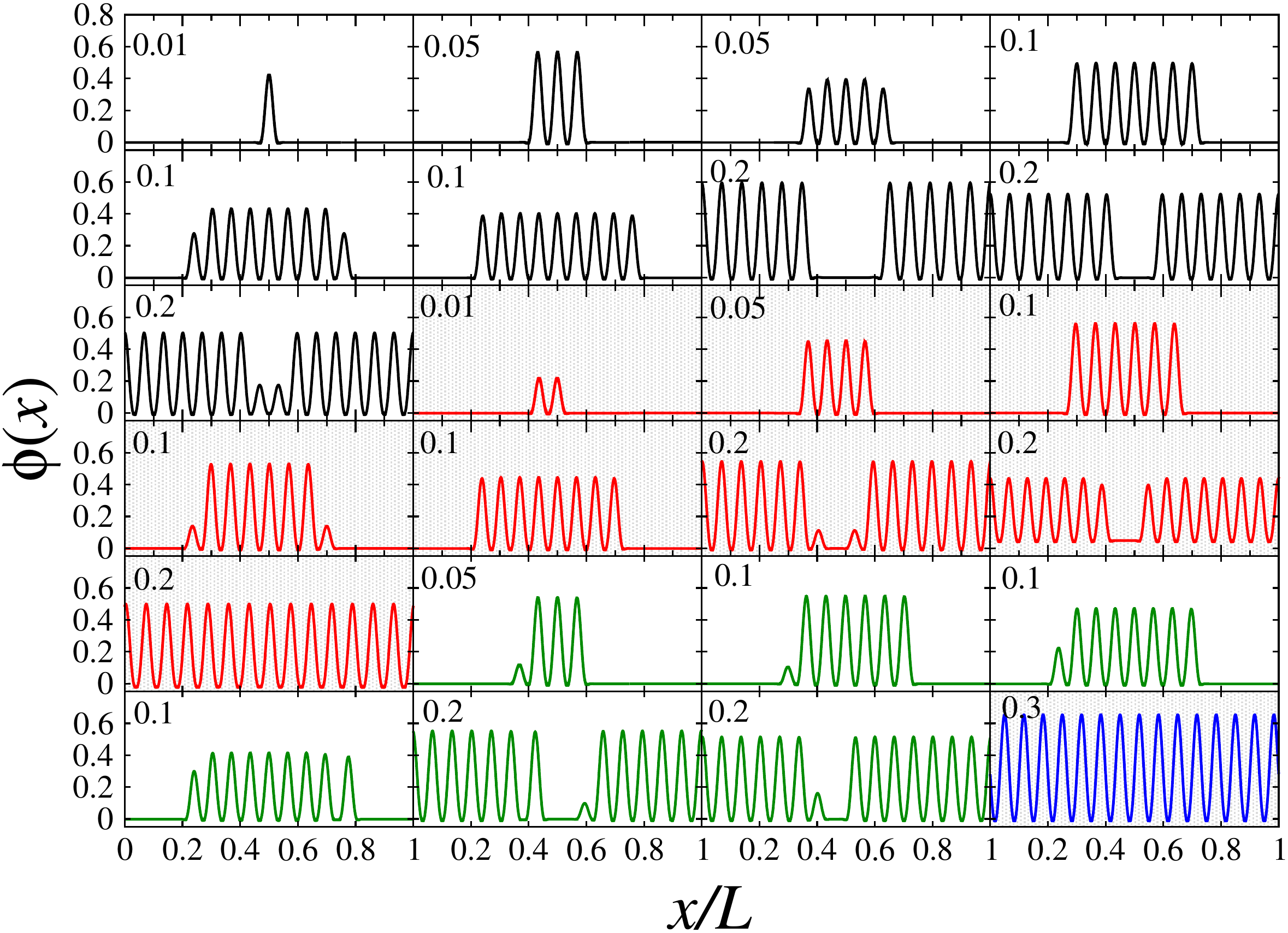}
	\caption{A selection of steady state profiles $\phi(x)$ for $r=-0.9$ at $\bar{\phi}=0.01$, $0.05$, $0.1$, and $0.2$.  From top left to bottom right we show first nine type (i) solutions, i.e., symmetric localised states with an odd number of maxima (in black), then eight type (ii) solutions, i.e., symmetric localised states with an even number of maxima (in red), followed by six type (iii) solutions, i.e., asymmetric localised states (in green). The final image is the $n=15$ periodic solution at $\bar{\phi}=0.3$ (in blue). The number in each panel indicates the corresponding value of $\bar{\phi}$. The solutions from the symmetric branches are shown in the sequence that follows the respective branch in Fig.~\ref{fig:loc-fam-rm09}(a), starting from the left. The asymmetric states for identical $\bar{\phi}$ are shown in the order of decreasing norm.} 
	\label{fig:loc-prof-rm09}
\end{figure}

Examples of order parameter profiles of types (i)--(iii) are presented in Fig.~\ref{fig:loc-prof-rm09}, 
corresponding to the various solution branches displayed in Fig.~\ref{fig:loc-fam-rm09}. This sequence of
profiles expands upon the few examples shown in Figs.~\ref{figPhaseD1D}(b)--(g).  Recall, however, that the
results in Figs.~\ref{figPhaseD1D}(b)--(g) are obtained starting from an order parameter profile with a small 
amplitude random noise and so they do not always exactly agree with the steady states at the same 
$\bar{\phi}$ resulting from the path continuation.  The $L^2$ norm, chemical potential $\mu$, mean free 
energy density $(F - F_0)/L$ and the mean grand potential $F/L - \bar{\phi}\mu$ have been calculated for 
the profiles obtained from time simulations [Figs.~\ref{figPhaseD1D}(b)--(f)] and are plotted as blue dots in 
Fig.~\ref{fig:loc-fam-rm09}.  A close inspection reveals that the energy of the time simulation results is 
often slightly higher than that from the continuation results, indicating that in these cases the time simulation 
converges to a local and not the global energy minimum. This is to be expected as the solutions shown in the 
bifurcation diagrams are only the `tip of the iceberg'. For instance, there exist many more solutions, where not 
all the inner distances between the bumps are identical. This is related to the fact that individual bumps have 
oscillatory tails and the `locking of these tails' allows for different equilibrium distances \cite{BK09}. The solutions presented 
in Figs.~\ref{fig:loc-fam-phi01} and \ref{fig:loc-fam-rm09} represent the solution having the lowest energy in 
the respective class. However, the energy differences between these and the `less symmetric' solutions are 
often tiny. Thus, it is not surprising that time simulations starting from random initial profiles often converge to 
solutions with greater disorder and energies than those shown in Fig.~\ref{fig:loc-fam-rm09}. For 
instance, the solution in Fig.~\ref{figPhaseD1D}(c) at $\bar{\phi}=0.1$ is a nine-bump solution similar to the 
odd symmetric localised states shown in the first two panels of the second row of Fig.~\ref{fig:loc-prof-rm09}. 
The amplitudes agree well and although the arrangements of the nine bumps are different, the free energy 
and norm still agree to $<1$\%.  However, at large average order parameter values $\bar{\phi}$ the time 
simulation results can converge to metastable states with energies quite different from the minimum energy states 
for domains of this size $L = 100$.  For example, the periodic solution obtained 
from the time simulation (shown in Fig.~\ref{figPhaseD1D}(f)) when $\bar{\phi} = 0.3$ has eighteen bumps.  
However, from Fig.~\ref{fig:loc-fam-rm09}(c) we observe that the energetic minimum is obtained by a periodic 
profile with fifteen bumps, as shown by the steady state solution in Fig.~\ref{fig:loc-prof-rm09}.  The 
convergence to a different number of bumps in the time simulation may be caused by discretisation effects or 
by the initial noise profile used.  As one would expect, the free energy associated with the eighteen bump 
periodic structure is significantly larger than the fifteen bumped profile.

In Fig.~\ref{fig:loc-fam-phi01} ($\bar{\phi}=0.1$) the localised states bifurcate subcritically from the periodic 
solution branch (that itself emerges from the trivial homogeneous solution that is displayed as the heavy black 
dotted line). Therefore, one expects hysteretic behaviour as encountered in the time simulations.  A 
magnification (not shown) allows us to determine the threshold values for the hysteretic transition. When 
decreasing $r$ in the region where periodic solutions are always found, one first passes 
$r_\mathrm{sn}=-0.685$ where the last 2 branches of localised solutions annihilate in a saddle-node
bifurcation [Fig.~\ref{fig:loc-fam-phi01}(a)]. Slightly below $r_\mathrm{sn}$, both the periodic solution and the localised state with a single 
bump are local energetic minima. Although the periodic solution represents the global minimum, particular
time simulations sometimes converge to the localised state. The differences in energy between the two is 
$< 1$\% in the case of Fig.~\ref{fig:loc-fam-phi01}. When $r$ is further decreased below 
$r_\mathrm{en}=-0.700$ the energy of the even symmetric states becomes smaller than the one of the 
$n=16$ periodic solution, that is however still linearly stable.  The situation changes at $r_\mathrm{c}=-0.749$ 
where both symmetric localised branches bifurcate from the $n=16$ branch, i.e., below $r_\mathrm{c}$ the 
latter is linearly unstable. Furthermore, below $r_\mathrm{c}$ the energy of \textit{all} localised states rapidly 
becomes much smaller than the energy of \textit{all} periodic states [Fig.~\ref{fig:loc-fam-phi01}(c)]. The 
hysteresis range displayed in Fig.~\ref{figPhaseD1D} provides a good approximation for the region between 
$r_\mathrm{c}$ and $r_\mathrm{sn}$. This region becomes larger as $\bar{\phi}$ is increased. 

The situation is very similar when $\bar{\phi}$ is changed for fixed $r$ (Fig.~\ref{fig:loc-fam-rm09}). The 
resulting hysteresis range is between $\bar{\phi}=0.150$ and $0.239$ for symmetric localised states with an odd number of maxima and between $\bar{\phi}=0.202$ and $0.265$ for symmetric states with an even number of maxima. Overall, one should therefore expect a wide hysteresis region roughly between $\bar{\phi}=0.15$ and $\bar{\phi}=0.25$. The hysteresis range obtained from the time simulations (indicated in Fig.~\ref{figPhaseD1D}) is roughly $0.19 < \bar{\phi} < 0.22$. This is narrower than the range deduced from the path continuation analysis of the localised steady states, but lies right in the middle of it.

Before we move on to discuss the two-dimensional case, we should comment on how our results fit into the 
wider context of research on localised states.  Much research on localised states focuses on the 
non-conserved Swift-Hohenberg equation \cite{BuKn06,BuKn07,BuKn07b}. There, such states can only exist 
if the primary bifurcation of periodic states from the homogeneous base state is subcritical.  The localised 
states exist in a sub-range of the existence range of the periodic states bounded on either side by the 
saddle-node bifurcations of the branches of symmetric localised states. In the non-conserved Swift-Hohenberg 
equation these accumulate exponentially rapidly towards the parameter values corresponding to first and
last tangencies between the unstable manifold of the homogeneous state in space and the stable manifold
of the periodic state. These tangencies define the pinning region containing the different localised
structures. In contrast, in the presence of a conserved quantity localised states may exist outside the 
existence region of periodic states, may occur even in the supercritical case and the saddle-node 
bifurcations of the localised states are no longer aligned, i.e., one finds slanted snaking~\cite{Dawe08}.
This is typically a finite size effect \cite{LBK11}. 


For the regular PFC (conserved Swift-Hohenberg equation) [Eq.~(\ref{eqPFCDyn}) with 
Eqs.~(\ref{eqPFCFree}) and (\ref{eqPFCOne})], localised states are briefly mentioned in Ref.~\cite{MaCo00}.  
However, no systematic results along the lines of those presented in Refs.~\cite{BuKn06} for non-conserved or 
\cite{Dawe08} for conserved order parameter fields are available. The model used here is a special case 
because it includes the non-analytic vacancy term (\ref{eqPFCVacTerm}). However, a similar bifurcation 
structure is found for the classical conserved Swift-Hohenberg equation i.e.,~the regular PFC model 
\cite{ARTK12}.

\subsection{Two dimensional model}

\begin{figure}[t] 
	\begin{minipage}[h]{0.67\linewidth}
		\includegraphics[width=\linewidth]{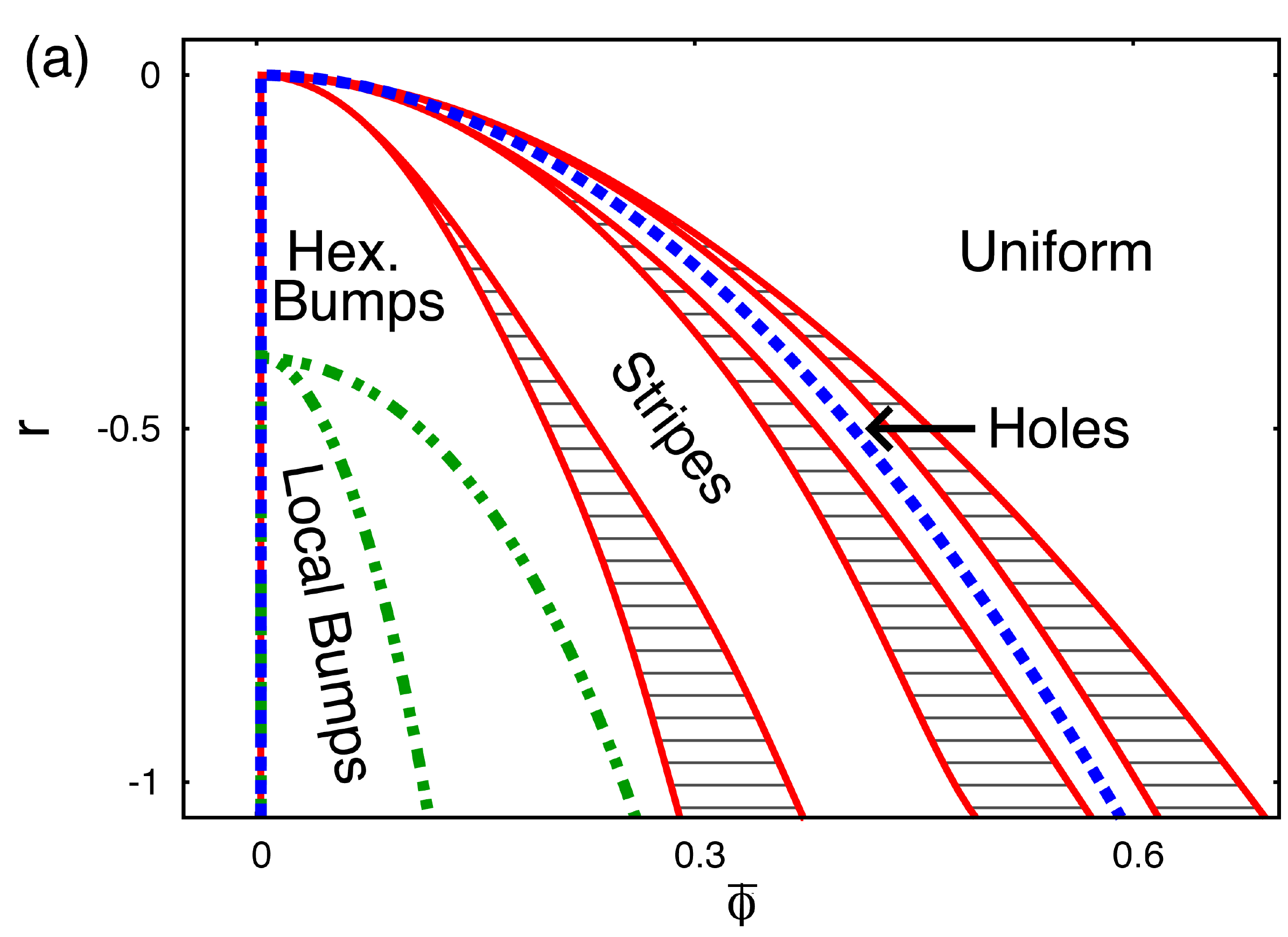}
	\end{minipage}
	\begin{minipage}[h]{0.3\linewidth}
		\includegraphics[width=\linewidth]{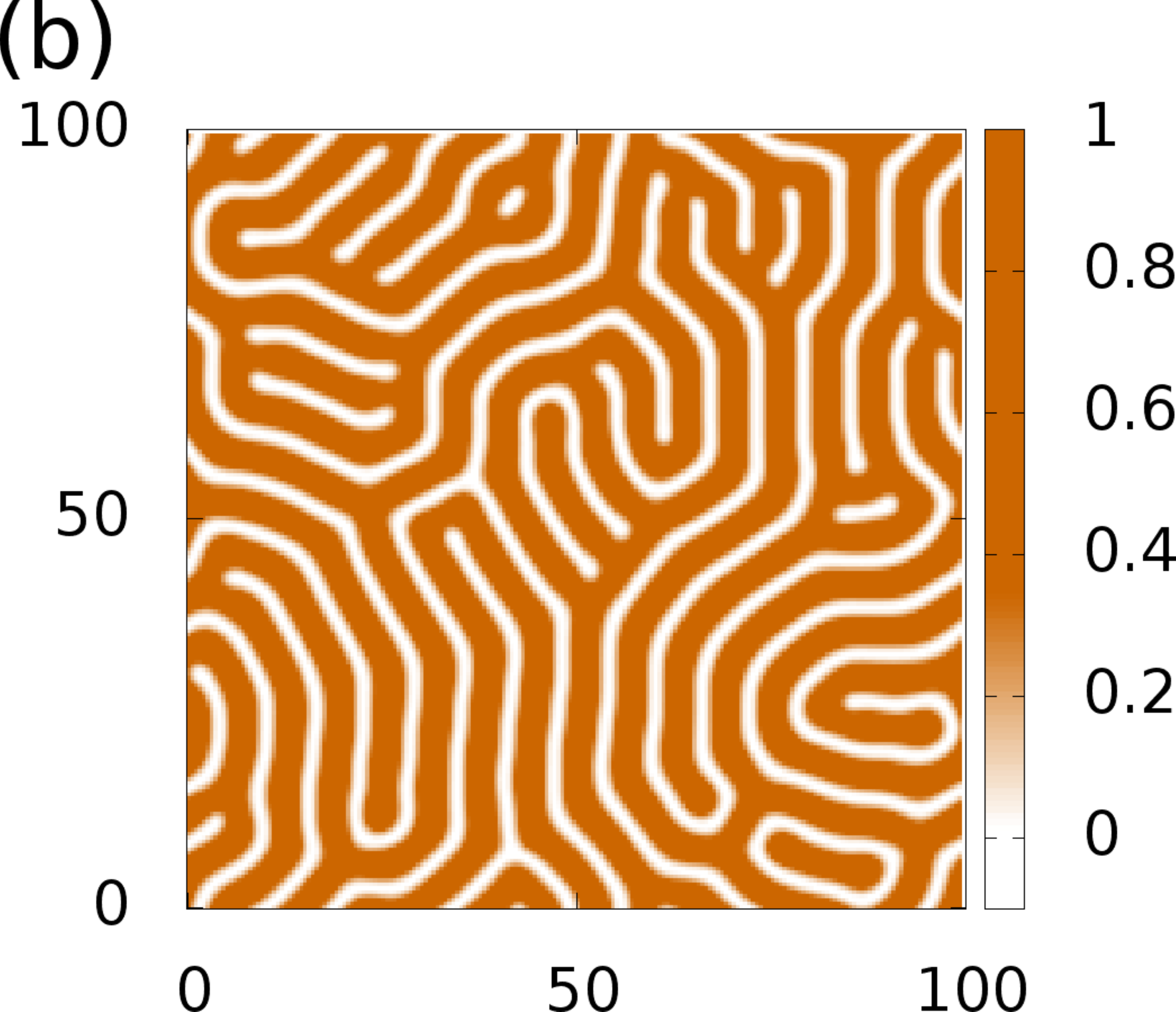}\\
		\includegraphics[width=\linewidth]{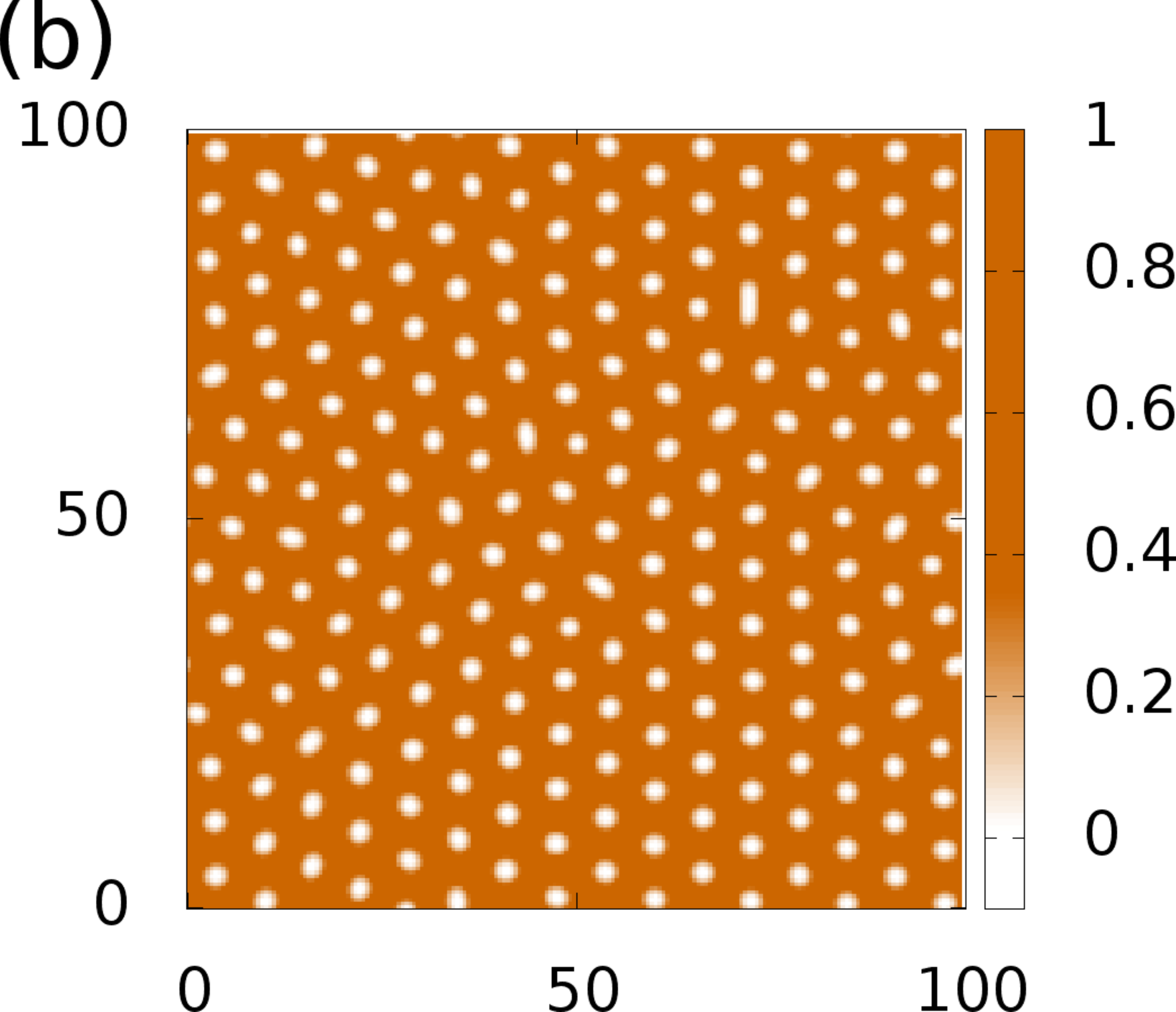}
	\end{minipage}
	\caption{The phase diagram for the 2D VPFC model (Eqs.~\eqref{eqPFCDyn} 
	and \eqref{eqVPFCOne}) is displayed in (a) for the case $q = 1$.  The red solid lines are the 
	coexistence curves between the various phases.  The blue dashed line is the limit of linear 
	stability for uniform profiles $\Delta = 0$.  The green dash-dotted lines indicate
	the region where localised and hexagonally ordered bump structures coexist. 
	Simulations of (b) stripes and (c) hexagonally ordered holes are also shown.
	The parameter values for these simulations are: $q = 1$, $r = -0.9$, $\alpha = 1$ and (b) 
	$\bar{\phi} = 0.4$ and (c) $\bar{\phi} = 0.53$.}
	\label{figPhaseD2D}
\end{figure}

We now move on to consider how the VPFC model behaves in two dimensions.  As with the regular 
PFC model \cite{ElGr04}, when we expand into two dimensions we observe stripes 
[see Fig.~\ref{figPhaseD2D}(b)] and hexagonally ordered bumps or holes 
[see Fig.~\ref{figPhaseD2D}(c)].  In Fig.~\ref{figPhaseD2D}(a) we display the phase diagram of the 
VPFC model in two dimensions and typical time simulation results from the striped \ref{figPhaseD2D}(b) 
and hole \ref{figPhaseD2D}(c) phases, calculated on a regular grid with grid spacing $dx = 0.5$.  Square 
ordering of bumps or holes does not appear in the phase diagram because these structures always have a 
higher free energy.  However, this can be changed through appropriate alterations to the free energy 
\cite{WPV10}.  Square ordering can also occur when extending to a two-component mixture 
(cf.~Sec.~\ref{ssOrdering} below).  Using the same method as outlined above, we calculate the regions of the 
phase diagram where there is coexistence between hexagonally ordered holes and the uniform distribution, 
between holes and stripes and between stripes and hexagonally ordered bumps.  The vacancy term 
\eqref{eqPFCVacTerm} shifts the modulated phases into the positive $\bar{\phi} > 0$ plane.  The section of the 
phase diagram where holes are observed is much smaller when compared to the regular PFC model and now 
extends beyond the limit of linear stability of the flat state (at $\Delta = 0$).  This means that for certain values 
of $\bar{\phi}$ (where $0 < \Delta \ll 1$), hexagonally arranged holes are energetically favourable but are only 
observed in time simulations for certain initial conditions -- i.e.,~when starting with an order parameter profile 
$\phi(\vx, t=0)$ which already has modulations which are sufficiently large in amplitude.  As $r$ is decreased 
(i.e.,~for larger $|r|$) it becomes increasingly difficult to obtain structures with holes up to and inside of the 
coexistence region between the hole and the uniform phases.  This is a direct consequence of the limit of 
linear stability occurring in the middle of the hole phase.  Therefore, the accuracy of results for the coexistence 
region between the hole and uniform phases decreases as $|r|$ becomes larger.  The stripe phase occurs in 
between the two hexagonal phases.  In the simulation order parameter profiles displayed in 
Fig.~\ref{figPhaseD2D}(b) and (c) we observe various defects and in (c) `grain' boundaries between regions 
with different orientations, which depend on the initial conditions (our initial profile was a flat state with 
additional small amplitude white noise).  The true minimum profile for case (b) is a series of parallel stripes 
which are identical to the periodic profiles in the 1D system (shown in Fig.~\ref{figPhaseD1D}(f)).

The most important portion of the phase diagram from the materials modelling point of view, is the 
bump phase because the basic assumption is that each bump represents a particle.  When $r \gsim -0.4$ or 
when $\bar{\phi}$ has a value close to that in the coexistence region between bumps 
and stripes, we observe hexagonally arranged bumps, similar to those in the regular PFC model.  
However, in a similar manner to the 1D system, we observe localised structures at small values of 
$\bar{\phi}$ when $r \lsim -0.4$.  In the phase diagram \ref{figPhaseD2D}(a) the green dot-dashed 
lines are numerically obtained estimates for the location in the phase diagram of the limits of linear stability of 
the uniform periodic states (lower curve) and the localised (vacancy) states (upper curve).  They are 
determined in the same manner as discussed above for the one dimensional system for a square system of 
side length $L = 25$.  It is important to note that the parameter range where localised bumps coexist with
regular periodically ordered bumps is much broader for the 2D system, implying a large amount of hysteresis.

\begin{figure}
	\includegraphics[width=0.31\linewidth]{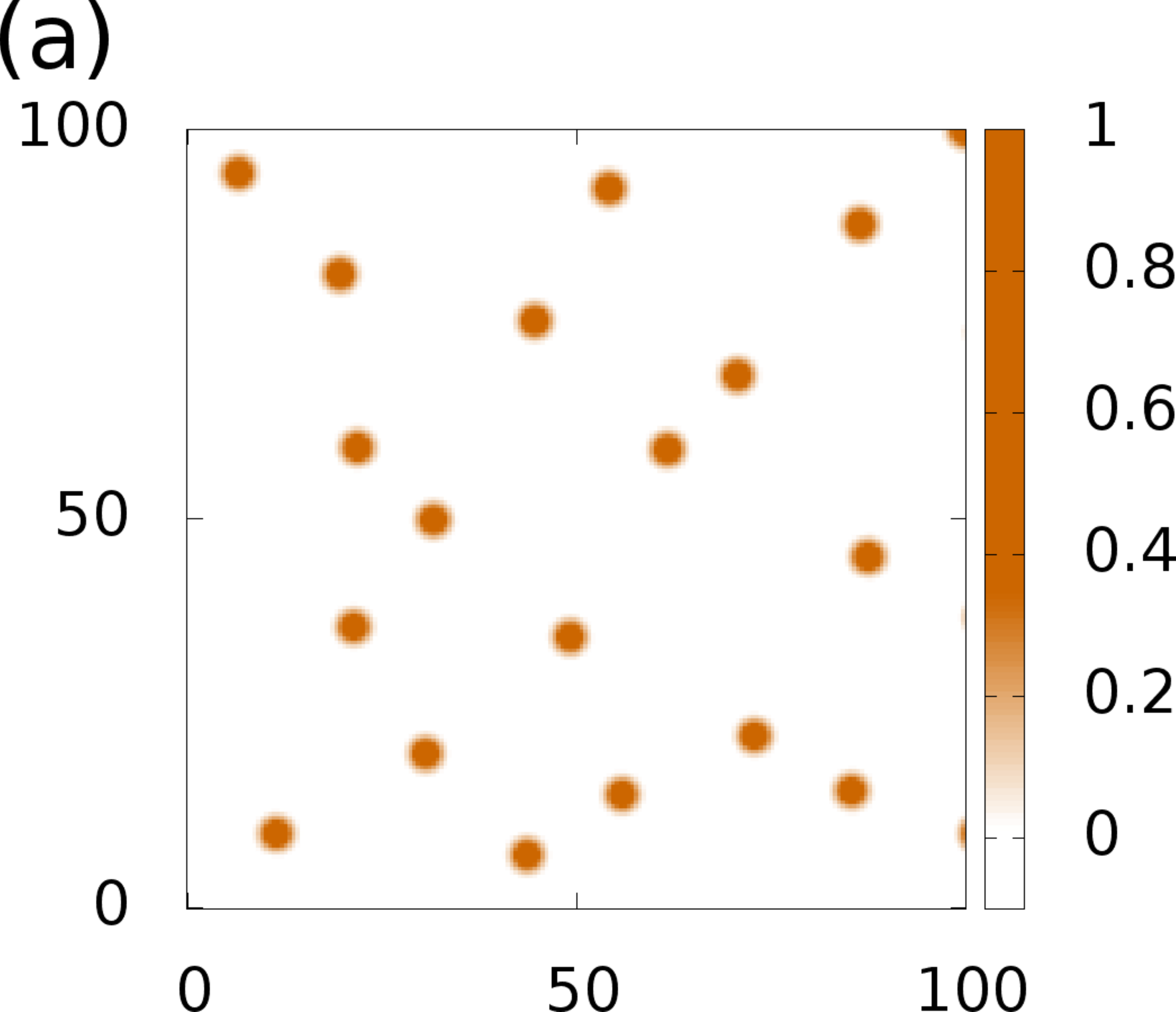} \hspace{1mm}
	\includegraphics[width=0.31\linewidth]{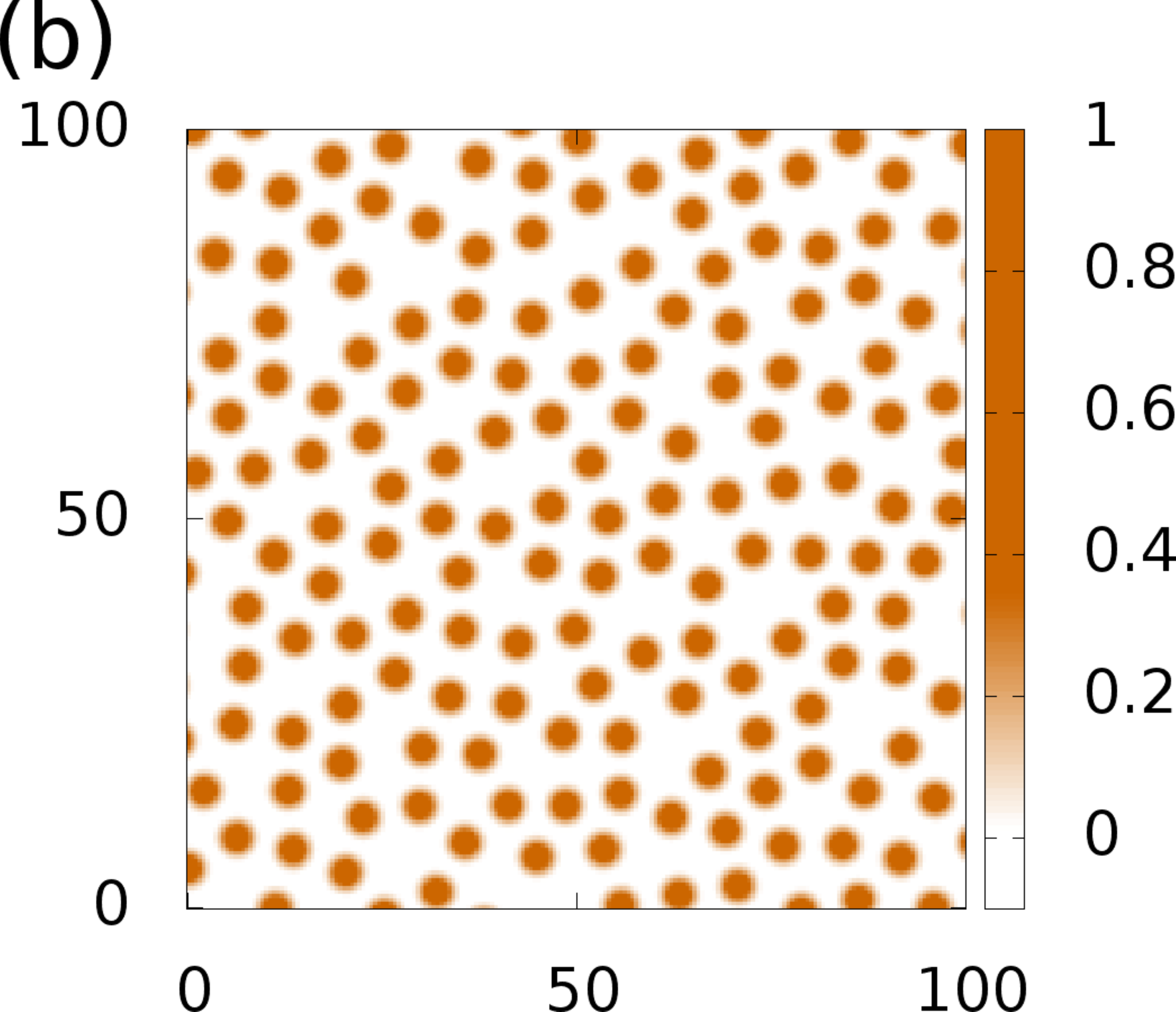} \hspace{1mm}
	\includegraphics[width=0.31\linewidth]{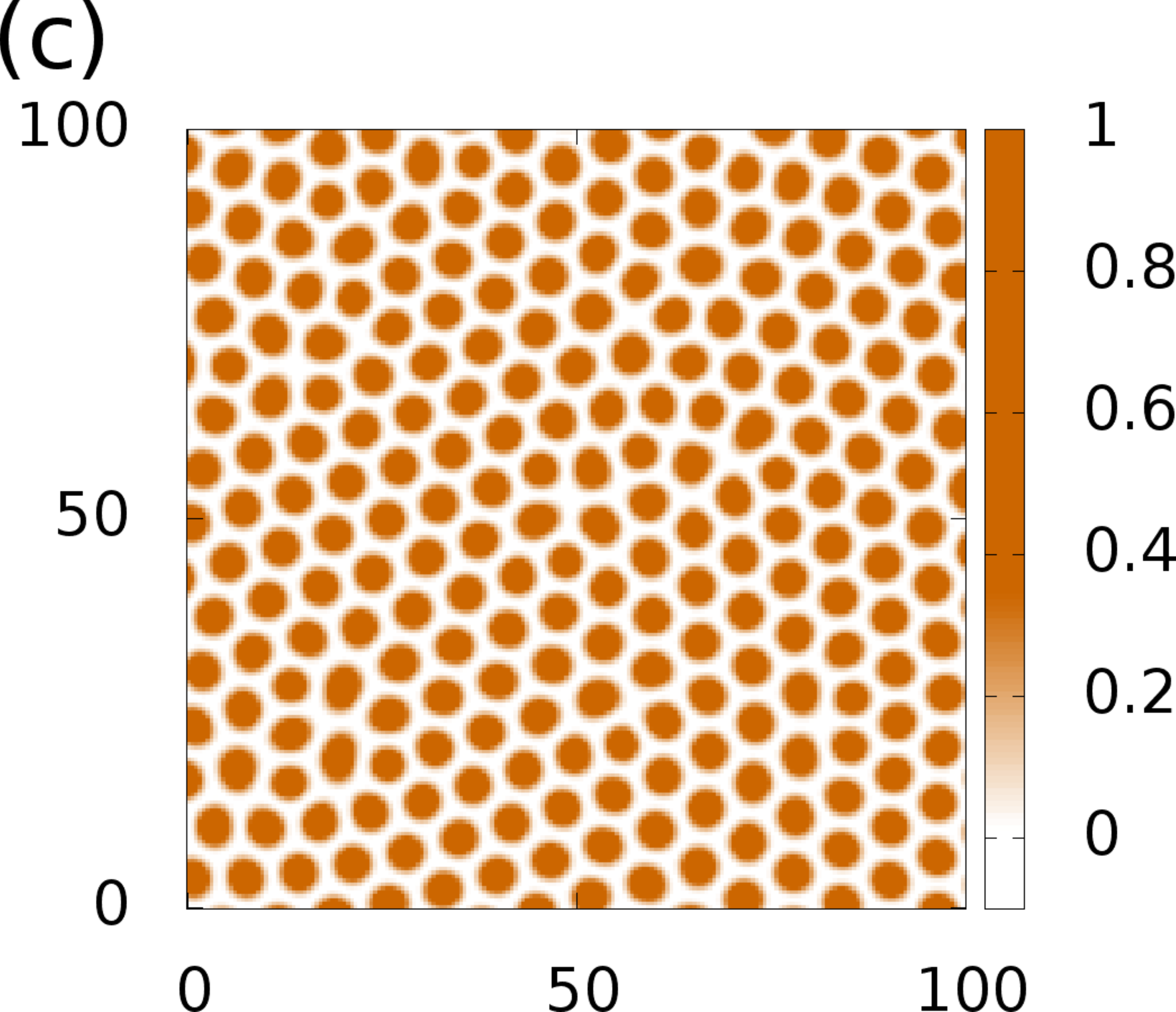}\\
	\includegraphics[width=0.32\linewidth]{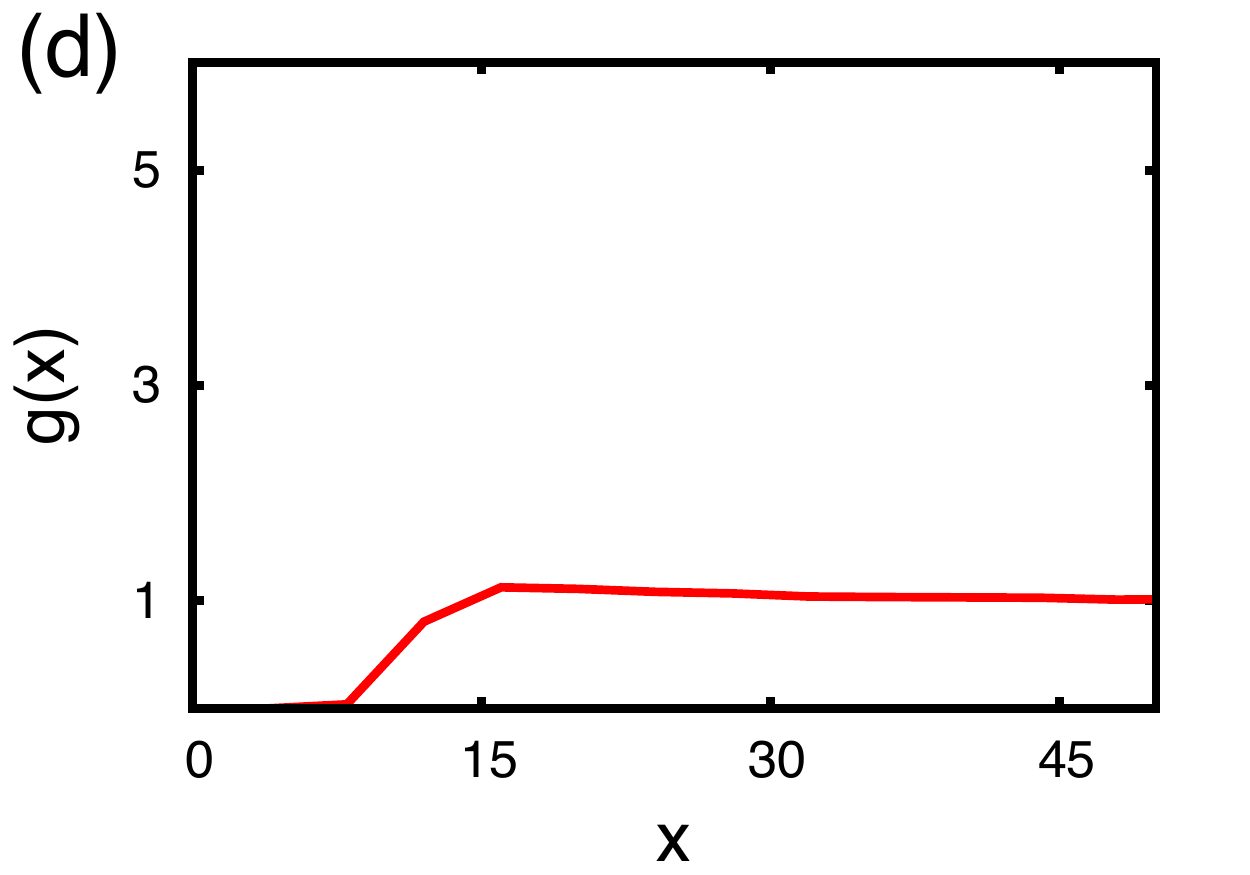}
	\includegraphics[width=0.32\linewidth]{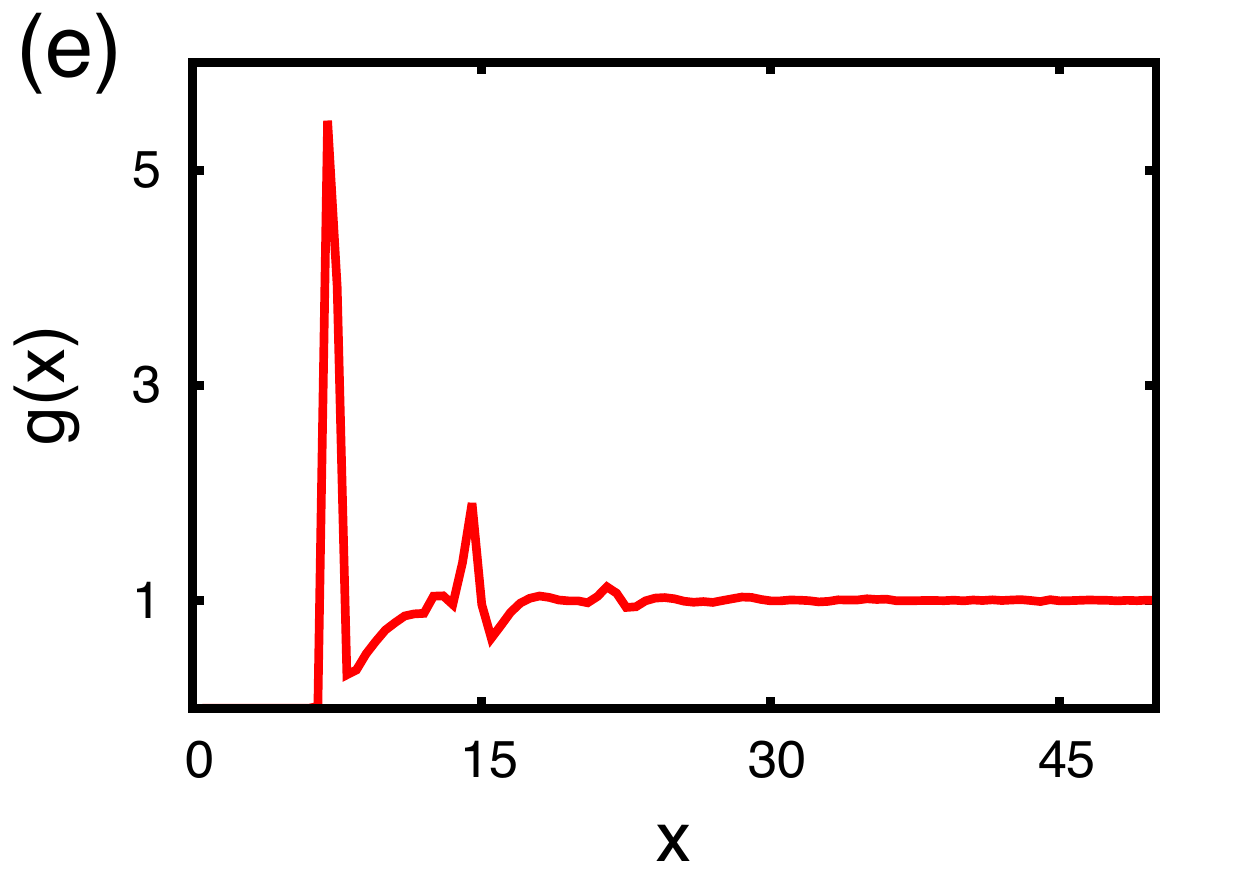}
	\includegraphics[width=0.32\linewidth]{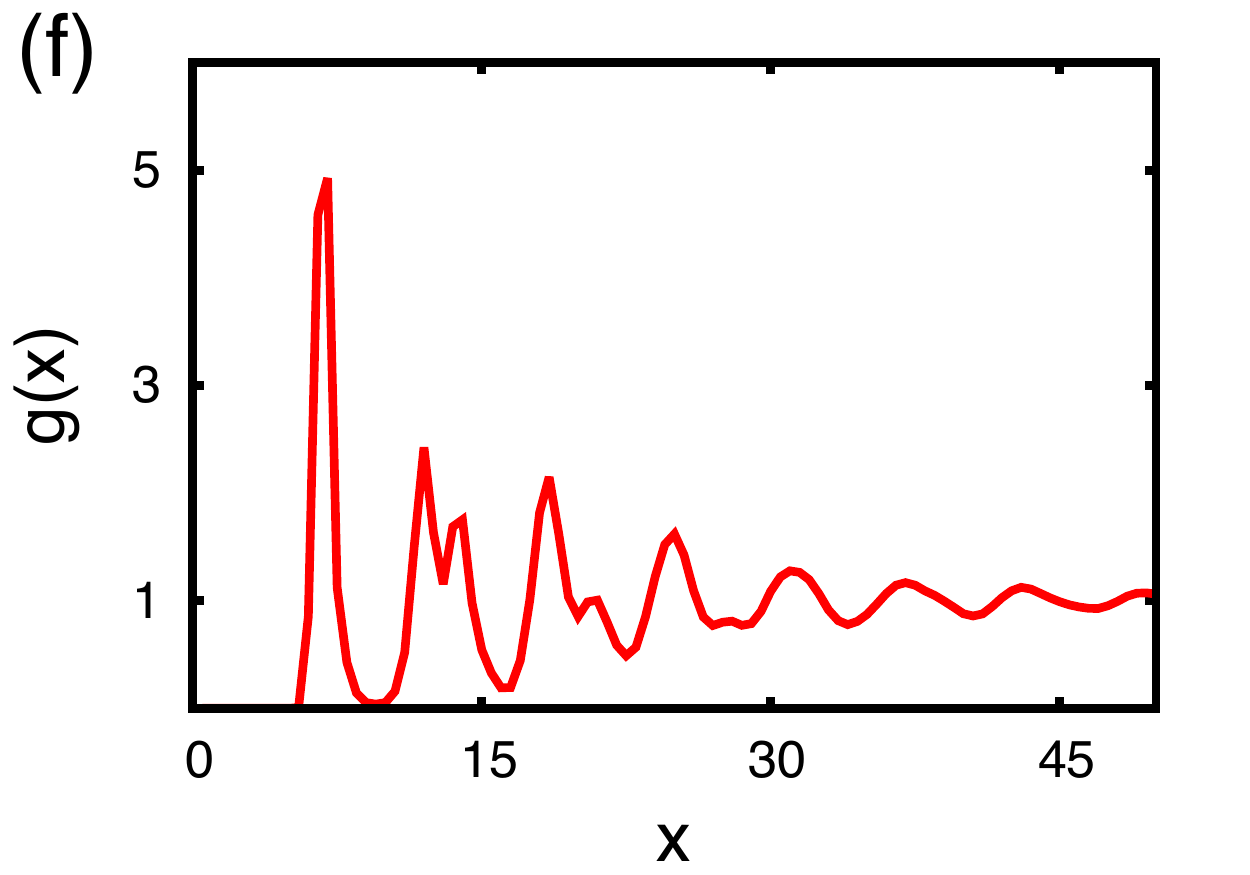}
	\caption{(a)-(c) Typical steady state order parameter profiles obtained in time simulations for
	            increasing $\bar{\phi}$. (d) - (f) show 
		      the corresponding radial distribution function $g(x)$ calculated from multiple simulations.  The 
		      parameter values are: $\alpha = 1$, $q = 1$, $r=-0.9$ and in (a), (d) $\bar{\phi} = 0.01$, in (b), 
		      (e) $\bar{\phi} = 0.1$ and in (c), (f) $\bar{\phi}=0.24$.}
	\label{figOneCompCorr}
\end{figure}

We now focus our discussion on the portion of the phase diagram where isolated bumps form.  As can be 
seen in Fig.~\ref{figOneCompCorr}, these profiles resemble particle configurations in gases, liquids and 
crystalline solids and so the VPFC may be a valuable model for describing materials on the microscale 
\cite{CGD09}.  This region of the phase diagram is full of complexity and many varied structures may be 
observed.  However, here we forego a full systematic study of this large region in parameter space and limit 
ourselves to showing representative results obtained for a single value of the undercooling parameter 
$r = -0.9$ for which there is a fairly large range in $\bar{\phi}$ with isolated bumps.  We set the initial order 
parameter profile to be a uniform state with a small amplitude noise 
$\phi(\vx,t=0) = \bar{\phi} + \lambda(Y-0.5)$, where $Y$ is a random real number uniformly distributed 
between $0$ and $1$ and $\lambda = 10^{-6}$ is the amplitude of the noise.  We consider three cases; 
i) $\bar{\phi} = 0.01$ and ii) $\bar{\phi} = 0.1$ where a disordered arrangement of localised bumps forms and iii) 
$\bar{\phi} = 0.24$ which is in the region where bumps are hexagonally ordered.  We average over many 
simulations to calculate the two point correlation function for each of these cases.  This is done by locating all 
the maxima in the equilibrium profile $\phi(\vx)$, for a given initial realisation of the noise; i.e.,~we locate the 
position of all the bumps.  From these sets of coordinates we calculate the radial distribution function $g(x)$ in 
the usual way \cite{AlTi99}.  We display a simulation result for case i) in Fig.~\ref{figOneCompCorr}(a) and the 
corresponding radial distribution function $g(x)$ in Fig.~\ref{figOneCompCorr}(d).  We find that there is almost 
no correlation between the bumps in this circumstance except for the core repulsion and  a very small peak at 
$x \approx 16$, indicating that there is a weak attraction between the bumps.  Therefore, simulations with 
these parameter values appear to qualitatively describe gas-like formations of particles/colloids.  In 
Fig.~\ref{figOneCompCorr}(b) and (e) we plot a typical order parameter profile and the corresponding $g(x)$ 
for case ii).  We observe a large increase in the number of bumps as compared to the previous case.  The 
radial distribution function shows that we have strong short range ordering, but without any long range order.  
This is very reminiscent of the ordering in liquids.  There is a very sharp peak in $g(x)$ at around $x = 7.5$ 
(which is approximately the diameter of the bumps) and a smaller peak around $x = 15$.  A similar example is 
also given in Ref.~\cite{CGD09}. If we further increase the value of $\bar{\phi}$ we eventually find the more 
familiar hexagonally structured array of bumps which is reminiscent of the ordering in simple crystalline solids.  
In Fig.~\ref{figOneCompCorr}(c) we display an example of the order parameter profile for case iii) and in 
Fig.~\ref{figOneCompCorr}(f) we show the corresponding $g(x)$.  For this case we observe that $g(x)$ is 
highly structured indicating the system has very strong short range correlations with a significant degree of 
long range ordering.  We observe the split second and third peaks, which is a classic sign of crystalline order.  
These results indicate that the VPFC model may be used to model crystalline structures, much like the regular 
PFC model.  The major difference between the two models is the existence of the fluid-like configuration of 
bumps observable in the VPFC model. In contrast, the fluid phase in the PFC model corresponds to the
homogeneous state.

\begin{figure}
	\includegraphics[width=0.45\linewidth]{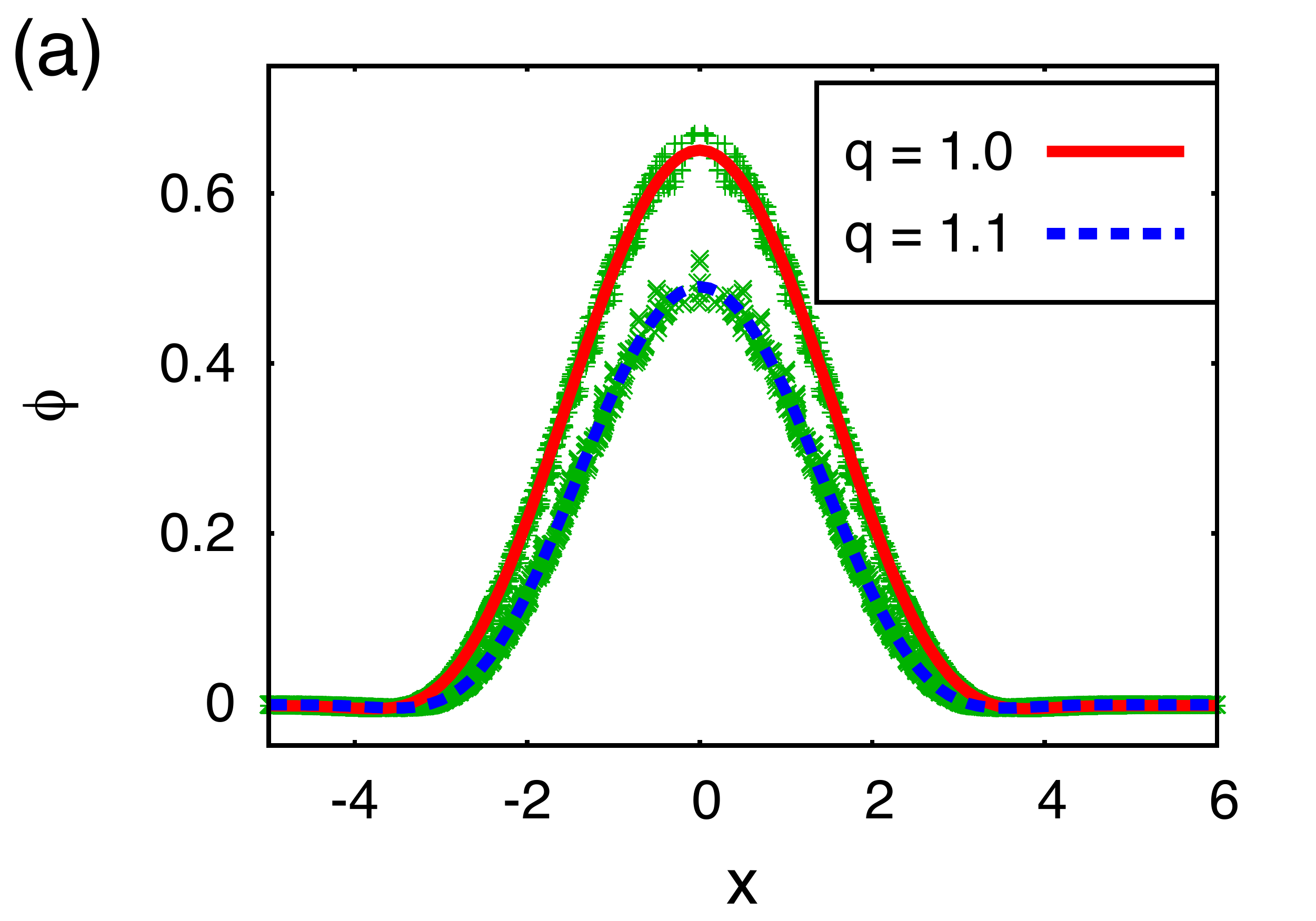}
	\includegraphics[width=0.5\linewidth]{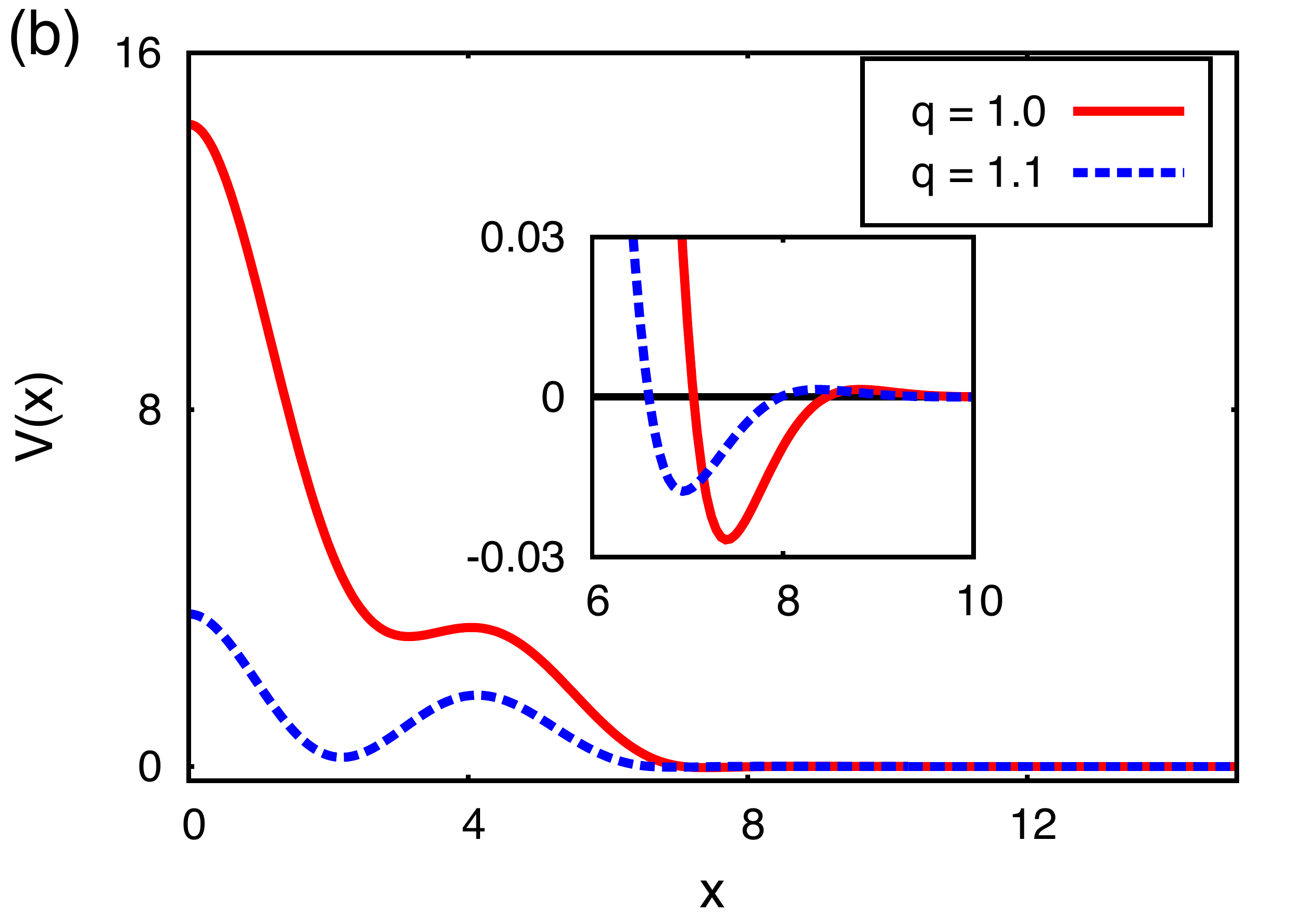}
	\caption{(a) Several sets of numerical results for the order parameter profile through the centre of a 
		       bump ($+$) for $q = 1$ and ($\times$) for  $q = 1.1$, together with fits to the data (solid red and 
		       blue dashed lines).  These fits are then used to calculate the effective pair potential between 
		       two bumps.  These pair potentials are displayed in (b).  The inset displays a magnification of the 
		       tails of $V(x)$.  The parameters values are: $r = -0.9$ and $\bar{\phi} = 0.01$.}
	\label{figOneCompPot}
\end{figure}

The variation in the size and shape of the bumps that are formed is fairly small.   In 
Fig.~\ref{figOneCompPot}(a) we display a selection of results for the order parameter profile through the 
centre of the bumps for the case when $r = -0.9$ and $\bar{\phi} = 0.01$.  We determine the shape of the 
bumps by plotting the value of the order parameter $\phi$ against the distance from the peak of each bump (as 
shown by the data points).  We can then fit functions which take the following form:

\begin{equation}
	\theta(x) = \beta_0 e^{-\beta_1x^2 - \beta_2 x^4 -\beta_3 x^6} \cos(\beta_4 x) + \beta_5.
	\label{eqFittedFunc}
\end{equation}  

We fit this form to the data using a least squares method.  The exponential part of $\theta(x)$ describes the 
decay of the modulation as the distance from the peak increases and the cosine function captures the 
oscillatory tail of the modulations which is an important factor in their interaction with other bumps 
\cite{TTP10, GTT11}.  Figure~\ref{figOneCompPot}(a) displays two cases; the ($+$) points and red solid line show 
the case $q = 1$ and the ($\times$) points and blue dashed line show the case where $q = 1.1$.  The size of 
the bump is reduced as we increase the value of $q$.  This is because increasing the value of $q$ increases 
the typical wave number which results in a smaller typical length scale (cf. Fig.~\ref{figDispRel} and 
Eq.~\eqref{eqTypWaveNum}). 

The curves obtained from fitting the bump profile can be used to obtain an approximation for the effective pair 
potential $V(x)$ between two isolated bumps, where $x$ is the distance between the centres of the bumps.  
We take a uniform system with the value of $\phi$ equal to that in the uniform areas between bumps found in 
simulations for $\bar{\phi} = 0.01$, corresponding to the results in Fig.~\ref{figOneCompCorr}(a).  We then 
impose upon this the profiles for two  bumps using the fitted curves shown in Fig.~\ref{figOneCompPot}(a).  We 
vary the distance between the superposed bumps and calculate the free energy of the system.  We assume thereby
that the two bumps retain their shape when they are close, despite the fact that in reality the bump shapes 
become distorted as bumps are pushed close together.

In Fig.~\ref{figOneCompPot}(b) we display the results for $q = 1$ (red solid line) and $q = 1.1$ (blue dashed 
line). We observe that there is a shallow minimum in the potential at the distance $x \approx 7.5$ when 
$q = 1$ and at $x \approx 7$ when $q=1.1$ (see inset of Fig.~\ref{figOneCompPot}(b)). The minimum is at a 
smaller distance when $q$ is larger because of the decreased diameter of the bumps - recall that $q$ 
determines the size of the bumps.  The resulting weak attraction between the bumps may also be 
inferred from the radial distribution function $g(x)$ calculated for the low density case $\bar{\phi} = 0.01$ when 
$q = 1$ displayed in Fig.~\ref{figOneCompCorr}(d). We observe a second minimum in the potentials at 
$x \approx 3.15$ when $q = 1$ and at $x \approx 2.3$ when $q = 1.1$, where the former rather appears like a 
`shoulder'. The order parameter profiles at the second minima resemble the elongated almost elliptical 
shapes which are observed in and around the coexistence region between bumps and stripes. See also 
Fig.~\ref{figPhaseD2D}(c) where we also observe elliptical holes along some of the grain boundaries.

\section{Two Component System}
\label{secTwoComp}

We now extend the model to consider a binary mixture in perhaps the most simple way possible, by adding 
together free energies like Eq.~(\ref{eqVPFCOne}) for two order parameter fields $\phi_a(\vx,t)$ and $\phi_b(\vx,t)$.  We introduce a 
simple coupling term which allows the two components to interact with each other.  This gives us the following 
expression for the free energy:

\begin{equation}
	F = \int d{\bf x} \hspace{2mm} \bigg[ f(\phi_a({\bf x},t)) + f_{vac}(\phi_a({\bf x},t)) + 
	 f(\phi_b({\bf x},t)) + f_{vac}(\phi_b({\bf x},t)) + \eta \phi_a \phi_b \bigg],
	\label{eqFreeEngTwo}
\end{equation}

where $\eta$ is the coupling coefficient and the functions $f$ and $f_{vac}$ are defined as before in 
Eqs.~\eqref{eqPFCOne} and \eqref{eqPFCVacTerm}.  The value of $r$ is set equal for both components.  
However, we allow the value of $q$ to be different for each species, so we now refer to these values as $q_a$ 
and $q_b$, where the subscript denotes the corresponding component.  Setting different values for $q$ in the 
two components (i.e.,~$q_a \ne q_b$) results in an asymmetrical system in which the size of the 
bumps/modulations in $\phi_a$ differs from that in $\phi_b$, as discussed further below in 
Secs.~\ref{ssInteract} and \ref{ssOrdering}.  Note that a different coupling term is used in Ref.~\cite{BeGr11}; 
a somewhat different two-component PFC model is presented in Ref.~\cite{MuHa10}.

Here,  just as for the one component model, we assume the dynamics of 
the system is governed by the following pair of equations (cf.~Eq.~\eqref{eqPFCDyn}):
\begin{eqnarray}
	\paDir{\phi_a}{t} &=& \alpha_a \nabla^2 \fnDir{F}{\phi_a}, \nonumber \\
	\paDir{\phi_b}{t} &=& \alpha_b \nabla^2 \fnDir{F}{\phi_b}.
	\label{eqDynTwo}	
\end{eqnarray}
We also assume that the two mobility coefficients are equal: $\alpha_a = \alpha_b = \alpha$.  The two 
components are coupled purely by the term $\eta \phi_a \phi_b$ in the free energy.  When $\eta > 0$,
this coupling term leads to a repulsion between the two species and so penalises structures which 
overlap or form on top of each other.  The value of the parameter $\eta$ determines the `strength' of the 
coupling, and so the two component model reduces to two disconnected one component models in the limit 
$\eta \to 0$.

\subsection{Phase behaviour}

When the coupling coefficient is fairly large $\eta \ge 0.1$, the coupling term has a significant impact on the 
phase behaviour of the model.  In particular, the limit of linear stability and the phase coexistence curves 
extend to much larger values of $\bar{\phi} = \barA + \barB$ than for the one component model.  We now 
determine the linear stability of a flat state in the model.  We assume that the order parameter profiles of both 
components take the form:
\begin{eqnarray}
	\phi_a &=& \barA + \delPhi = \barA + \xi e^{ik{\bf x}} e^{\beta t}, \nonumber \\
	\phi_b &=& \barB + \chi \delPhi = \barB + \chi \xi e^{ik{\bf x}} e^{\beta t},
\end{eqnarray}
where the amplitude $|\xi| \ll 1$ and the parameter $\chi$ is the ratio between the amplitude of the 
modulations in the two components.  The sign of $\chi$ indicates whether instabilities are in-phase 
($\chi > 0$) or anti-phase ($\chi < 0$) between the two coupled order parameter fields. From the magnitude of 
$\chi$ we can deduce whether the instability is initiated from species $a$ $(|\chi| \ll 1)$, species $b$ 
$(|\chi| \gg 1)$, or a combination of both $(|\chi| =O(1))$.  We make a Taylor series expansion
of the functional derivatives of the free energy with respect to the two order parameters $\phi_a$ and 
$\phi_b$, to obtain:
\begin{eqnarray}
	\fnDir{F}{\phi_a} &=& (r + q_a^4) \barA + 3H \barA (|\barA| - \barA) + \barA^3 + \eta \barB \nonumber \\ &&
	+ \big[(k^2 - q_a^2)^2 + \Delta_a +\chi \eta \big] \delPhi + O(\delPhi^2), \nonumber \\ 
	\fnDir{F}{\phi_b} &=& (r + q_b^4) \barB + 3H \barB (|\barB| - \barB) + \barB^3 + \eta \barA \nonumber \\ &&
	+ \big[\chi (k^2 - q_b^2)^2 + \chi \Delta_b +\eta  \big] \delPhi + O(\delPhi^2).
	\label{eqPFCFuncDerTwo}
\end{eqnarray}
where $\Delta$ is defined as before (Eq.~\eqref{eqDelta}) and the subscript denotes the corresponding
component.  We substitute these expressions into the dynamical equations \eqref{eqDynTwo}, yielding the matrix problem \cite{PBMT05, RAT11}:
%
%
%
\begin{equation}		
	\beta \chiArray = \mathbf{M} \chiArray ,
	\label{eqMatrixForm}
\end{equation}
where
\begin{equation}
	\mathbf{M} = -k^2\alpha \mat{(q_a^2 - k^2)^2 + \Delta_a}{\hspace{5mm}
	\eta}{\eta}{(q_b^2 - k^2)^2 + \Delta_b}. \nonumber
\end{equation}
We can now determine the dispersion relation $\beta(k)$ by calculating the eigenvalues of $\mathbf{M}$:
\begin{equation}
	\beta(k) = \frac{\mbox{Tr}(\mathbf{M})}{2} \pm 
	\sqrt{\frac{\mbox{Tr}(\mathbf{M})^2}{4} - |\mathbf{M}|}.
	\label{eqBetaTwo}
\end{equation}
The resulting dispersion relation $\beta(k)$ is a double-valued function.  However, since the growth rate along the $+$ branch is always larger than that along the $-$ branch, the limit of linear stability can be determined from the $+$ 
branch alone.  If we assume that $q_a = q_b = q$, the dispersion relation simplifies significantly, yielding:
\begin{equation}
	\beta(k) = - \frac{\alpha k^2}{2} \bigg[ 2(k^2-q^2)^2 + \Delta_a + \Delta_b 
			- \sqrt{ (\Delta_a - \Delta_b)^2 + 4 \eta^2} \bigg].
	\label{eqSimpBetaTwo}
\end{equation}   
There is a local maximum of this expression which occurs at the positive wave number:
\begin{equation}
	k_m = \frac{1}{6} \bigg[24q^2 + 6 \bigg( 4q^4 - 6 (\Delta_a + \Delta_b) 
	+ 6 \sqrt{(\Delta_a - \Delta_b)^2 + 4 \eta^2} \bigg)^{\frac{1}{2}} \hspace{1mm} \bigg]^{\frac{1}{2}}.
	\label{eqTypWaveNumTwo}
\end{equation}
Substituting this wave number back into the dispersion relation \eqref{eqSimpBetaTwo},
allows us to calculate the parameter values such that $\beta(k_m) = 0$ (i.e., the limit of linear
stability of a flat state).  We arrive at the following relation:
\begin{equation}
	\Delta_a \Delta_b = \eta^2.
	\label{eqLinStabTwo}
\end{equation}
When the system is linearly unstable it is possible for $\beta(k = 0)$ to be a minimum or maximum 
(this transition occurs at $\Delta = - q^2$ in the one component model).  This is equivalent to the 
coefficient of $k^2$ changing from a positive value (minimum) to a negative value (maximum).  The 
sign of the coefficient of $k^2$ is determined by the sign of the following quantity:
\begin{eqnarray}
	C_2 &=& \paDirH{g}{\phi_a}{2} \paDirH{g}{\phi_b}{2}  -  \bigg(\frac{\partial g}{\partial \phi_b 
	\partial \phi_a} \bigg)^2, \nonumber \\ &=& (q_a^4 + \Delta_a) (q_b^4 + \Delta_b) - \eta^2,
	\label{eqBetak0}
\end{eqnarray} 

\begin{figure}[t]
	\includegraphics[width = 0.7\linewidth]{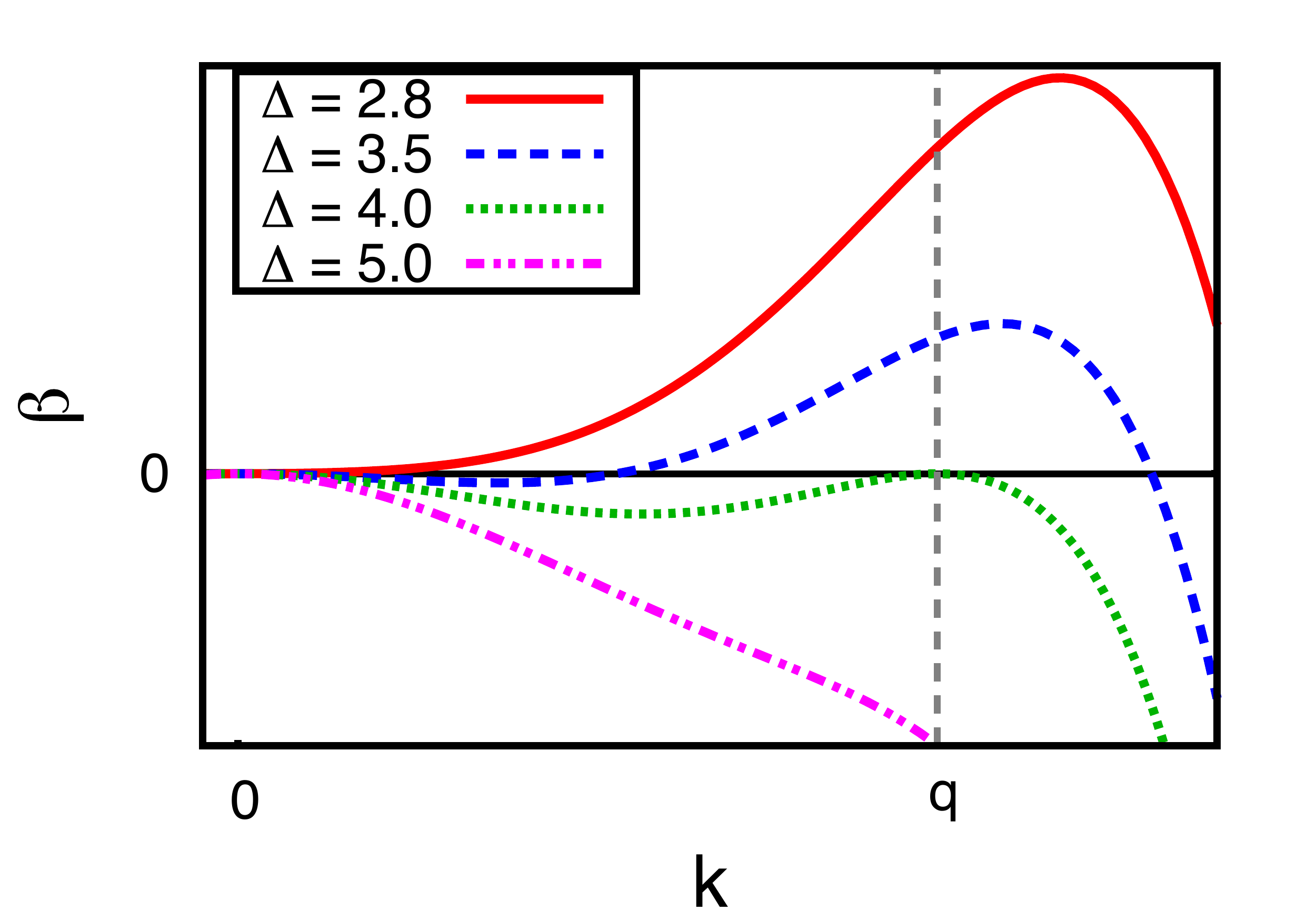}
	\caption{Dispersion relation curves for the two component VPFC model 
		      (Eqs.~\eqref{eqFreeEngTwo} and \eqref{eqDynTwo}), when $q_a = q_b = 1$, 
		      $\Delta_a = \Delta_b = \Delta$ and $\eta = 4$.  Four cases are shown: (i) $\beta(k_m) > 0$ with $\beta(k = 0)$ a minimum (red solid line), (ii) $\beta(k_m) > 0$ and $\beta(k = 0)$ a maximum (blue dashed line), (iii) $\beta(k_m) = 0$ and (iv) $\beta(k_m) < 0$ (magenta dash-dotted line).}
	\label{figDispRel2}
\end{figure}
 
where $g(\phi_a, \phi_b) = f(\phi_a) + f_{vac}(\phi_a) + f(\phi_b) + f_{vac}(\phi_b) + \eta \phi_a \phi_b$.
When $C_2$ is negative/positive $\beta(k=0)$ is a minimum/maximum, this relation also holds for asymmetric 
systems where $q_a \ne q_b$.  In 
figure \ref{figDispRel2} we display typical dispersion relations when $q_a = q_b = 1$, $\Delta_a = 
\Delta_b = \Delta$ and $\eta = 4$.  We show the case when i) the system is linearly unstable and $C_2$
[Eq.~\eqref{eqBetak0}] is negative (red solid line), ii) the system is linearly unstable and 
Eq.~\eqref{eqBetak0} is positive (blue dashed line), iii) the system is at the limit of linear stability 
(i.e.,~Eq.~\eqref{eqLinStabTwo} holds) (green dotted line) and iv) the system is linearly stable (magenta 
dash-dotted line).  We observe that when $q_a = q_b$, the typical wave number $k_m \to q_a$ as we 
approach the limit of linear stability $\Delta_a \Delta_b - \eta^2 \to 0$.  In the more general 
case with $q_a \ne q_b$ the dispersion relation may have two maxima at positive values of $k$ neither of which
occurs at $q_a$ and $q_b$. In this case the stability boundary is defined by the vanishing of growth rate 
$\beta(k_m) = 0$ of the larger of the two possible maxima of $\beta$.

\begin{figure}
	\includegraphics[width = 0.3\linewidth]{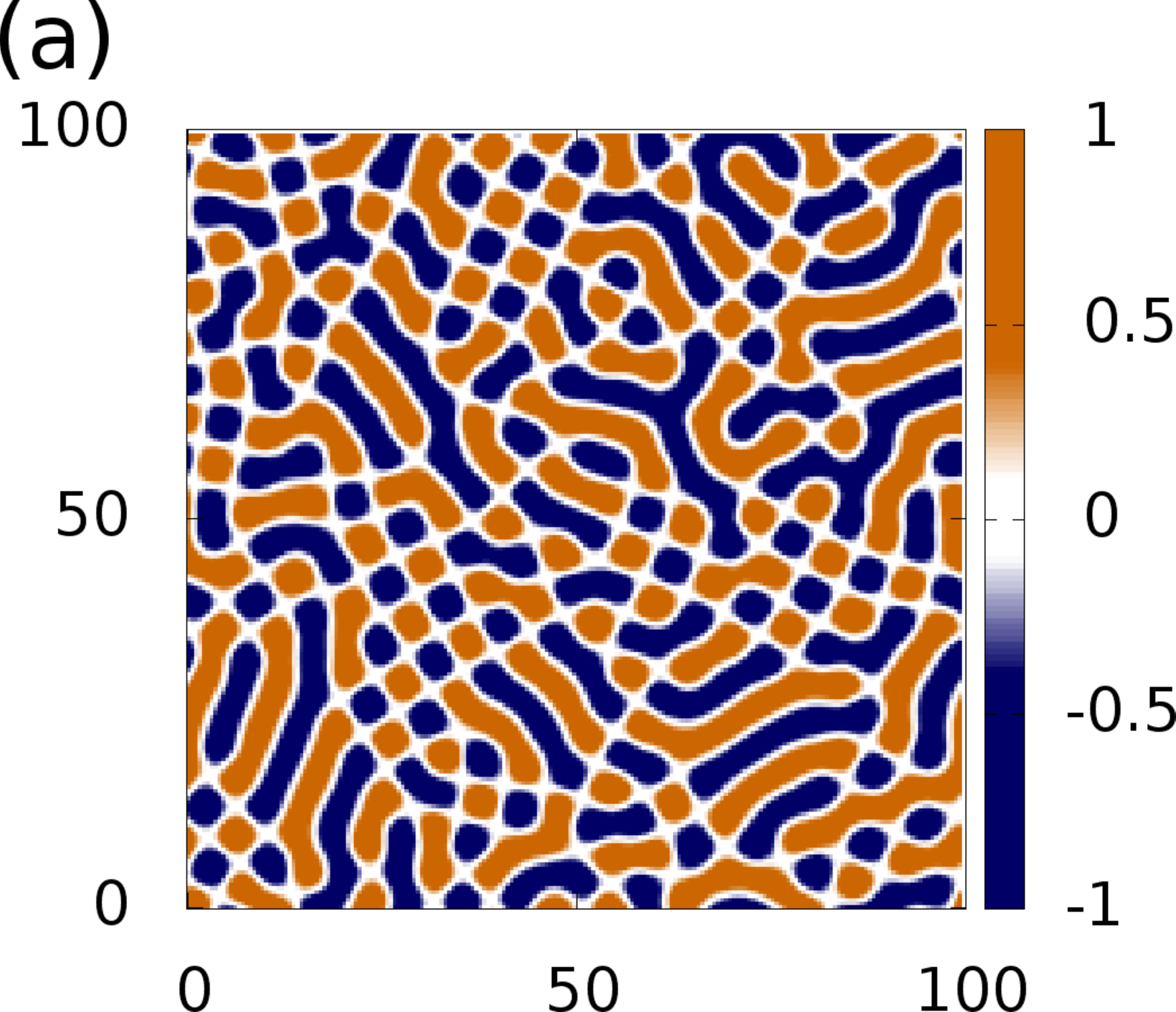}
	\includegraphics[width = 0.3\linewidth]{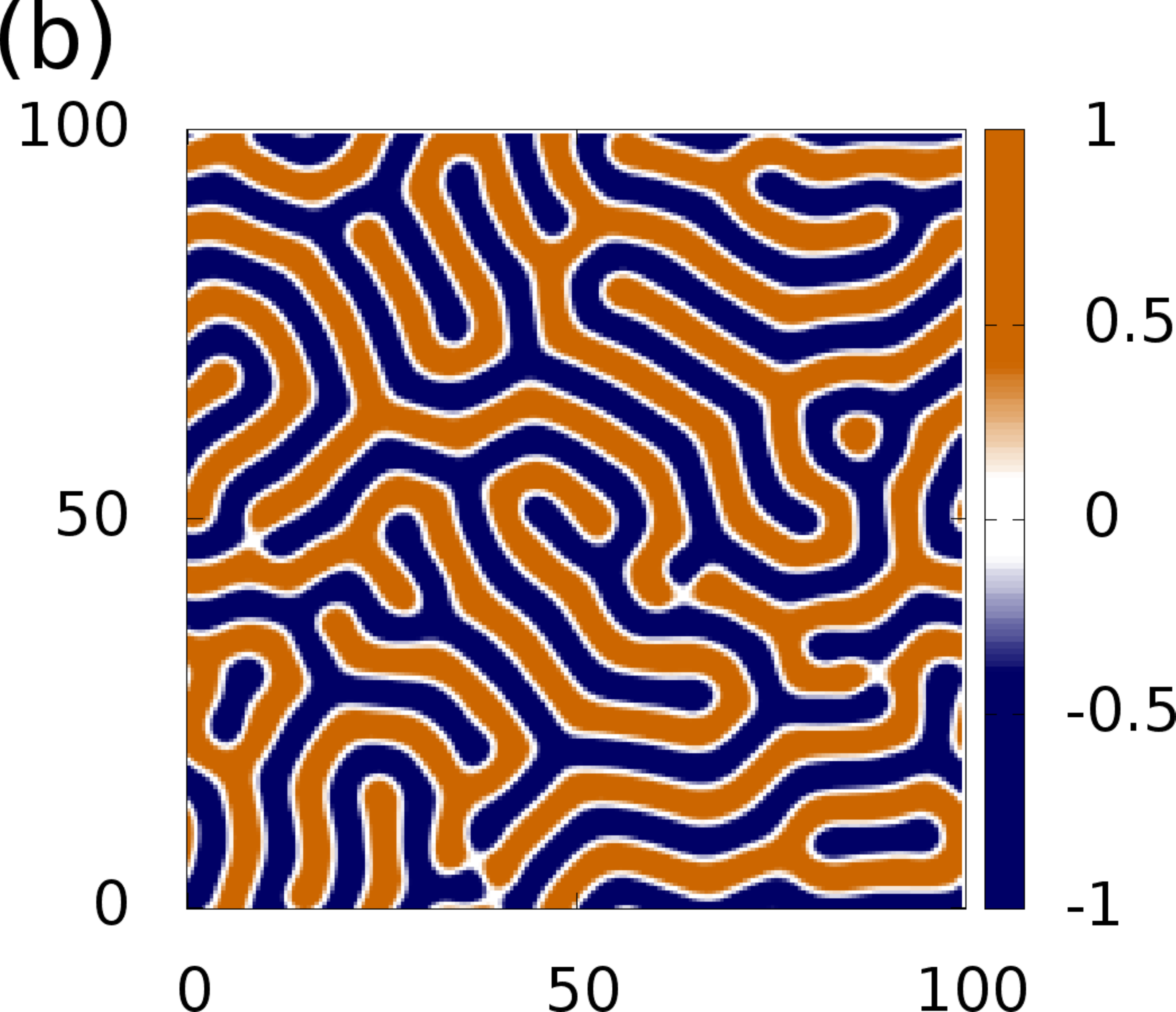}
	\includegraphics[width = 0.3\linewidth]{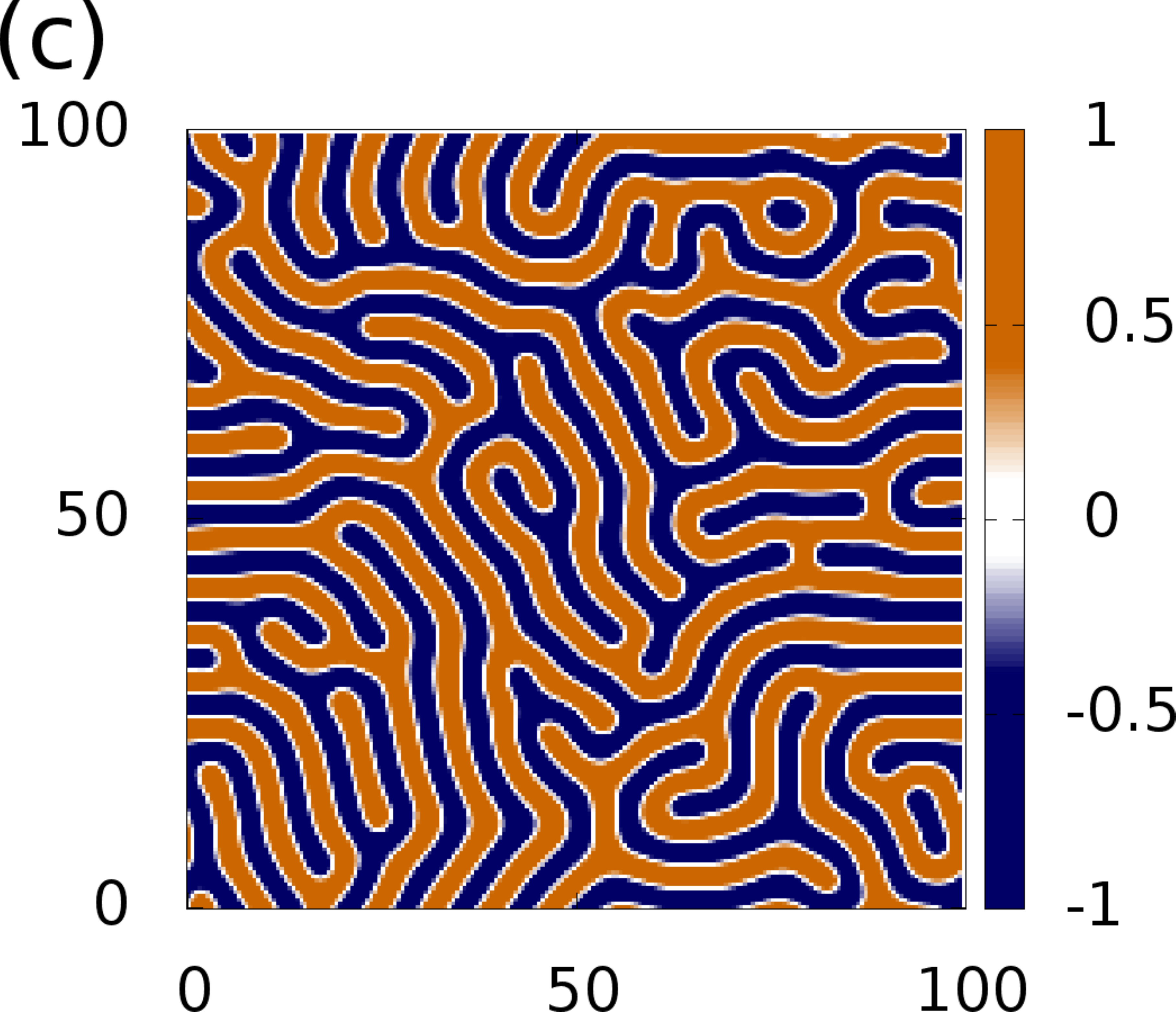}\\
	\includegraphics[width = 0.3\linewidth]{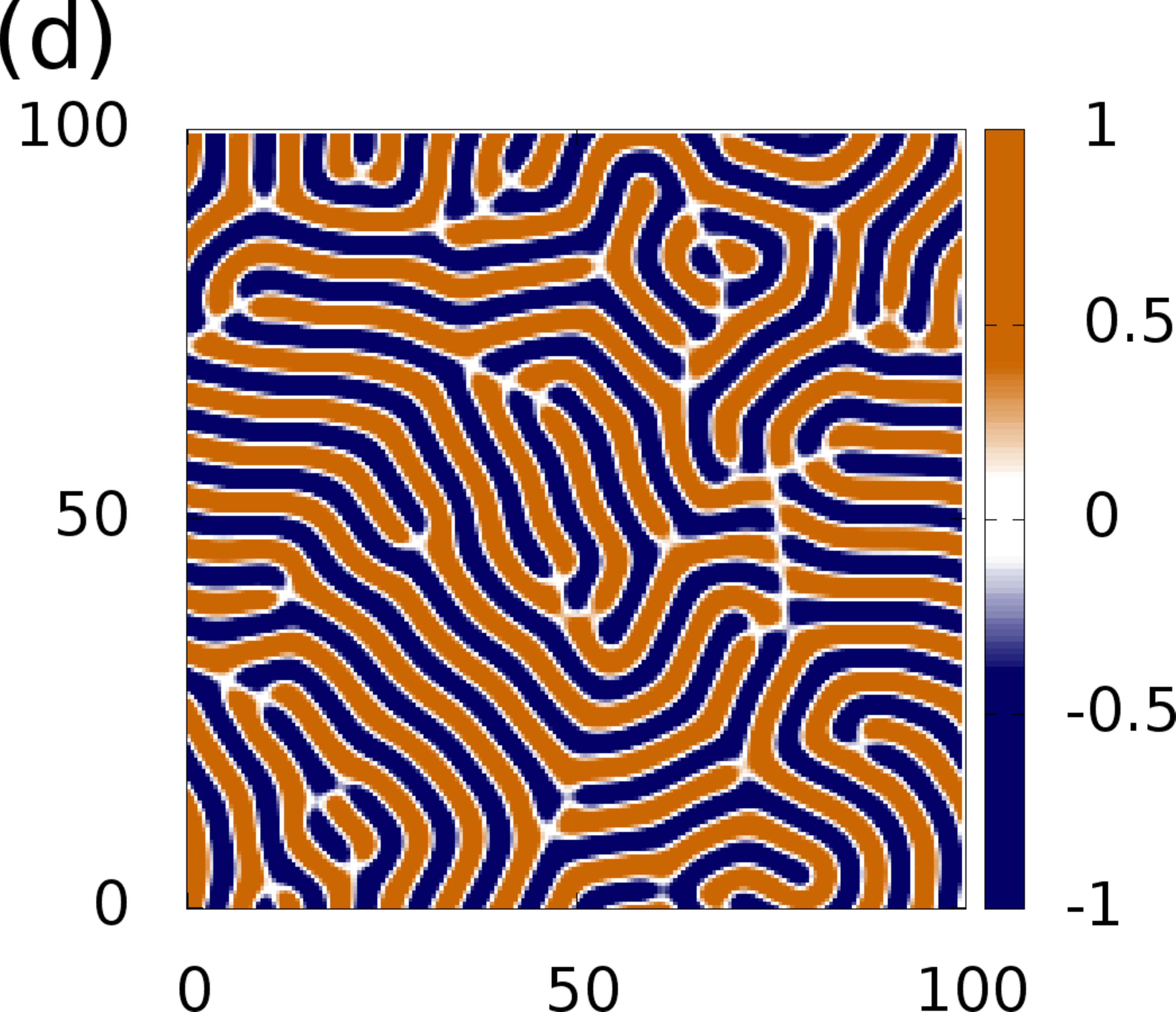}
	\includegraphics[width = 0.3\linewidth]{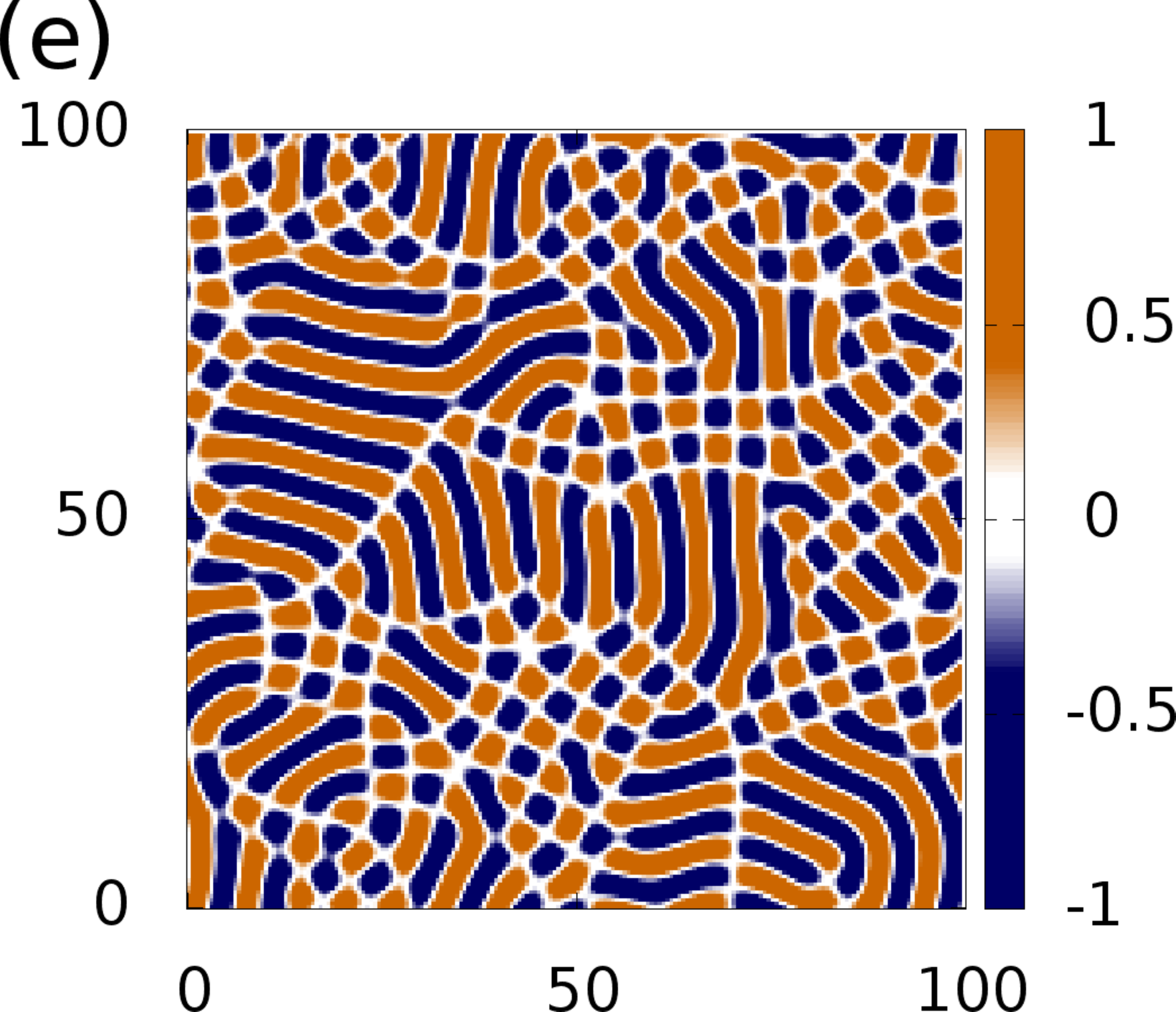}
	\includegraphics[width = 0.3\linewidth]{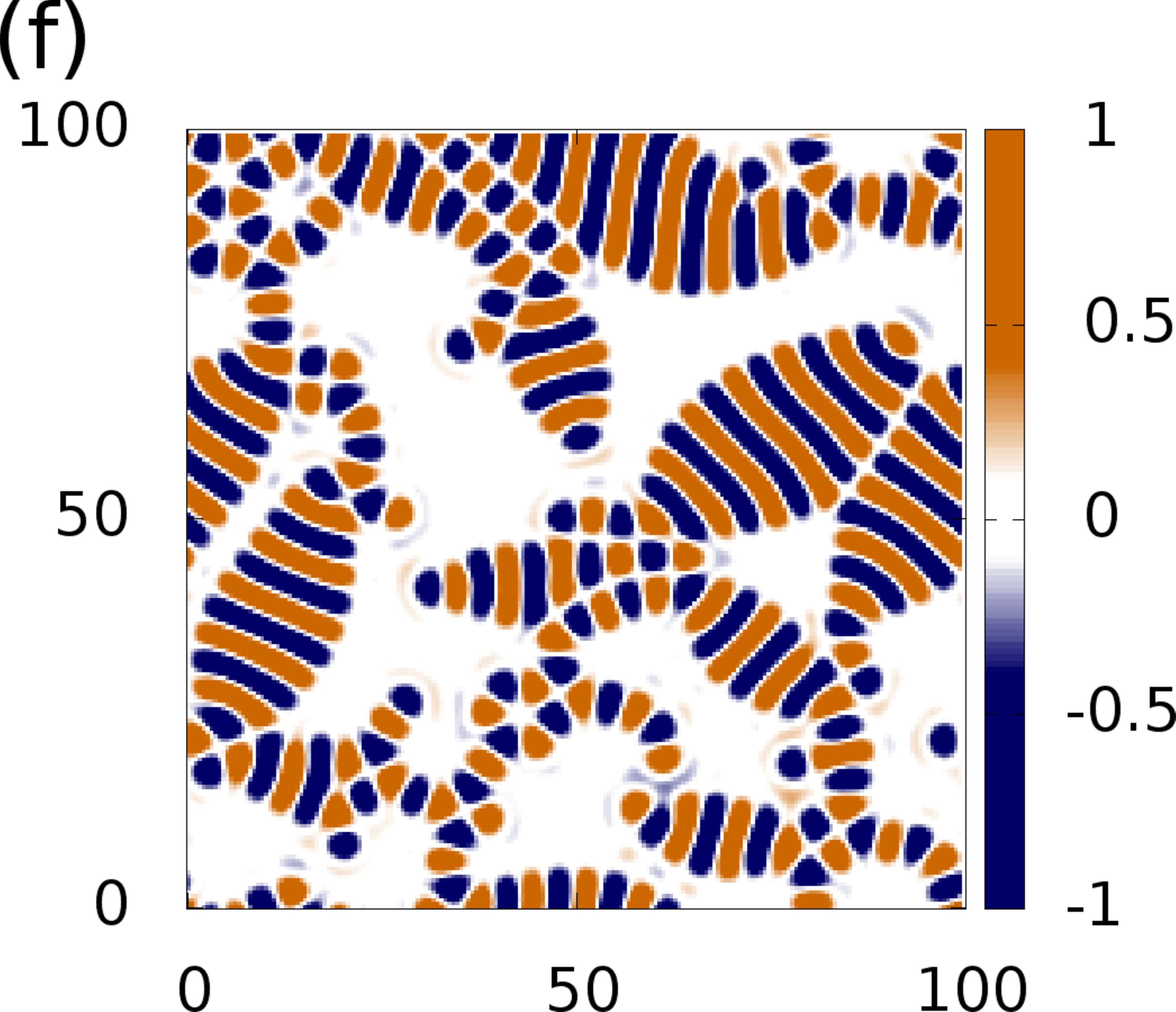}
	\caption{The scaled order parameter $\Delta \hat{\phi}$ for the two-component model, corresponding to
		      minima of the free energy.  The peaks in species $a$ are shown in orange, peaks in species $b$ 
		      are shown as blue and white areas show regions where $\phi_a \approx \phi_b$.  The 
		      parameter values are: $\eta = 4$, $r = -0.9$, $q_a = q_b = 1$, $\barA = \barB = \bar{\phi}$, 
		      where (a) $\bar{\phi} = 0.25$, (b) $\bar{\phi} = 0.3$, (c) $\bar{\phi} = 1$, (d) $\bar{\phi} = 1.15$, 
		      (e) $\bar{\phi} = 1.2$ and (f) $\bar{\phi} = 1.27$.}
	\label{figHighPhive}
\end{figure}

From Eq.~\eqref{eqLinStabTwo} it is clear that depending on the value of the coupling coefficient $\eta$, the 
region of parameter space where the system is linearly unstable can be greatly larger than that for the one 
component system.  For example, picking the value $\eta = 4$ when $r = -0.9$ and setting the average value 
of both order parameters to be equal $\barA = \barB = \bar{\phi}$, we find that the limit of linear stability 
increases from $\bar{\phi} = 0.548$ (for the one component case) to $\bar{\phi} = 1.278$.  As one would 
expect, this also increases the region of the phase diagram where modulated structures are formed.  Our focus 
here is on the regions of parameter space where bumps are formed as this is the regime relevant to modelling 
crystalline solids.  However, before proceeding to this, we make a brief survey of some of the structures which 
may be observed for larger values of $\phi_a$ and $\phi_b$ which lie outside of the bump phase.  For the 
parameter values $r = -0.9$, $\eta = 4$, $q_a = q_b = 1$ and $\barA = \barB = \bar{\phi}$ we show in 
Fig.~\ref{figHighPhive} a sequence of order parameter profiles with increasing $\bar{\phi}$, for values of 
$\bar{\phi}$ that lie above the region where bumps are observed (see later sections for a detailed analysis of 
the bump structures found in the two component model).  In Fig.~\ref{figHighPhive} we display scaled plots of 
order parameter profiles which are stationary states obtained from time simulations.  We plot an order 
parameter defined as the normalised difference between the $\phi_i(\vx)$ values of the two components 
$\Delta \hat{\phi}(\vx) \equiv \phi_a(\vx)/\hat{\phi}_a - \phi_b(\vx)/\hat{\phi}_b$ where
\begin{equation}
\hat{\phi}_i = \frac{\phi_a^{\mx} \phi_b^{\mx} - \phi_a^{\mn} \phi_b^{\mn}}{\phi_i^{\mx} + \phi_i^{\mn}}
	\label{eqhatPhi}
\end{equation}   
and where $\phi_i^{\mx}$ and $\phi_i^{\mn}$ are, respectively, the maximum and minimum values of 
$\phi_i(\vx)$.  $\Delta \hat{\phi}$ is defined so as to take a value in the range $[-1, 1]$.  When 
$\Delta \hat{\phi} \approx + 1$, then the local value of $\phi_a$ is high whilst the value of $\phi_b$ is low.
Conversely, when $\Delta \hat{\phi} \approx -1$, then the local $\phi_a$ is low and $\phi_b$ is high. The 
average order parameter values in Fig.~\ref{figHighPhive} are: (a) $\bar{\phi} = 0.25$, (b) $\bar{\phi} = 0.3$, 
(c) $\bar{\phi} = 1$, (d) $\bar{\phi} = 1.15$, (e) $\bar{\phi} = 1.2$, (f) $\bar{\phi} = 1.27$.  The most palpable 
change from the one component model is that the phase diagram is largely dominated by the striped profiles, 
with stripes appearing in the range $0.22 \lsim \bar{\phi} \lsim 1.28$.  Just outside the range of $\bar{\phi}$ 
where bump structures are formed, we observe order parameter profiles which contain a mixture of bumps 
and stripes -- see Fig.~\ref{figHighPhive}(a) -- this value of $\bar{\phi}$ must lie inside the coexistence region 
between the bump and stripe phases.  As we increase the value of $\bar{\phi}$ we enter the large region of
parameter space where stripe structures are formed [Fig.~\ref{figHighPhive}(b) and (c)], the only significant
change as we increase $\bar{\phi}$ from $0.3$ to $1$ is the decrease in the width of the stripes; this is due to 
the fact that the typical length scale in the system is $2\pi/k_m$, where the wave number $k_m$ given by 
Eq.~\eqref{eqTypWaveNumTwo}, is inversely proportional to the average order parameter values $\barA$ and 
$\barB$.  Increasing the value of $\bar{\phi}$ further, we continue to observe striped profiles, but now there are 
points where the stripes of one species `connect' to stripes of the other species -- see Fig.~\ref{figHighPhive}(d) 
(these `connections' appear as white lines in Fig.~\ref{figHighPhive}(d)).  Increasing $\bar{\phi}$ further, we 
observe a mixture of holes and stripes [Fig.~\ref{figHighPhive}(e)].  Close to the instability curve 
Eq.~\eqref{eqLinStabTwo} we find interesting profiles where we observe a mixture of stripes, holes and 
regions where the profile is approximately uniform $\phi_a \approx \phi_b \approx \bar{\phi}$ 
[Fig.~\ref{figHighPhive}(e)].  Various modulated structures are observed over a large range of parameter 
values.  It would be possible to consider the structures formed for different values of the coupling coefficient 
$\eta$ and different values of the average order parameters, where $\barA \ne \barB$.  However, here we 
do not make a systematic study of the entire parameter space and instead focus on the various bump 
formations.  These structures closely resemble the configurations of particles/colloids in condensed matter 
systems and we believe that in this regime the model may be useful to understanding the fluid and solid 
phases of such systems.

\subsection{Intermolecular interactions}
\label{ssInteract}
 
For the remainder of this paper, we pursue the idea that the bumps in this two component model 
represent two different types of molecules or colloidal particles suspended in a fluid medium.  We perform time 
simulations of the two component model choosing parameter values which result in the formation of bump 
structures.  We run these simulations until the order parameter profiles reach an (almost) stationary state, 
which corresponds to being at (near) an energetic minimum.  We then determine the coordinates of the 
particles by locating the position of the maximum of each of the peaks.  The radial distribution functions are 
calculated by analysing these coordinates.  We also calculate the effective pair potentials between the bumps.  
Later in Sec.~\ref{ssOrdering} we consider the nearest neighbour bond angles and the ordering in crystalline 
configurations.
 
The bump phase in the two component model appears to behave in a similar manner to that of the one 
component model (cf. Fig.~\ref{figPhaseD2D}).  We observe bump structures when the average value
of the order parameters $\barA$ and $\barB$ are small.  In particular, when $\barA = \barB = \bar{\phi}$ 
and $r = -0.9$ we observe bumps within the range $0 \lsim \bar{\phi} \lsim 0.15$.  We study and compare 
two different systems: the symmetric case where $q_a = q_b = 1$ and the asymmetric case where
$q_a = 1$ and $q_b = 1.1$.  In the symmetric case, interactions between bumps of the same type ($aa$ and 
$bb$) are identical in both components, but the nature of the interaction between a bump in $\phi_a$ and a 
bump in $\phi_b$ ($ab$) is determined by the coupling term in the free energy.  In the asymmetrical case, the 
different values of $q$ mean that the size of the bumps are different in $\phi_a$ and $\phi_b$, hence, all 
possible interactions $aa$, $bb$ and $ab$ are different. 

\begin{figure}
	\includegraphics[width=0.29\linewidth]{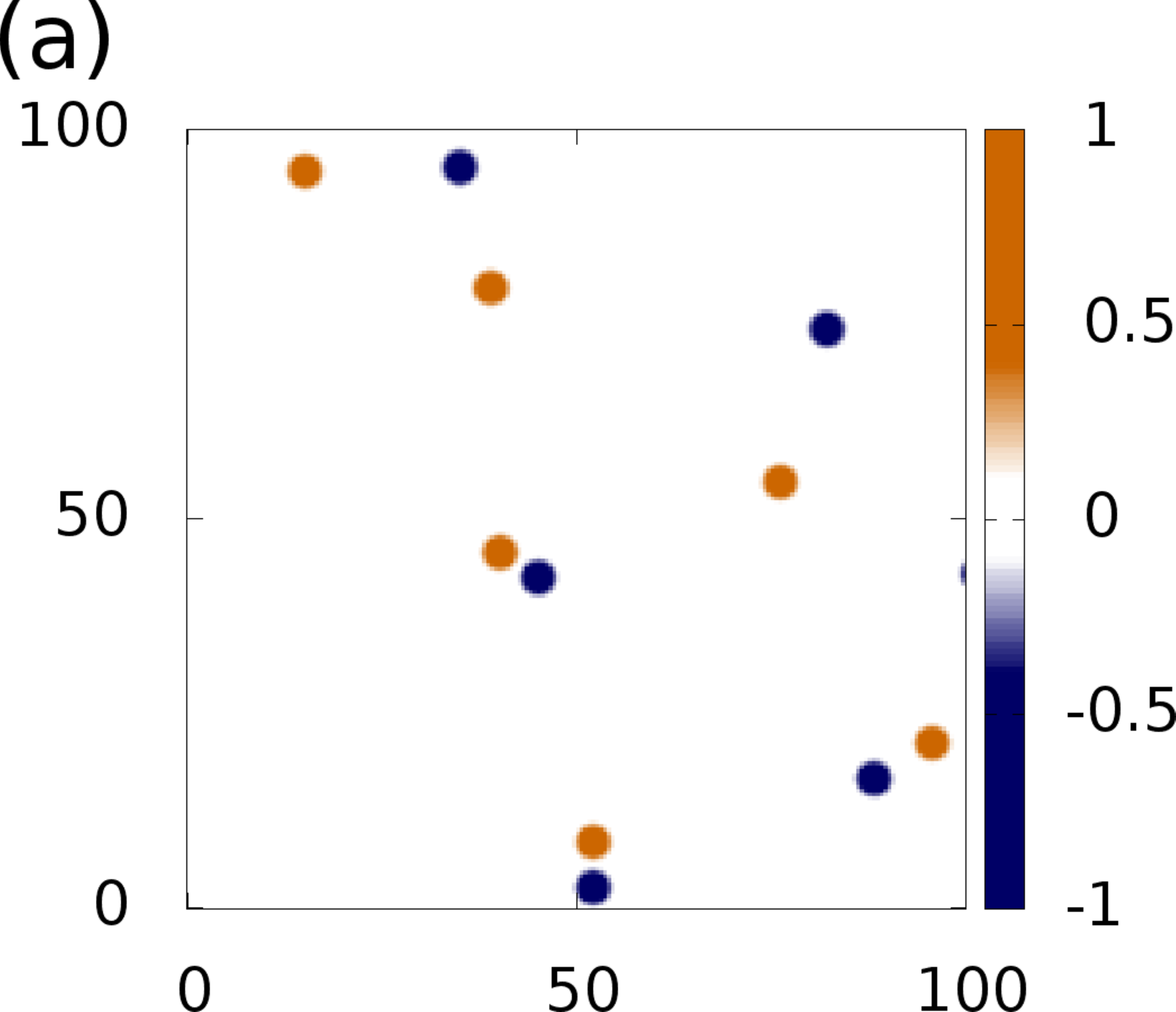}
	\includegraphics[width=0.29\linewidth]{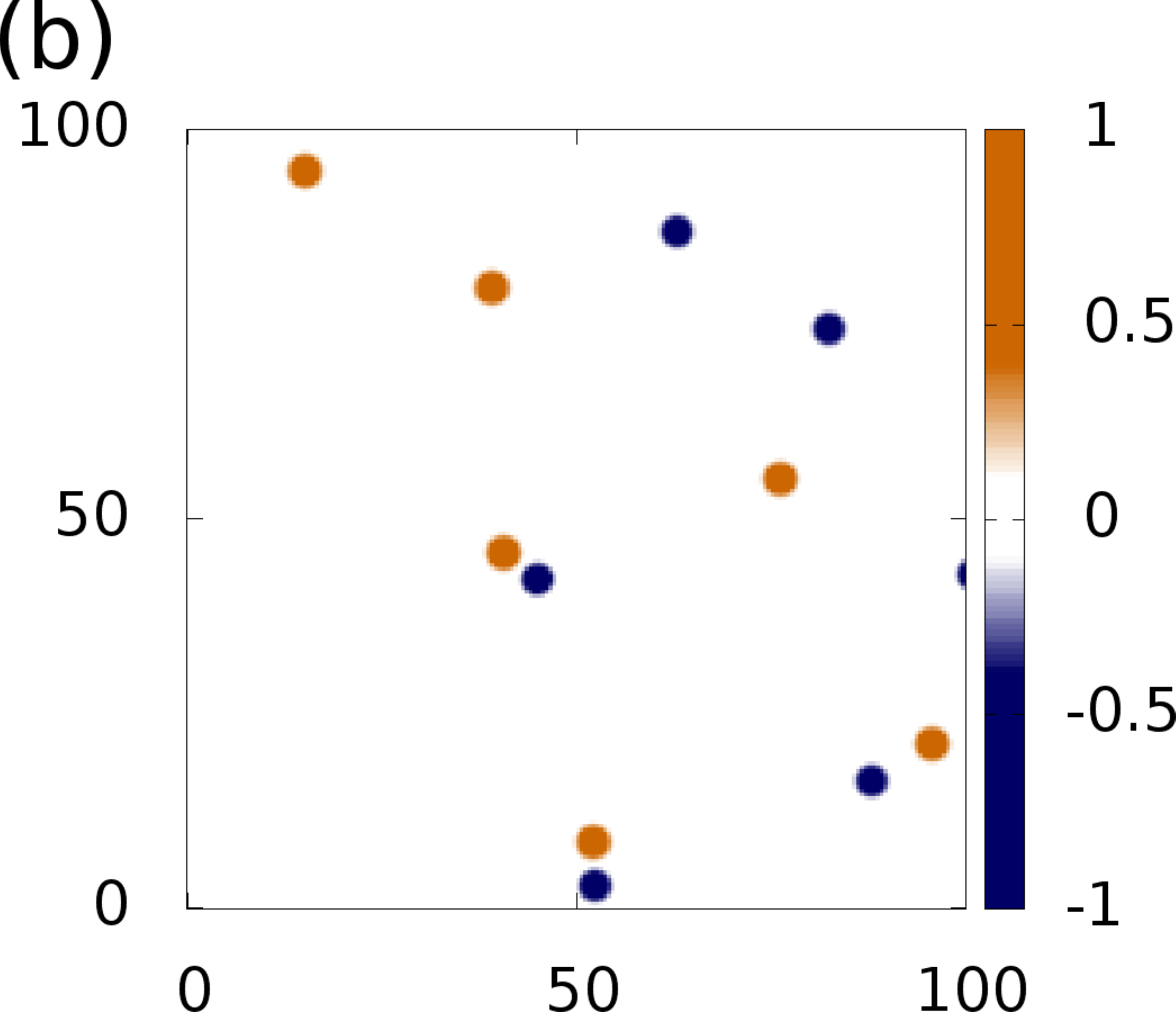}
	\includegraphics[width=0.4\linewidth]{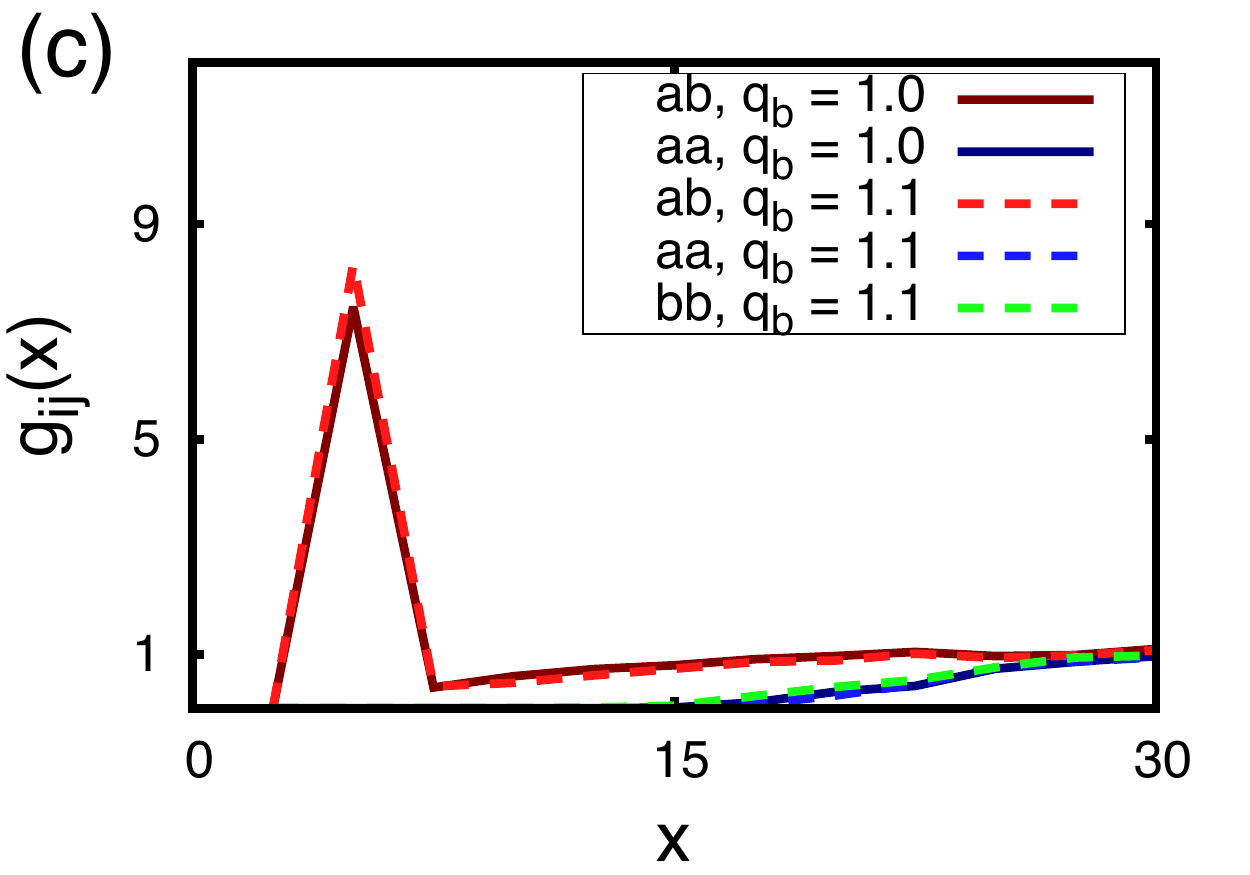}\\
	\includegraphics[width=0.29\linewidth]{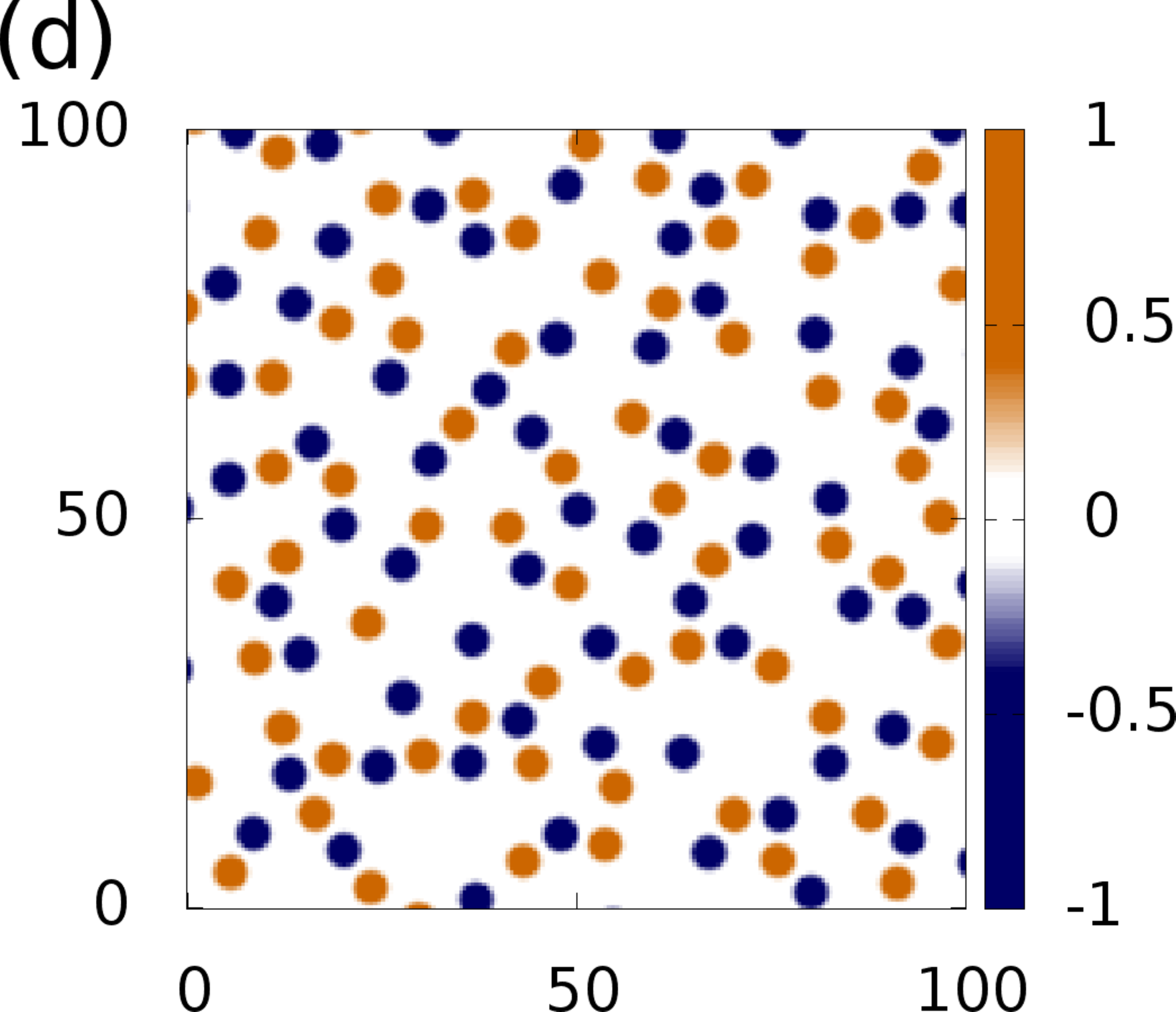}
	\includegraphics[width=0.29\linewidth]{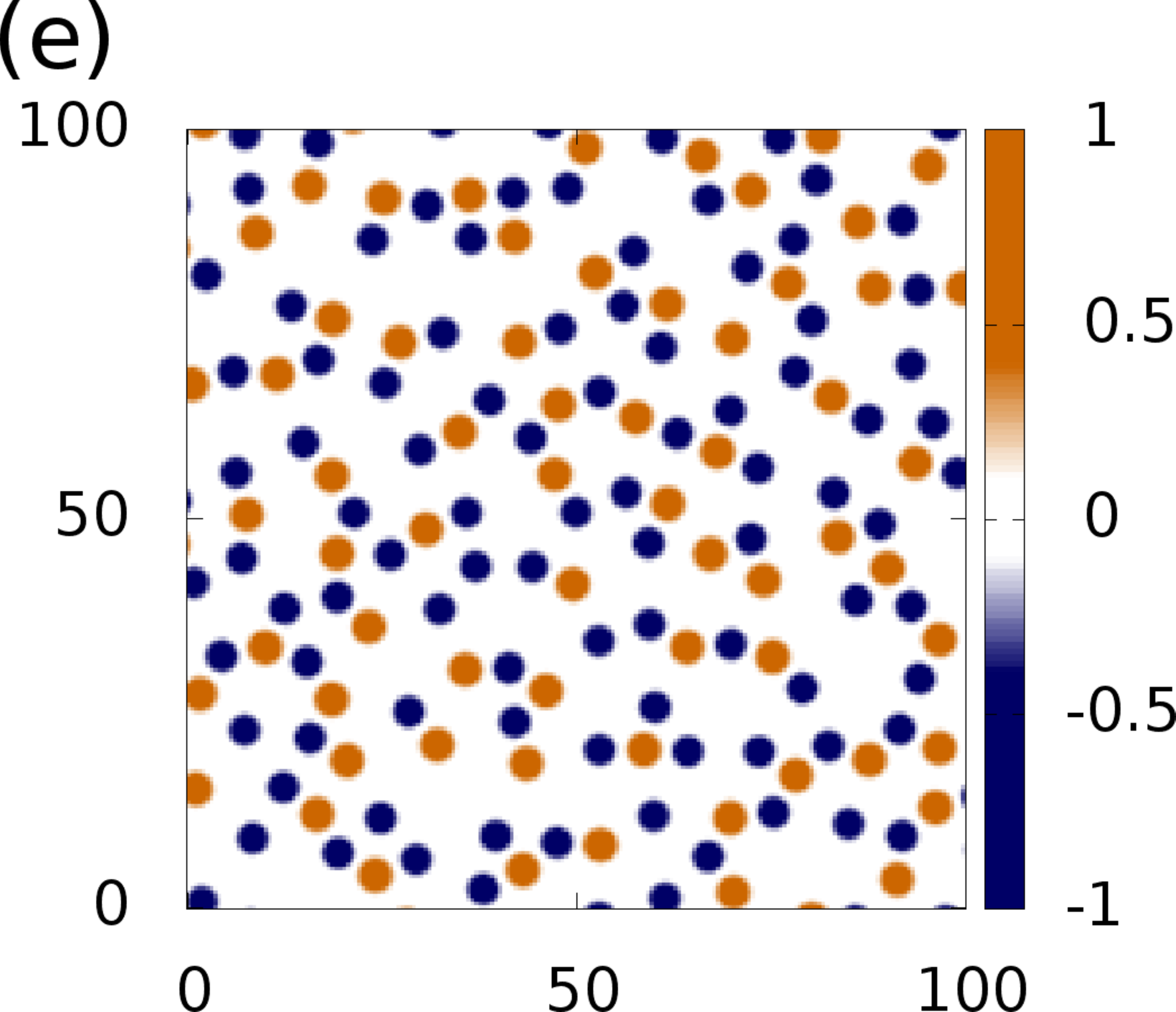}
	\includegraphics[width=0.4\linewidth]{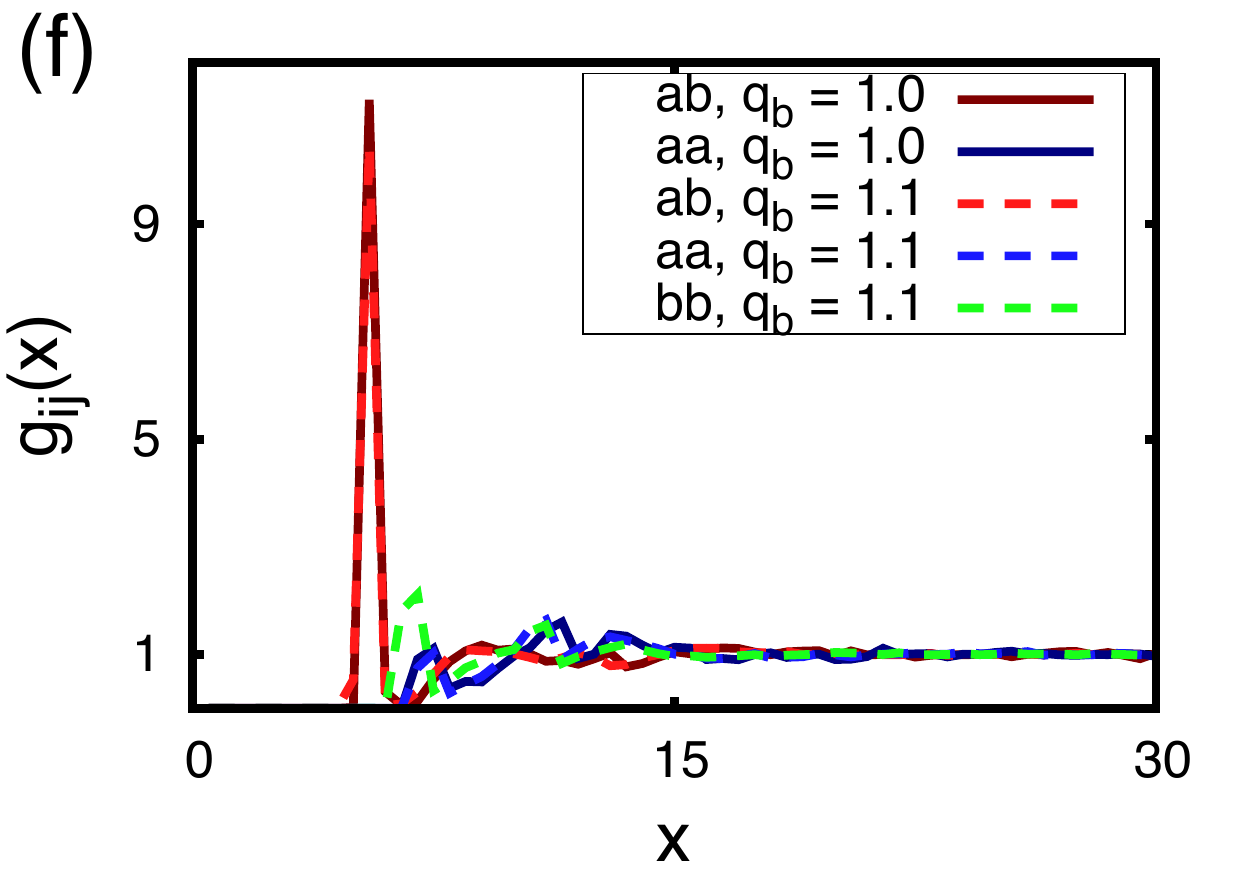}\\
	\includegraphics[width=0.29\linewidth]{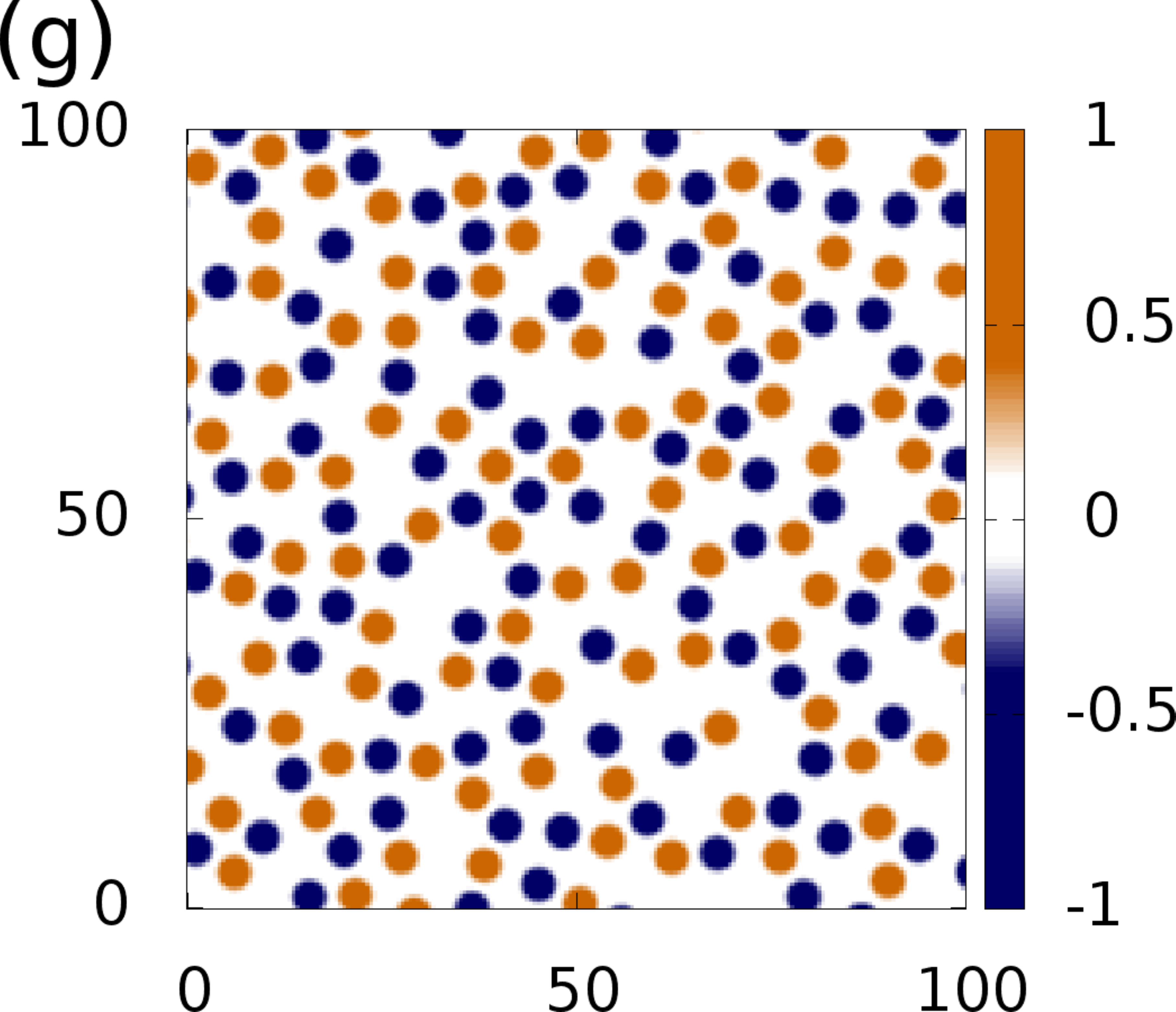}
	\includegraphics[width=0.29\linewidth]{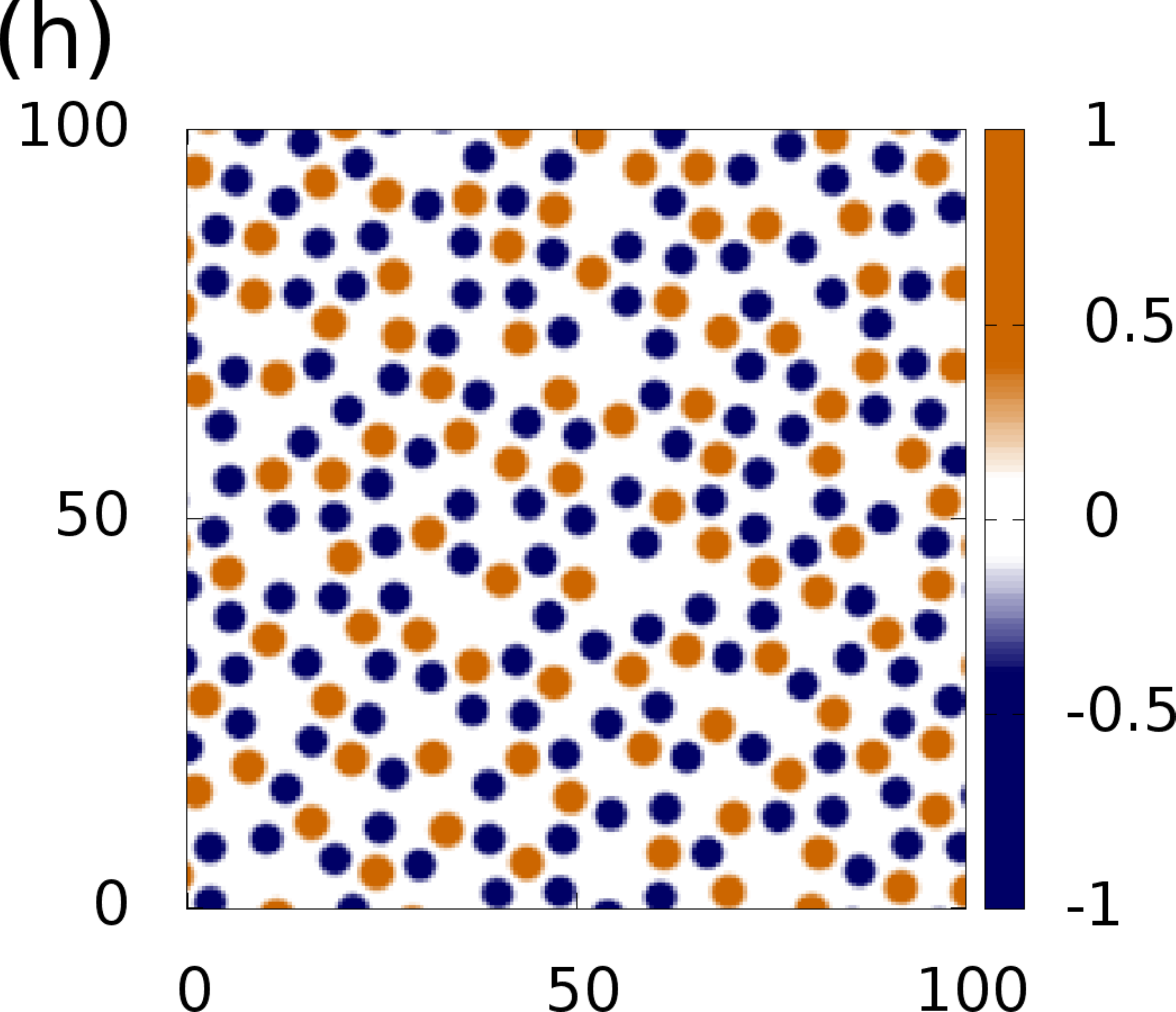}
	\includegraphics[width=0.4\linewidth]{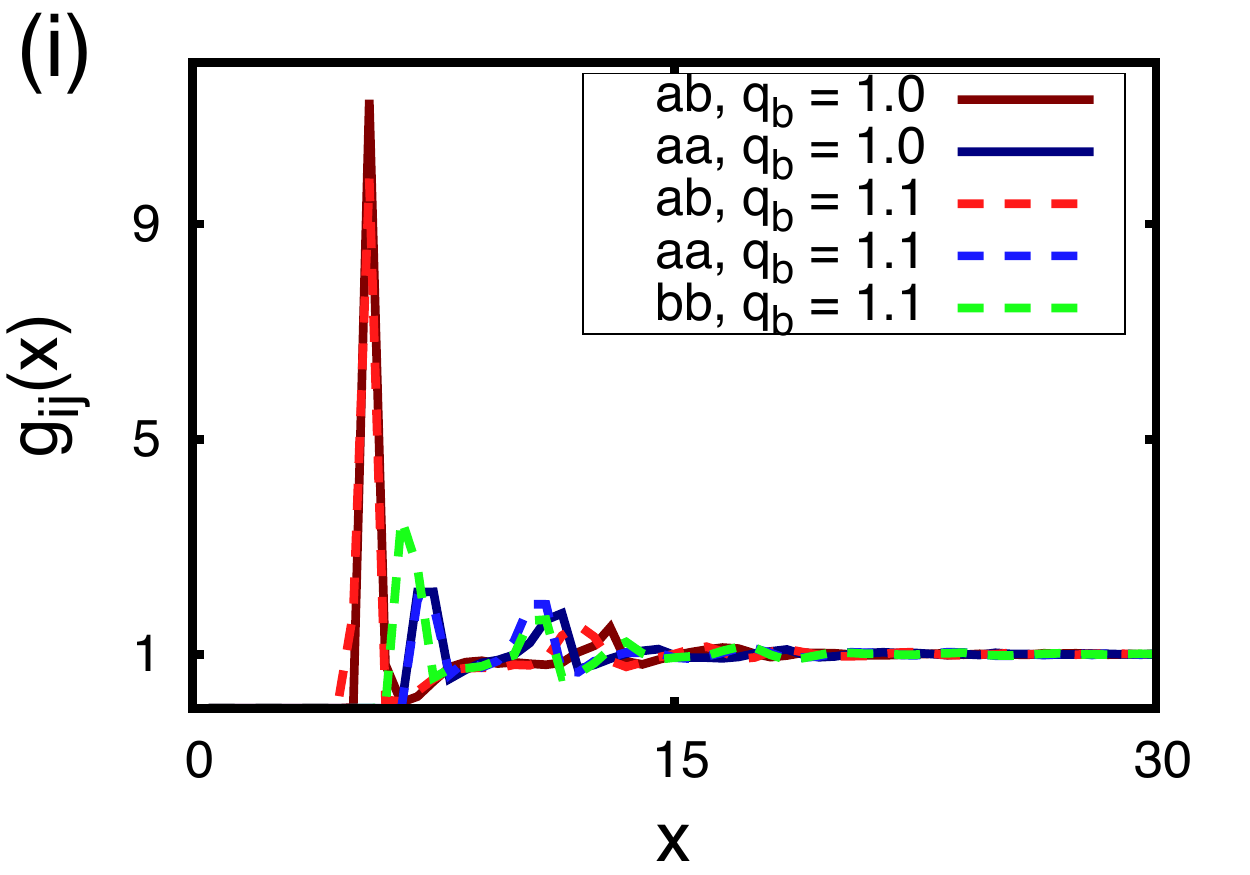}\\
	\includegraphics[width=0.29\linewidth]{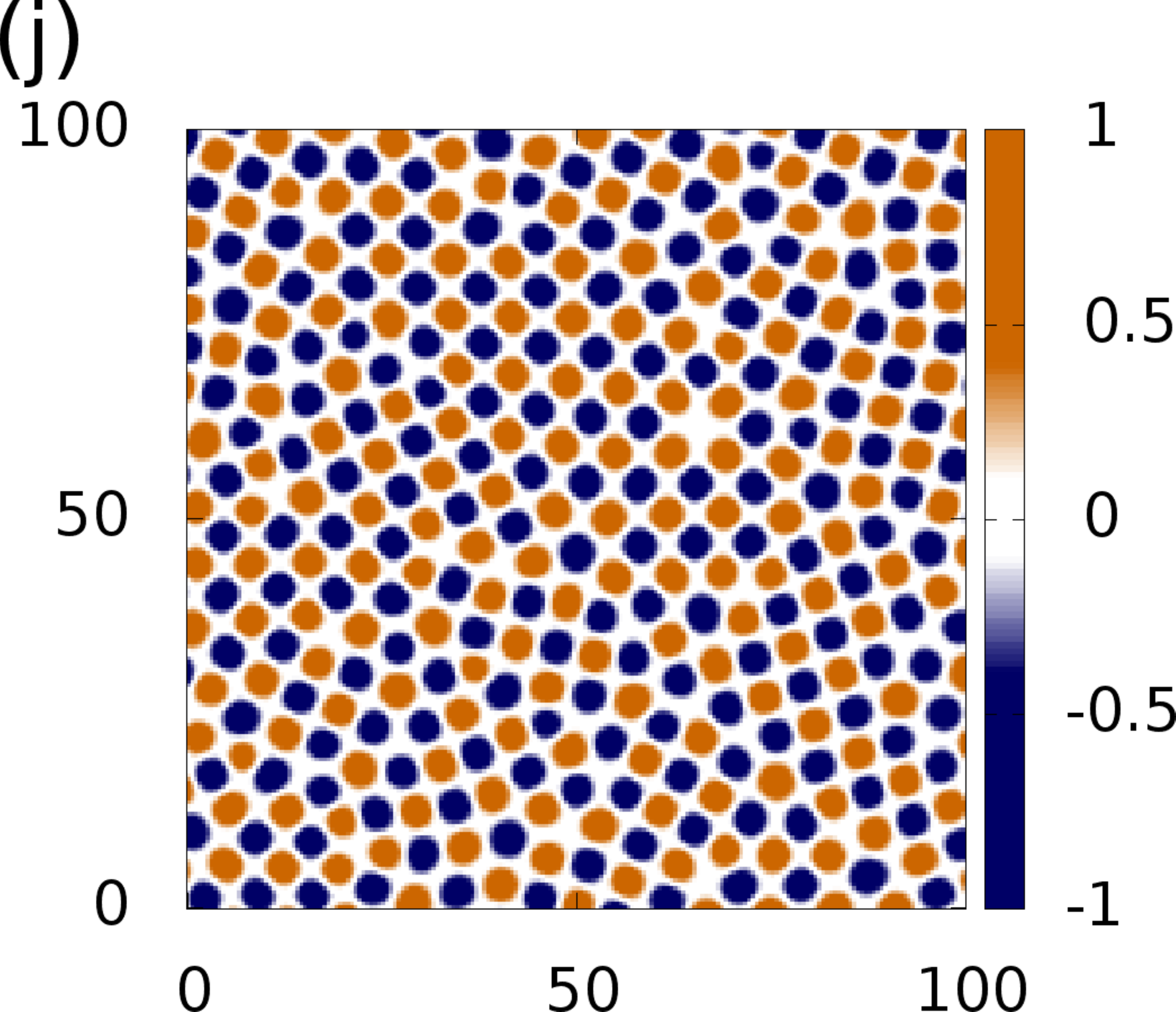}
	\includegraphics[width=0.29\linewidth]{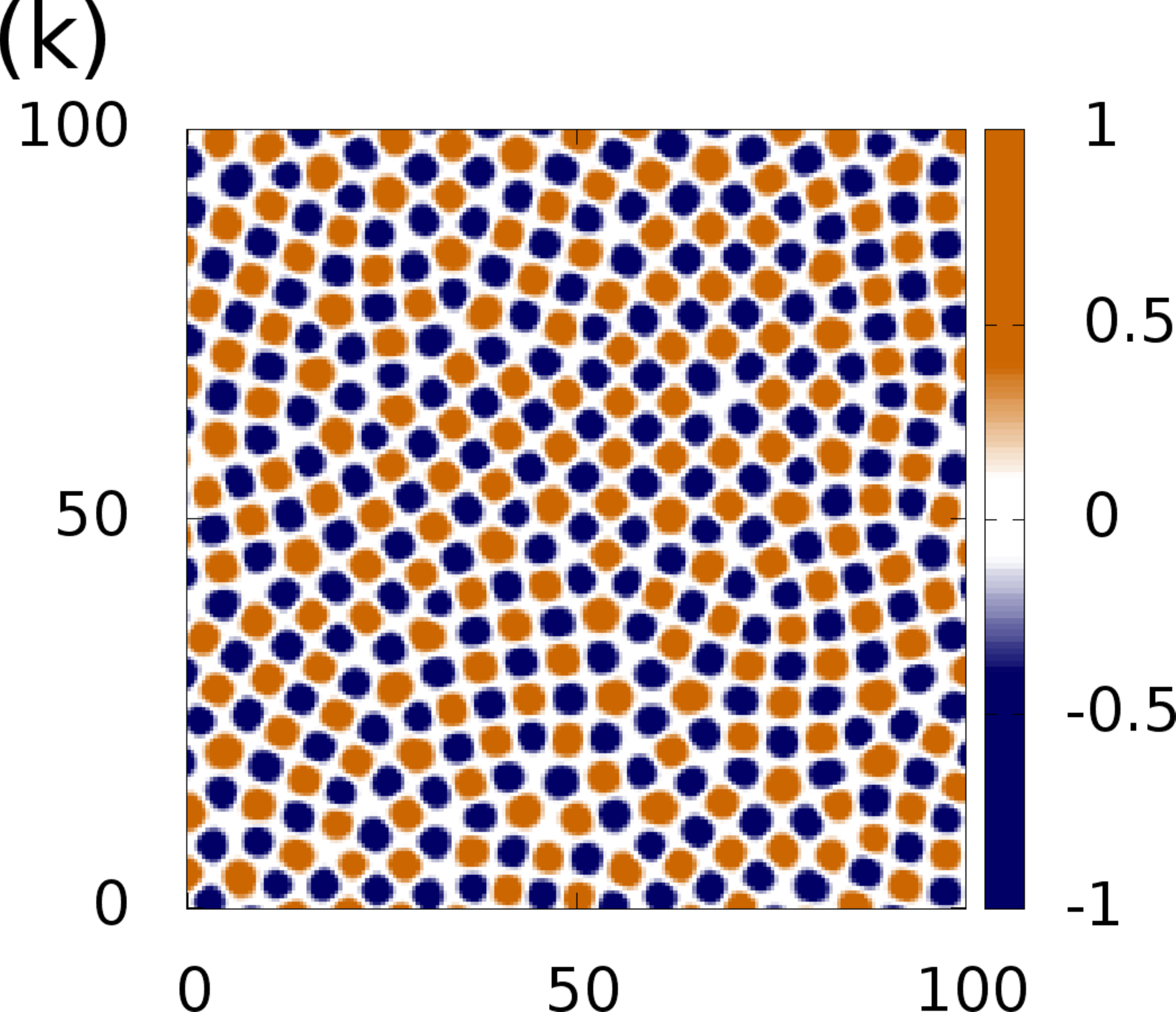}
	\includegraphics[width=0.4\linewidth]{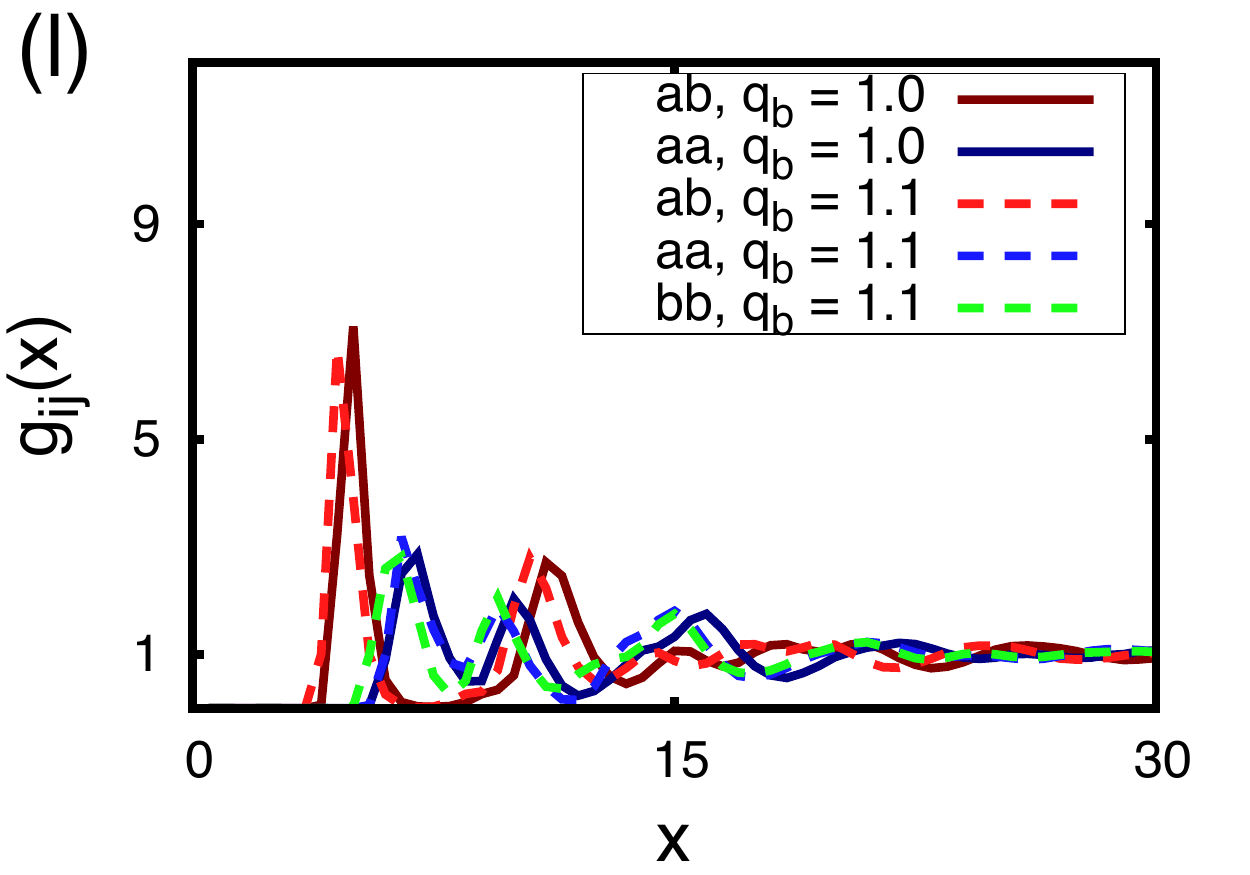}
	\caption{In the left hand column we display typical simulation results for $\Delta \hat{\phi}$ for the 
		      symmetrical case where $q_a = q_b = 1$.  In the middle column we display results from the 
		      asymmetrical case where $q_a = 1$ and $q_b = 1.1$.  The orange regions show where there is 
		      a $\phi_a$ bump, while the blue show the $\phi_b$ bumps which are slightly smaller in the 
		      asymmetric mixture.  In the right hand column the radial distribution functions $g_{ij}(x)$ are 
		      shown for the symmetrical case (solid lines) and the asymmetrical case (dashed lines).  The 
		      parameter values are: $\alpha = 1$, $\eta = 4$, $r = -0.9$, $\barA = \barB = \bar{\phi}$, where 
		      (a) - (c) $\bar{\phi} = 0$, (d) - (f) $\bar{\phi} = 0.04$, (g) - (i) $\bar{\phi} = 0.06$ and (j) - (l) 
		      $\bar{\phi} = 0.15$.}
	\label{figTPCTwo}
\end{figure} 

We begin by considering how the two order parameter profiles change as we alter their average values
$\barA = \barB = \bar{\phi}$.  Thus we keep the concentration of the mixture fixed at $c = 0.5$, where

\begin{equation}
	c = \frac{\barA}{\barA + \barB}.
	\label{eqConc} 
\end{equation}

We set the other parameter values to $\alpha = 1$, $r = -0.9$ and $\eta = 4$.  In Fig.~\ref{figTPCTwo} we 
display typical results.  We plot the normalised difference between the two order parameters $\Delta \hat{\phi}$ 
(as defined in Eq.~(\ref{eqhatPhi})).  In the left column we show profiles from the symmetric case and in the 
middle column we display the profiles from the asymmetric system.  In the right column we present the radial 
distribution functions, which are obtained by averaging over at least fifty runs, each with different realisations 
of the initial noise.  The solid lines show the radial distribution functions for the symmetric case and the dashed 
lines show the asymmetric case.  It is very apparent that this region of the parameter space shares many 
similarities with the one component model in both one and two dimensions.  If we select a small value of 
$\bar{\phi}$ we find localised peaks surrounded by vacant areas, as shown in Fig.~\ref{figTPCTwo} (a) and 
(b).  We observe a tendency for bumps in $\phi_a$ and in $\phi_b$ to sit pairwise next to each other 
resembling configurations occurring in mixtures of oppositely charged colloidal particles 
\cite{LCH05,HCR06,HLB06}.  When $\barA \approx \barB$, the arrangement of the bumps also resembles 
snapshots of monovalent salts.  It is very difficult to differentiate between the structures formed by the 
symmetric and asymmetric models for small values of $\bar{\phi}$.  This is because structurally, there is very 
little difference between the two cases.  If we examine the radial distribution functions $g_{ij}$ (where 
$i, j = a, b$) for the symmetric and asymmetric systems [Fig.~\ref{figTPCTwo}(c)] we observe that the average 
distance between the different bumps seems to be independent of the $q$ values (any differences between 
the curves is of the same order of magnitude as the statistical error).  This is due to the large vacant areas, 
which means that there are not many bumps which are close to one another, especially between bumps in the 
same species ($aa$ and $bb$).

As we increase the values of $\bar{\phi}$ we find that the number of bumps of both species increases.  In 
Fig.~\ref{figTPCTwo}(d)--(f) we show the case where $\bar{\phi} = 0.04$ and in Fig.~\ref{figTPCTwo}(g)--(i) 
we show the case where $\bar{\phi} = 0.06$.  There is now a clear difference between the symmetric (d), (g) 
and the asymmetric (e), (h) cases.  We observe a larger number of bumps in $\phi_b$ when $q_b = 1.1$.  This 
is because the larger value of $q$ reduces the length scale of the modulations, meaning that more bumps can 
be created before the value of $\phi_b$ becomes small (and negative) in vacant areas.  There is an optimum 
value of $\phi_a$ and $\phi_b$ in the vacant (uniform) areas which depends on the parameter values. This 
explains why increasing the value of $\bar{\phi}$ increases the number of bumps (i.e.,~more modulations are 
needed in order to reach the optimum value of $\phi$ in the vacant regions).  These intermediate values of 
$\bar{\phi}$ produce profiles with bump configurations that resemble real fluid structures.  However, in stark 
contrast to the one component system (shown in Fig.~\ref{figOneCompCorr}(b)), we now find the formation of 
chains of alternating bumps reminiscent of structures observed in charged fluids.  The radial distribution 
functions in Fig.~\ref{figTPCTwo}(f) and (i) show that the asymmetry induced by the different values of $q$ 
begins to take effect at these intermediate values of $\bar{\phi}$.  We observe that statistically the bumps sit 
closer together in the asymmetrical case, especially when two bumps in $\phi_b$ are next to each other 
($bb$, shown by green dashed line).  This is due to the decreased size of the bumps in $\phi_b$, allowing 
them to sit slightly closer to their neighbours.

Increasing the average order parameter values $\bar{\phi}$ further we begin to observe the formation of 
crystalline structures as shown by Fig.~\ref{figTPCTwo}(j)--(l).  The interesting thing is that now we observe 
square ordering of the particles instead of the hexagonal ordering which is present in the regular PFC model 
and the one component VPFC model.  This implies that as we increase the concentration of one of the species 
from $c = 0$ (almost a pure one-component system) to $c = 0.5$ there must be a transition from hexagonal to 
square ordering of bumps, this is something we return to below in Sec.~\ref{ssOrdering}.  Just as for the one 
component system, we find that there are more modulations in $\phi_b$ when $q_b = 1.1$.  The profiles 
obtained with these parameter values resemble a compound crystal structure with vacancies and grain 
boundaries.  The radial distribution functions in Fig.~\ref{figTPCTwo}(l) show that the smaller size difference of 
the $\phi_b$ bumps in the asymmetric mixture has a large impact on the average position of the bumps in the 
structure compared to the symmetric mixture.  This is because the higher concentration of particles forces them 
all closer together resulting in all pairs of bumps $aa$, $bb$ and $ab$ being closer together. 

\begin{figure}
	\begin{minipage}[h]{0.54\linewidth}
		\includegraphics[width=0.49\linewidth]{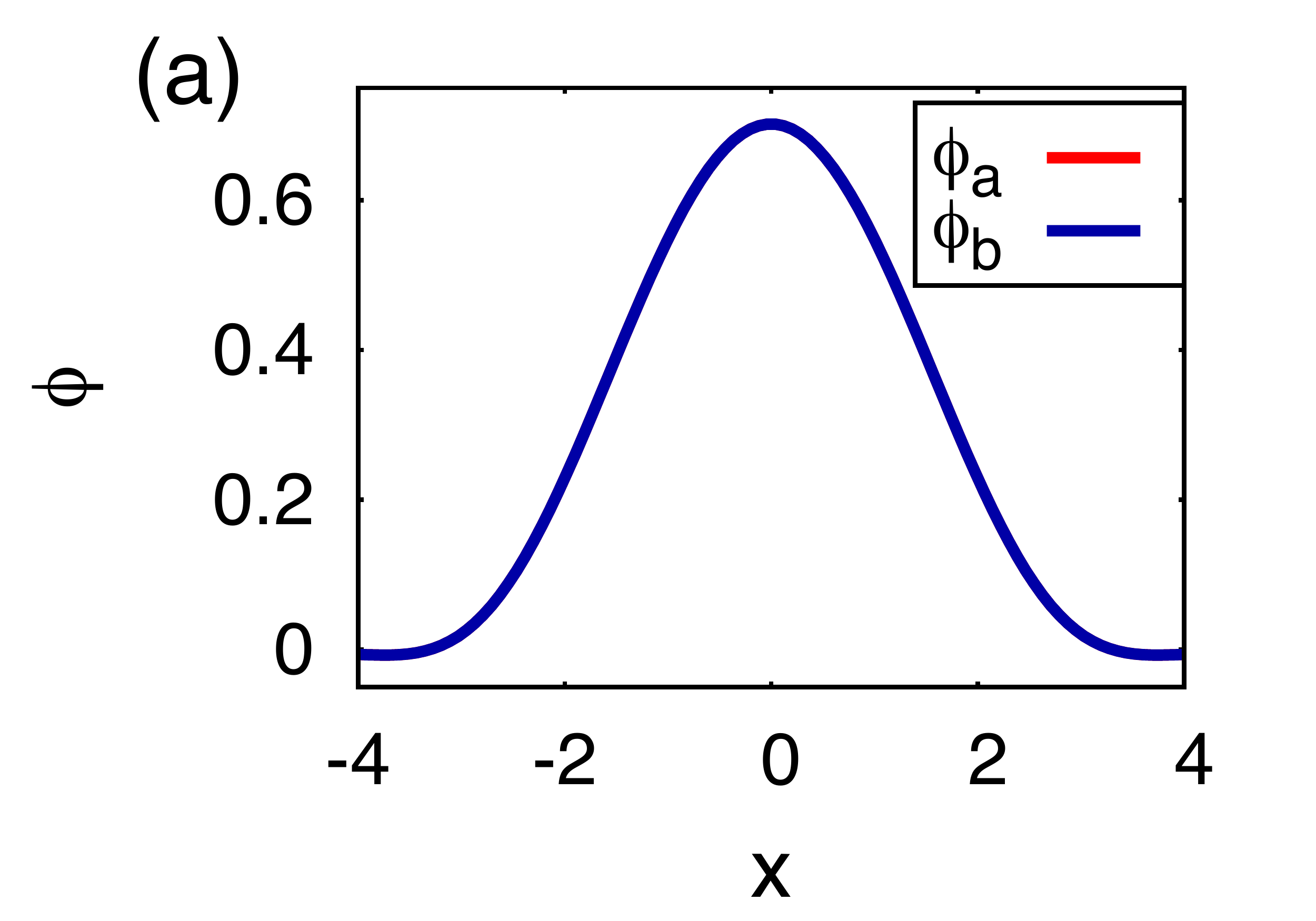}
		\includegraphics[width=0.49\linewidth]{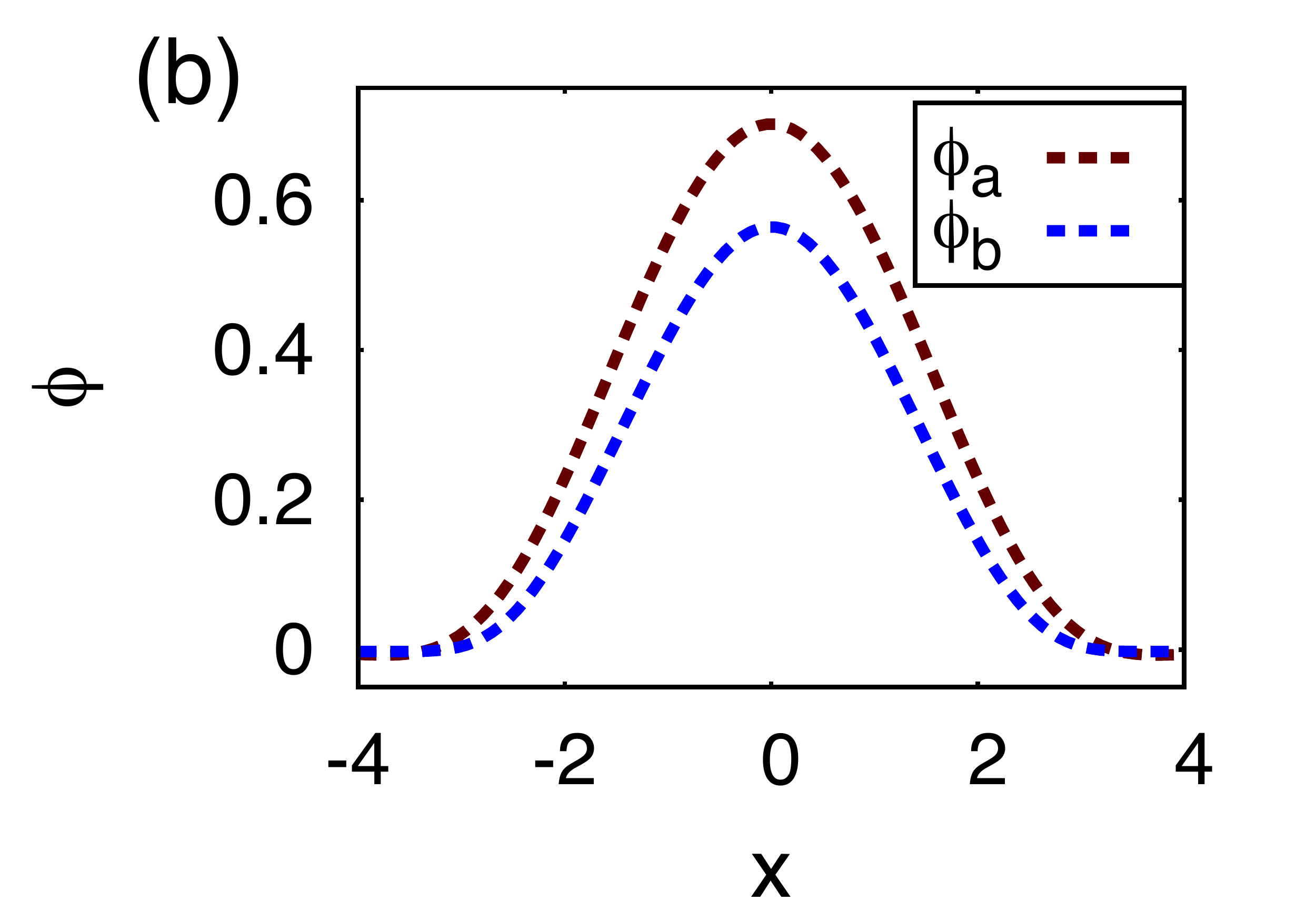}\\
		\includegraphics[width=0.49\linewidth]{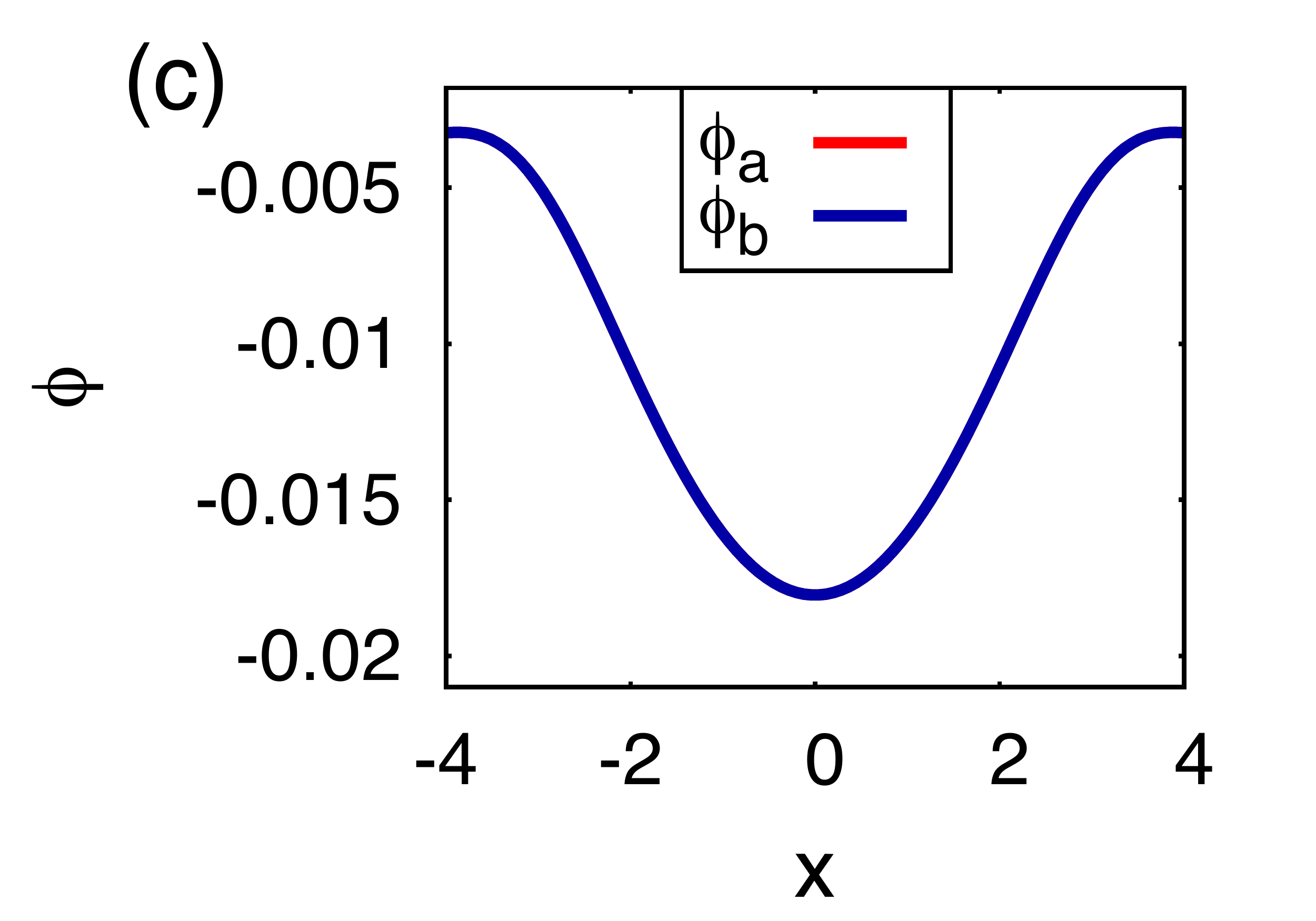}
		\includegraphics[width=0.49\linewidth]{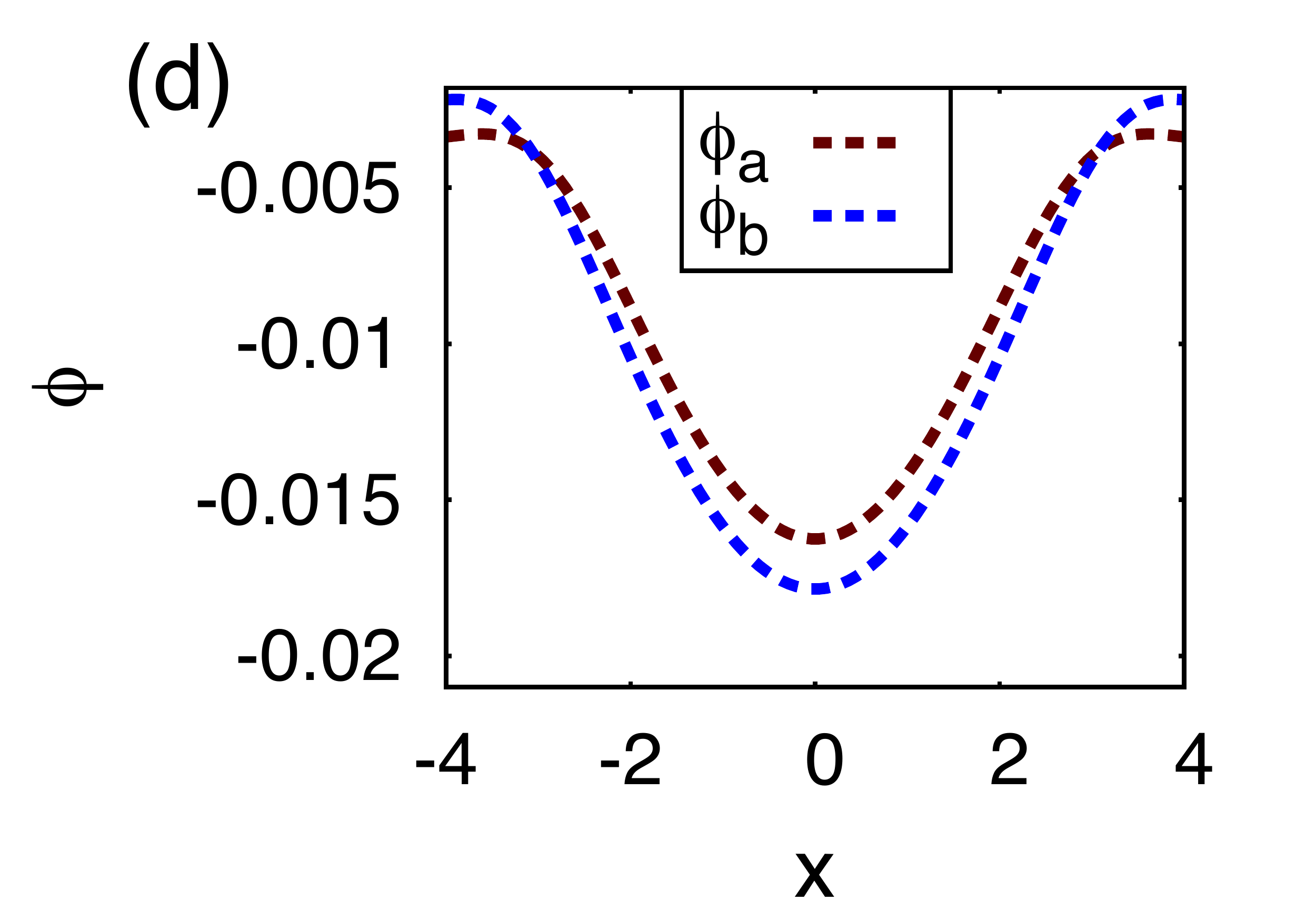}
	\end{minipage}
	\begin{minipage}[h]{0.44\linewidth}
		\includegraphics[width=\linewidth]{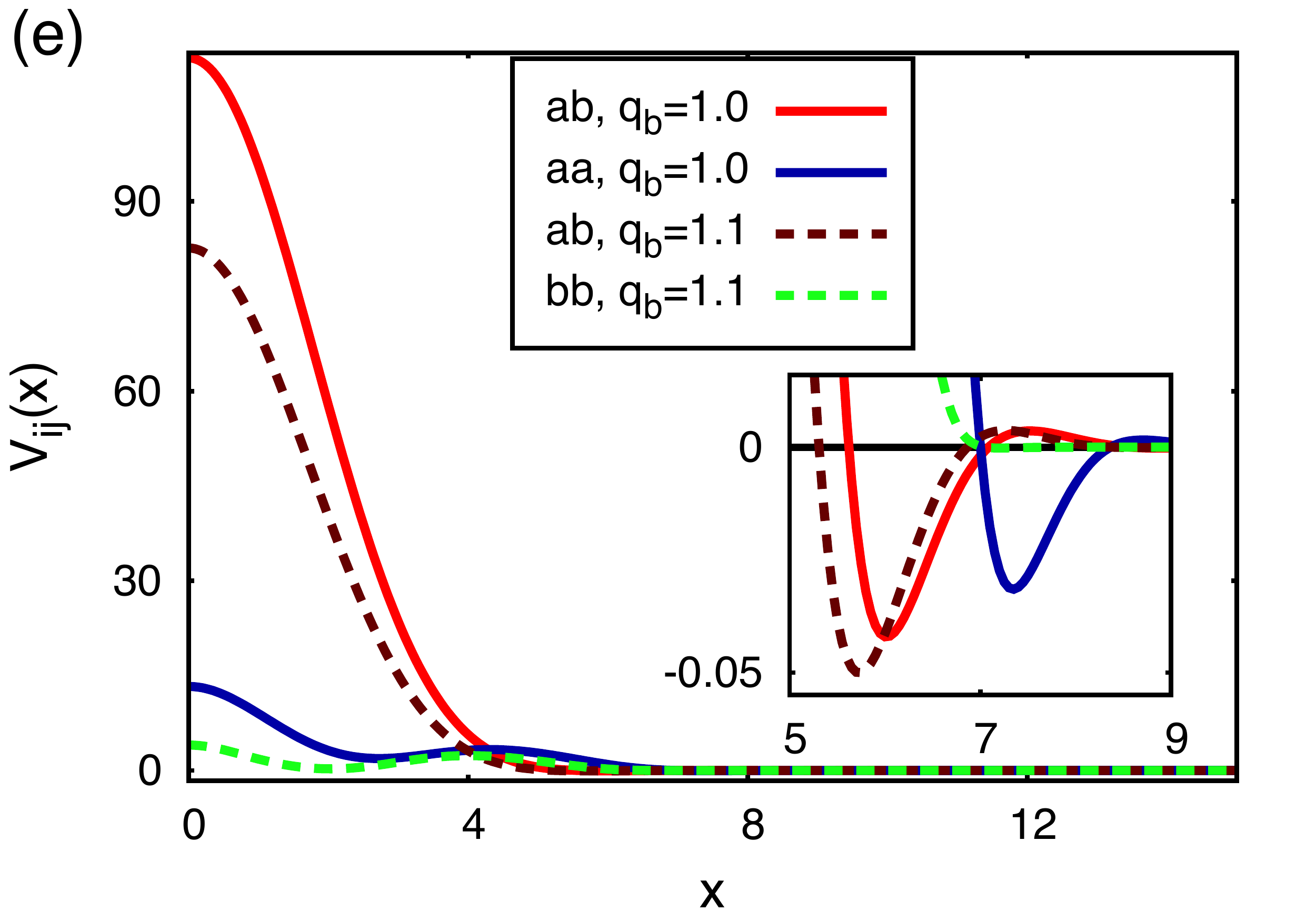}
	\end{minipage}
	\caption{Fits to the shape of individual bumps (cf.~Fig.~\ref{figOneCompPot}) in the (a) symmetric, when
		      $q_b = 1$, and (b) asymmetric, when $q_b = 1.1$, systems and the corresponding dips in the 
		      other order parameter profile which occur under the bumps in (c) the symmetric and (d) the 
		      asymmetric systems.  In the symmetric profiles (a) and (c) the $\phi_a$ and $\phi_b$ curves 
		      are equal everywhere. The bump profile in $\phi_a$ is virtually identical in the cases where 
		      $q_b = 1$ and $q_b = 1.1$. These fits are then used to calculate the effective pair potential 
		      between bumps, which are displayed in (e).  The inset displays a magnification of the tails of 
		      $V_{ij}(x)$.  The resulting pair potential $V(x)$ between two bumps in $\phi_a$ when 
		      $q_b = 1.1$ lies on top of the $aa, q_b = 1$ curve.  The parameter values are: $\alpha = 1$, 
		      $r = -0.9$, $\eta =4$, and $\barA = \barB = 0$.}		
	\label{figPotTwo}
\end{figure} 

In Fig.~\ref{figPotTwo}(a)--(b) we show the shape of the individual bumps in $\phi_a$ and $\phi_b$ 
obtained in the low density limit $\bar{\phi} \to 0$.  To determine these radially symmetric profiles we fit 
functions of the form $\theta(x)$ as defined above in Eq.~\eqref{eqFittedFunc}.  The bumps in $\phi_a$ are 
virtually identical for both the symmetrical and asymmetrical systems.  The $\phi_a$ bump in the symmetric 
system and the $\phi_b$ bump in the asymmetric system decay to different values due to the different 
values of $\phi_a$ and $\phi_b$ in the vacant areas of the asymmetrical system.  We observe that in this two 
component model, a bump in one order parameter profile coincides with a small depression in the other order 
parameter profile.  This is caused by the coupling term, which means that the combination of a bump in one 
order parameter and a hole in the other order parameter reduces the free energy of the system.  In 
Fig.~\ref{figPotTwo}(c)--(d) we show the shape of the `holes' which form in one profile under the bumps in 
the other order parameter field.  These are determined the same way as the bump profiles: by fitting a function 
of the form shown in Eq.~\eqref{eqFittedFunc} to data points obtained from simulations.  The depth of the holes 
is much smaller in size than the height of the bumps.  This is because the vacancy term prevents the hole from 
reaching large negative values of $\phi_a$ or $\phi_b$.

Using the fitted functions shown in Fig.~\ref{figPotTwo}(a)--(d) we calculate effective pair potentials 
$V_{ij}(x)$ between the different particles in the system ($i, j = a, b$).  We do this by determining the free energy 
for a system containing two bumps and their corresponding holes at various distances apart.  In 
Fig.~\ref{figPotTwo}(e) we display the effective pair potentials for both the symmetric (solid lines) and the 
asymmetric (dashed lines) systems.  The results show that there is an attraction between all of the bumps, just 
as we found for the one component system [Fig.~\ref{figOneCompPot}(b)].  In both the symmetric and 
asymmetric cases we find that the attraction between two bumps from different species ($ab$) is stronger and 
occurs at a smaller value of $x$ than that of two bumps of the same species ($aa$ and $bb$).  This explains 
the tendency for the bumps to form chains at intermediate values of $\phi_a$ and $\phi_b$ 
[Figs.~\ref{figTPCTwo}(d), (e), (g) and (h)] and square ordered crystalline structures at larger values of 
$\phi_a$ and $\phi_b$ [Fig.~\ref{figTPCTwo}(j)--(k)].  This is also consistent with the appearance of the 
large peak in $g_{ab}(x)$ which occurs at a smaller $x$ value than the main peaks in $g_{aa} (x)$ and 
$g_{bb}(x)$ -- see Figs.~\ref{figTPCTwo}(c), (f), (i) and (l).  The effective pair potential $V_{aa}(x)$ is almost 
identical in the symmetric and the asymmetric systems.  This suggests that the small hole which appears in 
$\phi_b$ has little effect on the interaction between the bumps.  The major difference between the symmetrical 
and asymmetrical systems is that in the asymmetric mixture the minimum of the pair potentials $V_{ab}(x)$ 
and $V_{bb}(x)$ are at smaller values of $x$ than in the symmetric mixture.  This is due to the reduced size of 
the $\phi_b$ bumps in the latter.  The minimum in $V_{bb}(x)$ is at a slightly larger value of $x$ than the 
minimum in $V_{aa}(x)$ and the attraction is also much weaker (in fact it is so much weaker that the minimum 
is barely visible in this plot).  This to some extent explains why the effect of the asymmetry is not visible for 
smaller values of $\phi_a$ and $\phi_b$, but becomes apparent for larger values of $\phi_a$ and $\phi_b$, 
where the vacant areas become smaller and we observe a close packing of the particles.

\subsection{Bond angles and the transition between hexagonal and square ordering}
\label{ssOrdering}

In the two dimensional one component model (Eqs.~\eqref{eqPFCDyn} and \eqref{eqVPFCOne}) we 
observe hexagonally ordered structures for certain parameter values [Fig.~\ref{figPhaseD2D}(a)
and Fig.~\ref{figOneCompCorr}(c)].  However, in the two component model when $\barA = \barB$, we 
instead observe a square ordered crystalline structure which alternates between species $a$ and 
species $b$ {Figs.~\ref{figTPCTwo}(j)--(k)].  Thus, as the composition of the mixture is varied we should 
see a transition/crossover from hexagonal to square ordering.  The number of bumps observed in each field 
$\phi_i$ depends on the respective average value $\bar{\phi_i}$.  When the concentration $c \approx 0$ or 
$c \approx 1$, where $c$ is defined in Eq.~\eqref{eqConc}, (i.e.,~when either $\barB \gg \barA$ or 
$\barA \gg \barB$) then the resulting order parameter profile $\Delta \hat{\phi}(x)$ has many more bumps of 
one type than of the other, and in these two limits we again observe hexagonal ordering.  Note that $c$ in 
Eq.~\eqref{eqConc} is not a bump concentration, but instead is a ratio between the two average order 
parameter values.  As the $\phi_i$ may take a negative value, for $c = 0$ there are still a few bumps of $a$ 
and similarly there are still some species $b$ bumps when $c = 1$.  When $c = 0.5$ the number of bumps is 
roughly the same in both species for the symmetrical case ($q_a$ = $q_b$), but this is not necessarily true for 
the asymmetrical system ($q_a \ne q_b$).  When $\barA = \barB$ and $q_a < q_b$ there are more $b$ bumps 
than $a$ bumps.

\begin{figure}
	\includegraphics[width=0.3\linewidth]{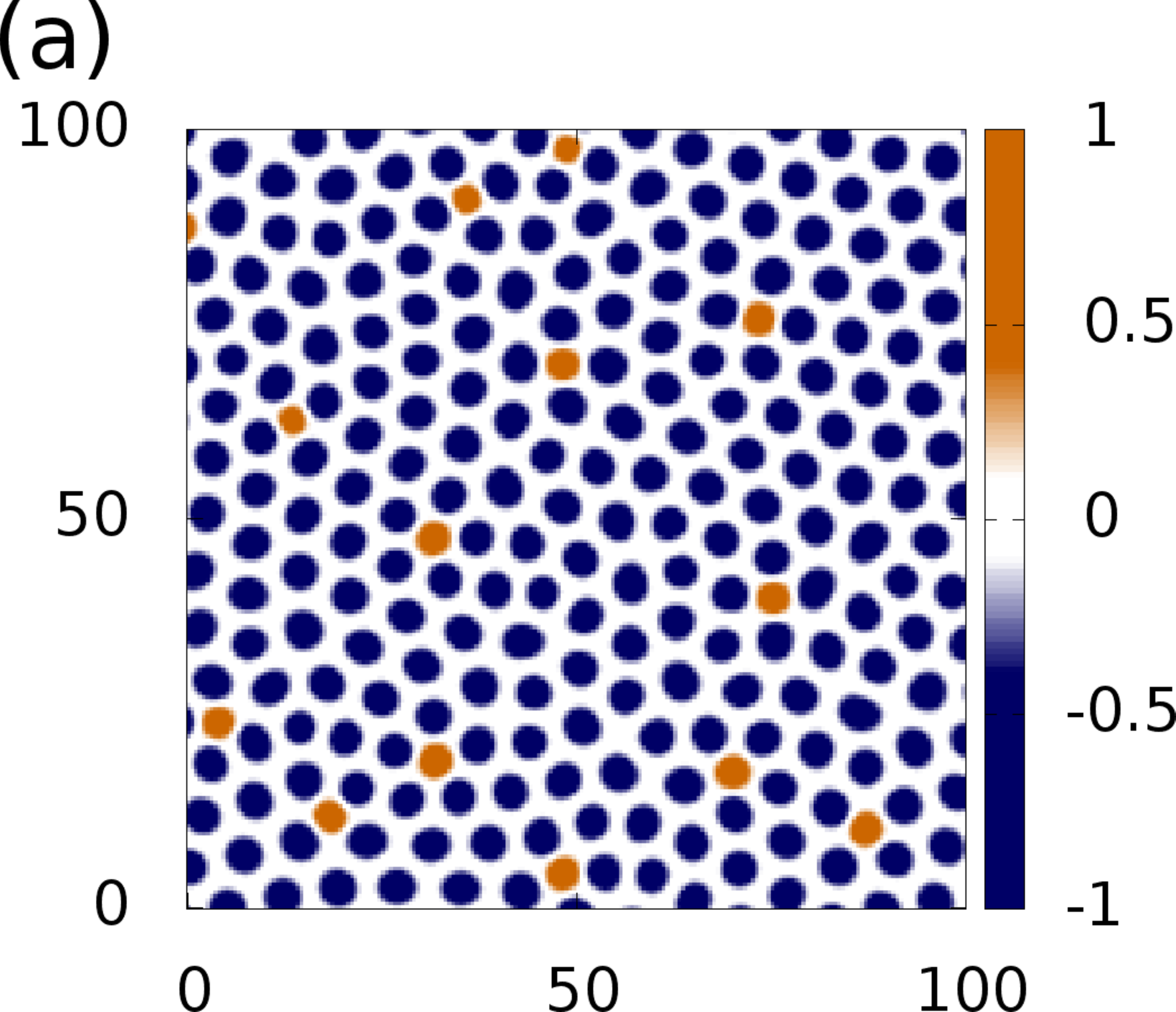}
	\includegraphics[width=0.3\linewidth]{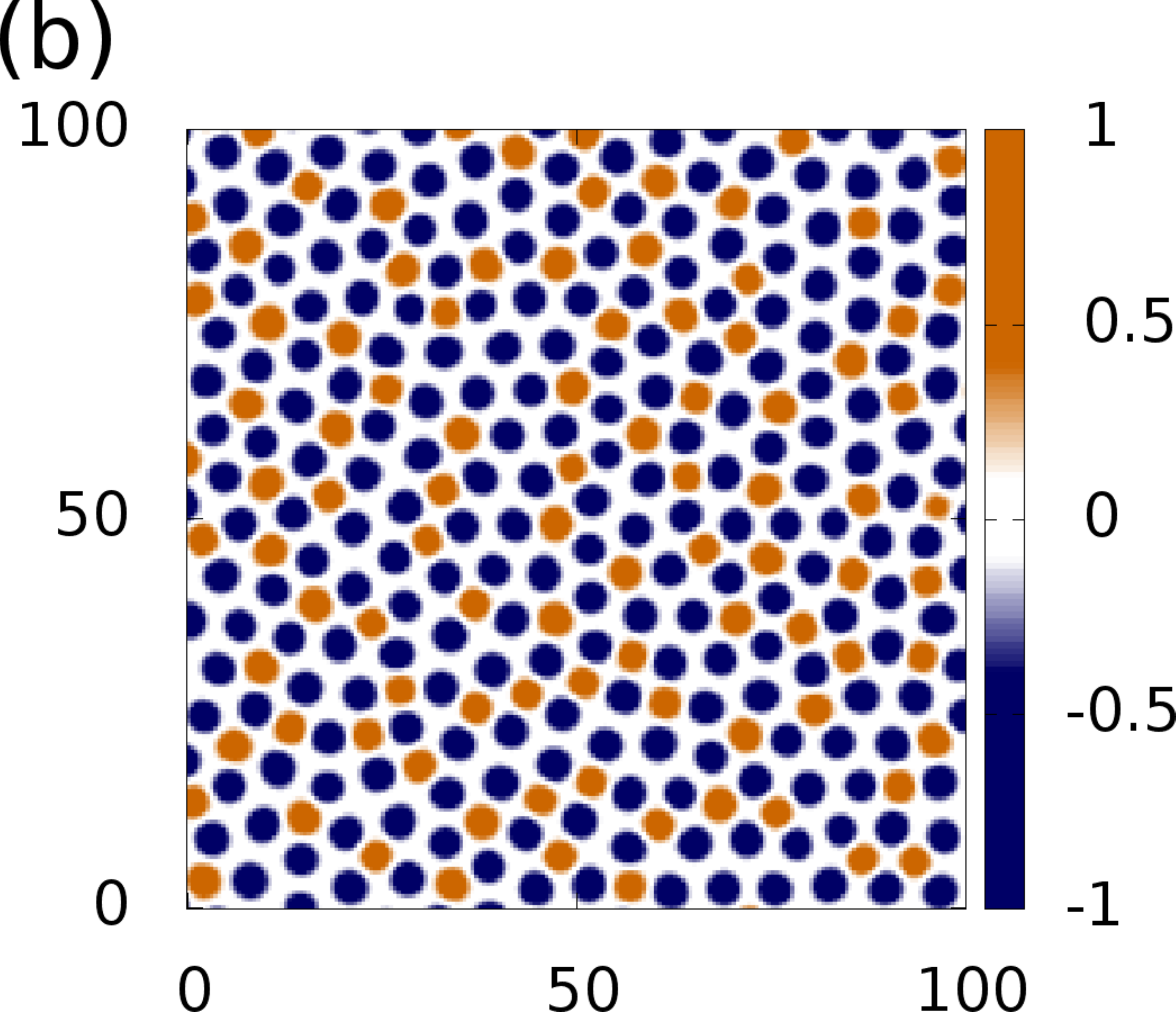}
	\includegraphics[width=0.3\linewidth]{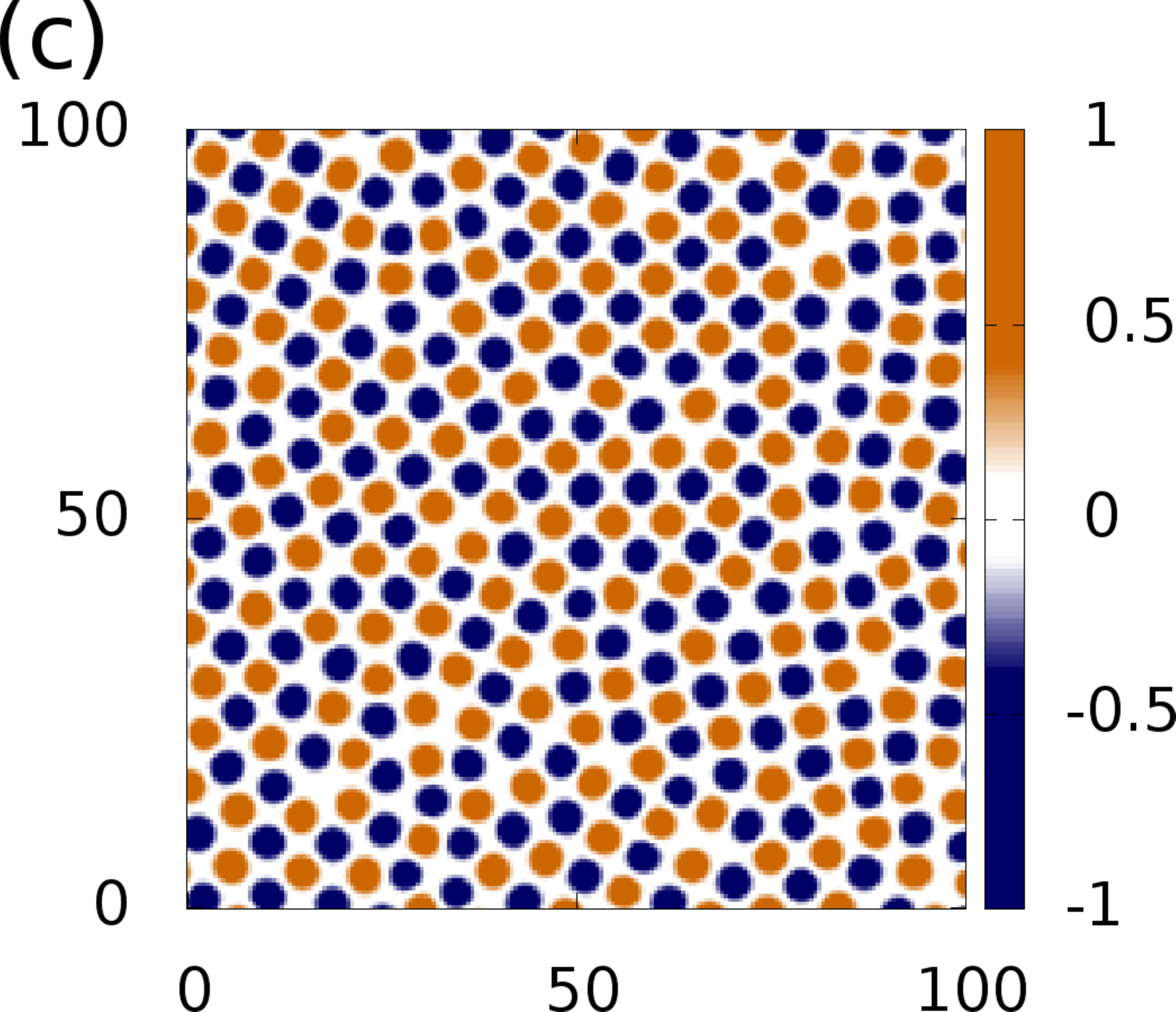}\\
	\includegraphics[width=0.3\linewidth]{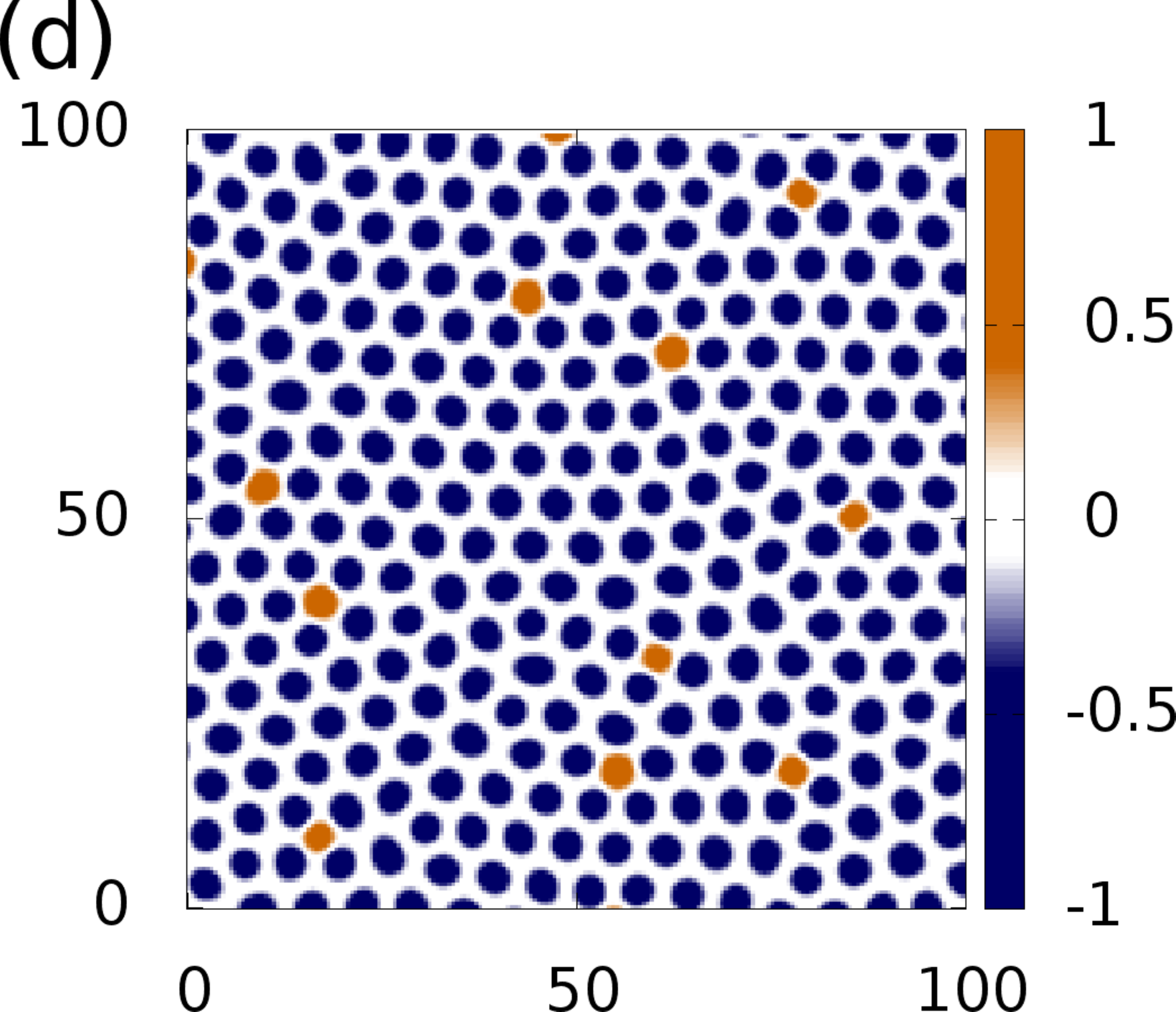}
	\includegraphics[width=0.3\linewidth]{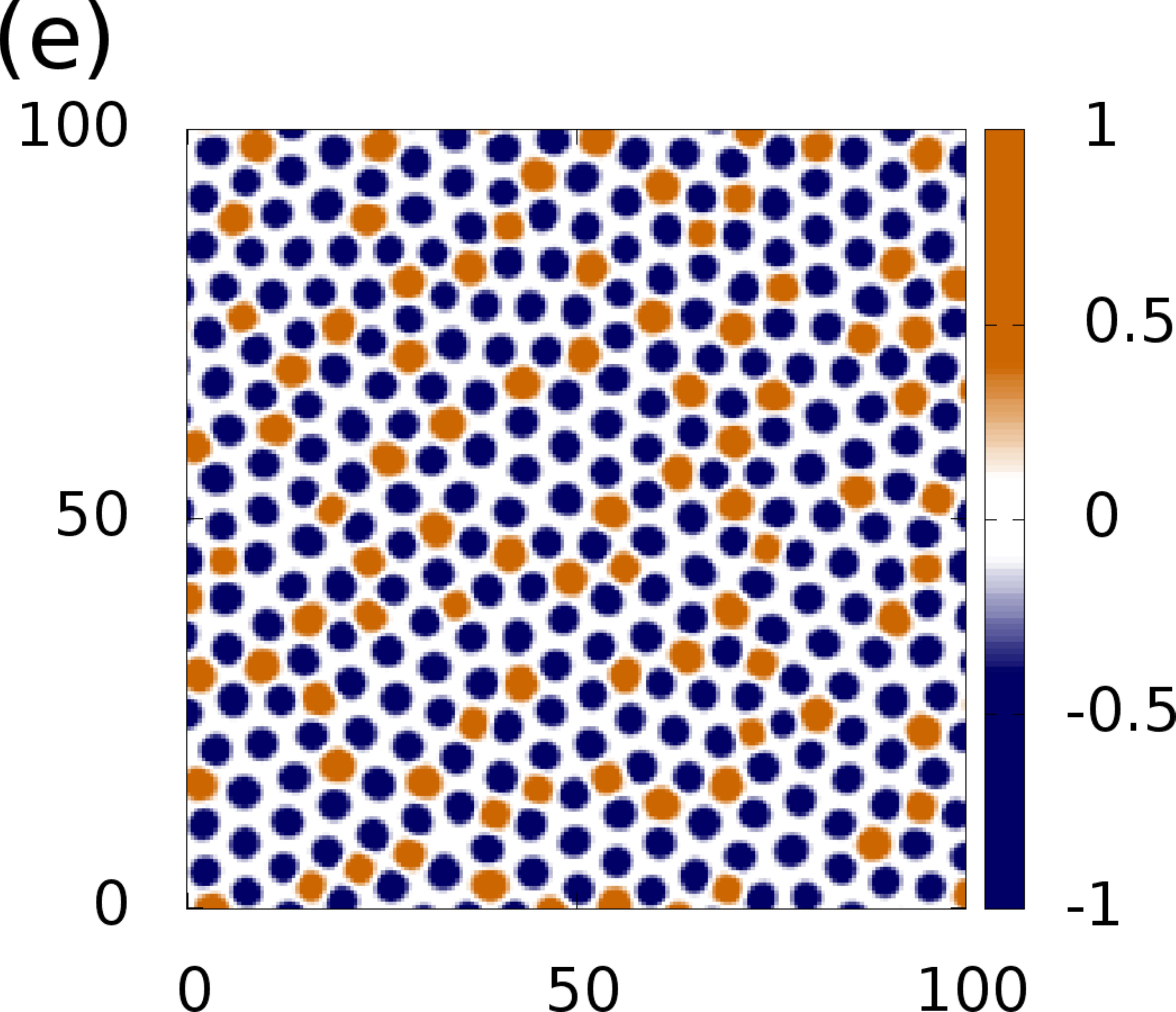}
	\includegraphics[width=0.3\linewidth]{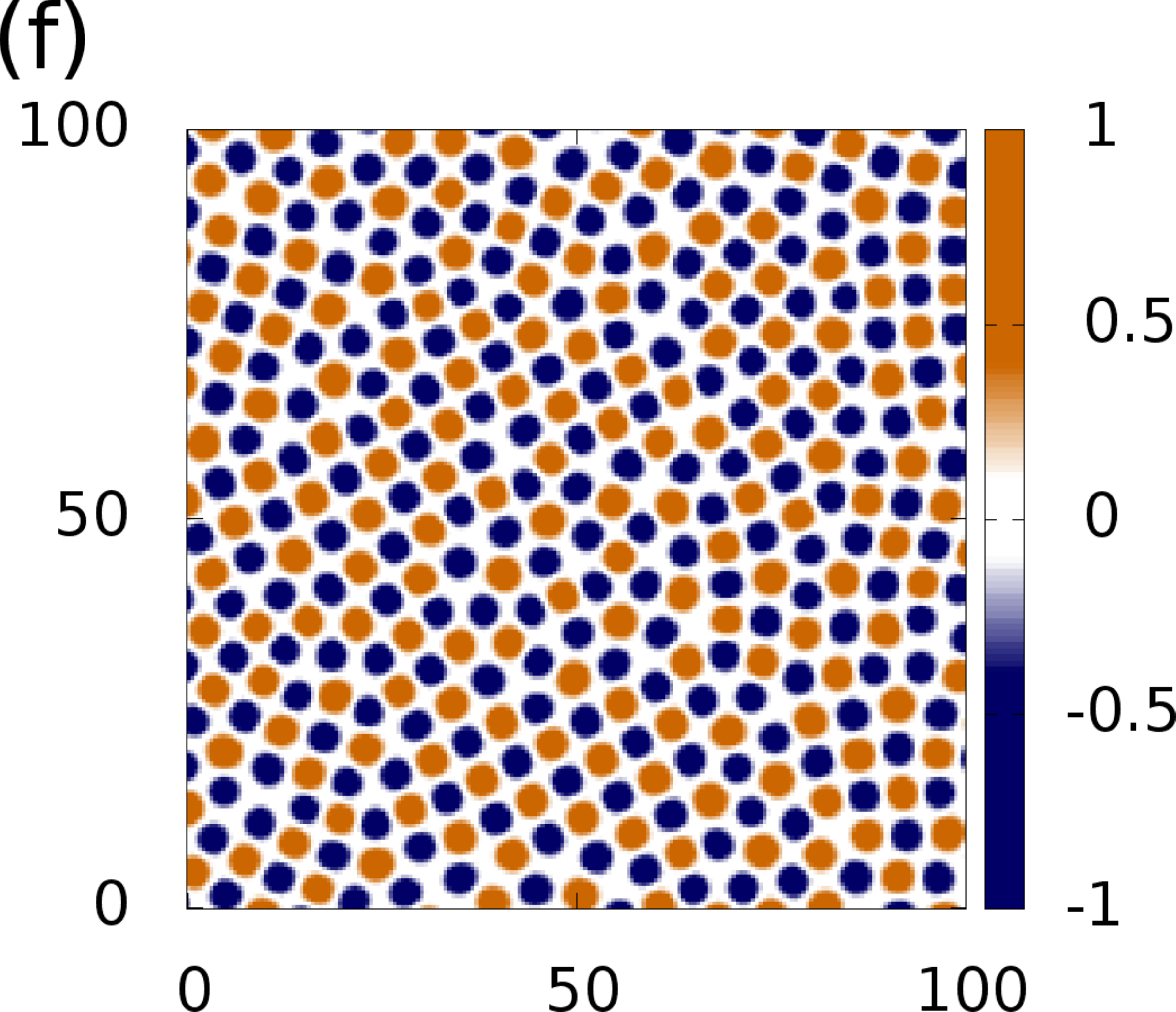}\\
	\includegraphics[width=0.3\linewidth]{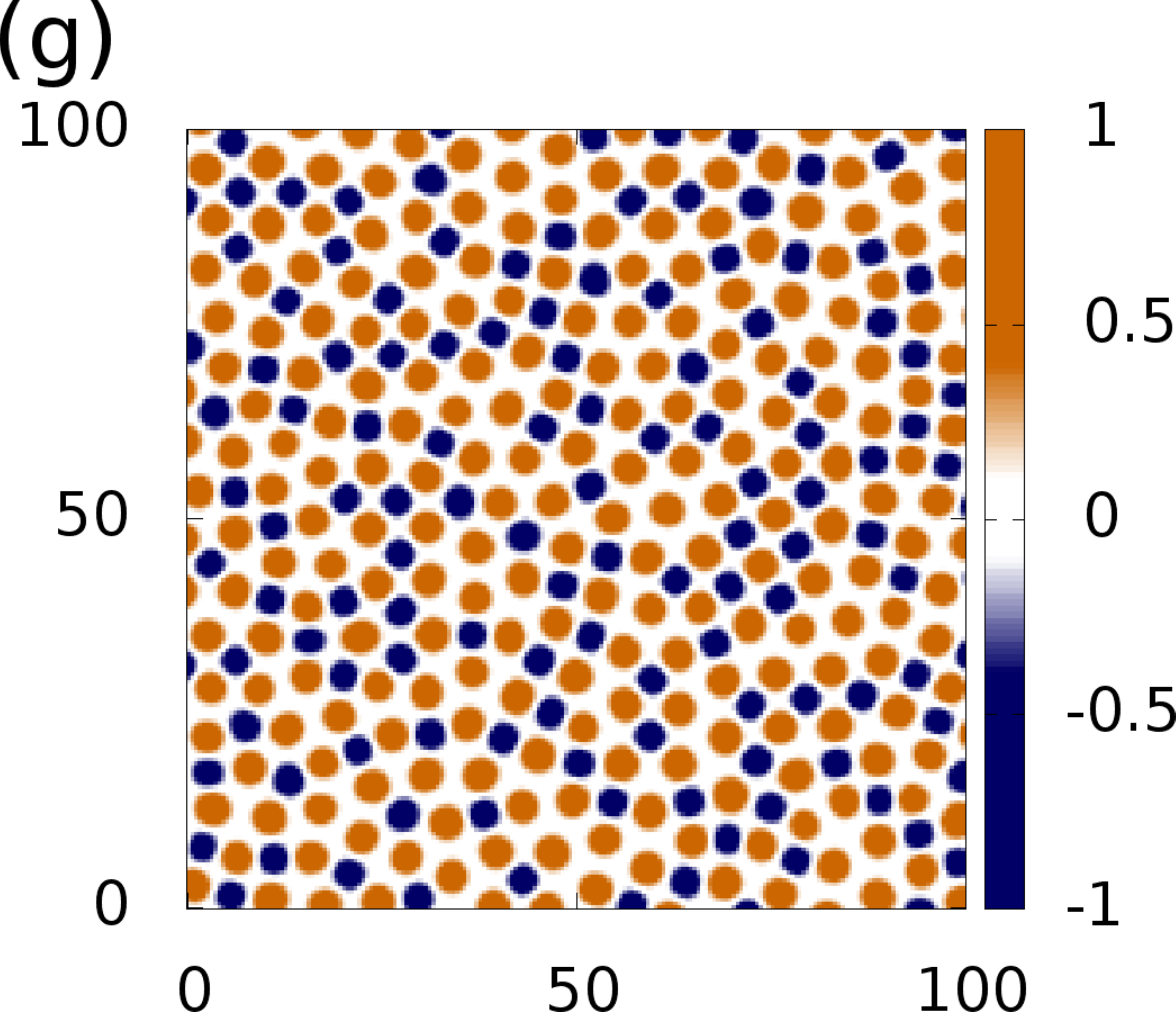}
	\includegraphics[width=0.3\linewidth]{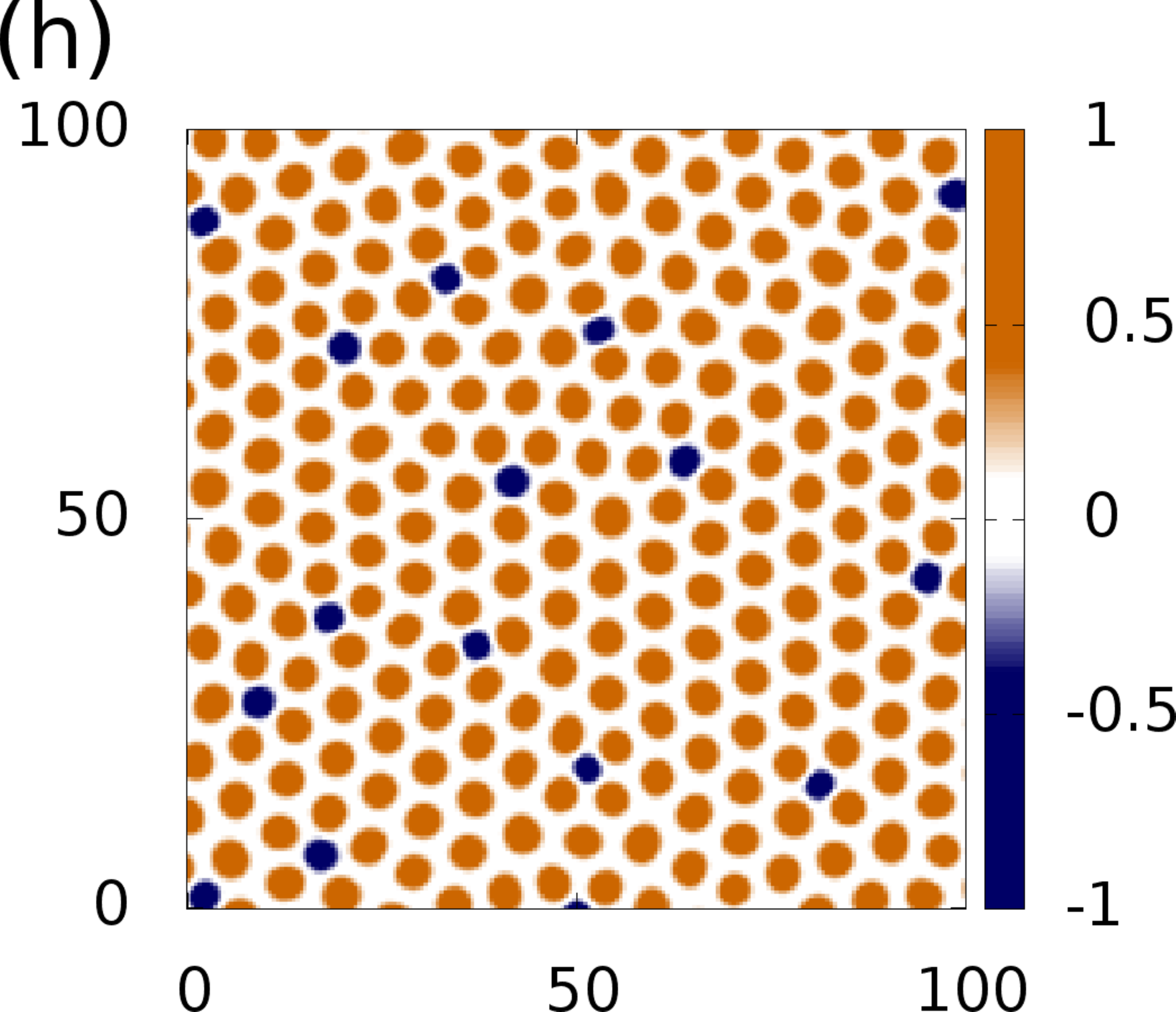}
	\caption{Plots of the order parameter $\Delta \hat{\phi}$, in which bumps in species $a$ appear in orange 
		      and bumps in species $b$ appear in blue, for a constant total order parameter value 
		      $\barA + \barB = 0.24$.  In (a) - (c) we show results from the symmetric mixture ($q_a = q_b = 1$) 
		      where the concentration of species $a$ is (a) $c = 0$, (b) $c = 0.25$ and (c) $c = 0.5$.  In 
		      (d) - (h) we show the results from the asymmetric mixture ($q_a = 1$ and $q_b = 1.1$)  where 
		      (d) $c = 0$, (e) $c = 0.25$, (f) $c = 0.5$, (g) $c = 0.75$ and (h) $c = 1$.  The parameter values 
		      are: $\eta = 4$ and $r = -0.9$.}
	\label{figTransProf}
\end{figure}

In Figs.~\ref{figTransProf}(a)--(c) we show the order parameter $\Delta \hat{\phi}$ for varying values of $c$ for 
the symmetric mixture ($q_a = q_b$).  We fix the total `density' $\barA + \barB = 0.24$ and investigate how the 
crystalline structures change as the concentration $c$ is varied.  In Fig.~\ref{figTransProf}(a), when $c = 0$ we 
observe a profile which is dominated by species $b$ bumps.  The crystal is hexagonally ordered with some 
defects (these tend to occur in the vicinity of the $\phi_a$ bumps).  There are only a few $\phi_a$ bumps, 
which means the bumps in $b$ are usually sitting next to each other, resulting in them ordering themselves in 
a similar manner to that observed in the one component model Fig.~\ref{figOneCompCorr}(c). Increasing the 
value of $c$ from $0$ to $0.25$, we observe a loss of crystalline structure, as shown in Fig.~\ref{figTransProf}(b).  The loss of  long range order is clearly visible in the associated radial distribution functions (not shown).  
The profile in Fig.~\ref{figTransProf}(b) shows a somewhat amorphous structure which appears to include 
both square and hexagonal ordering in equal measure.  Increasing the concentration further to $c = 0.5$, we
observe a similar square ordering of bumps as in Figs.~\ref{figTPCTwo}(j)--(k). (in Fig.~\ref{figTPCTwo}(j): 
$q_b = 1$, whereas in Fig.~\ref{figTPCTwo}(k): $q_b = 1.1$, all other parameter values are the same).  The 
crystalline structure in Fig.~\ref{figTransProf}(c) at $\barA = \barB = 0.12$ contains more vacancies and defects 
than the one in Fig.~\ref{figTPCTwo}(j) at $\barA = \barB = 0.15$, as both average order parameter values 
are smaller.  Owing to the symmetry induced by choosing $q_a = q_b$ (i.e., $\phi_a \to \phi_b$ as $c \to 1 - c$), a 
case with concentration $c$ is equivalent to the case with concentration $1 - c$. Thus~Fig.~\ref{figTransProf}(b) also shows the case $c = 0.75$ if one interchanges the orange and blue bumps. For this reason values $c>0.5$ are not 
shown.

For the asymmetric system the $c\to 1-c$ symmetry does not exist and we therefore show five cases 
for $c$ varying from $0$ to $1$ in Figs.~\ref{figTransProf}(d) $c = 0$, (e) $c = 0.25$, (f) $c = 0.5$, (g) 
$c = 0.75$ and (h) $c = 1$.  We again observe a transition from hexagonal ordering in Fig.~\ref{figTransProf}(d) 
to square ordering in Fig.~\ref{figTransProf}(f) and back to hexagonal ordering in Fig.~\ref{figTransProf}(h)
as the value of $c$ is increased from 0 to 1.  In between the highly structured states we observe the
mixed ordered states [Figs.~\ref{figTransProf}(e) and (g)] that were also present in the symmetrical system.  By 
eye, it is very difficult to pick out the differences between the symmetrical and the asymmetrical cases.  As 
previously discussed, the different value of $q_b$ in the asymmetrical system changes the shape, size
and quantity of $b$ bumps.  In order to characterise and better understand the organisation of the crystalline 
structures that are formed, we require a measure which may be used to quantify the structures and distinguish 
between hexagonal and square ordering in both the symmetric and the asymmetric systems.  To do this, we 
use Delaunay triangulation \cite{BKOS97,HjDa06} to calculate the distribution of the bond angles $p(\Theta)$ 
between nearest neighbours.  We could have used other measures from stochastic geometry \cite{SKM95}, 
which were used to characterise the hexagon-square transition in B\'enard convection \cite{ThEc98}.

\begin{figure}
	\includegraphics[width=0.46\linewidth]{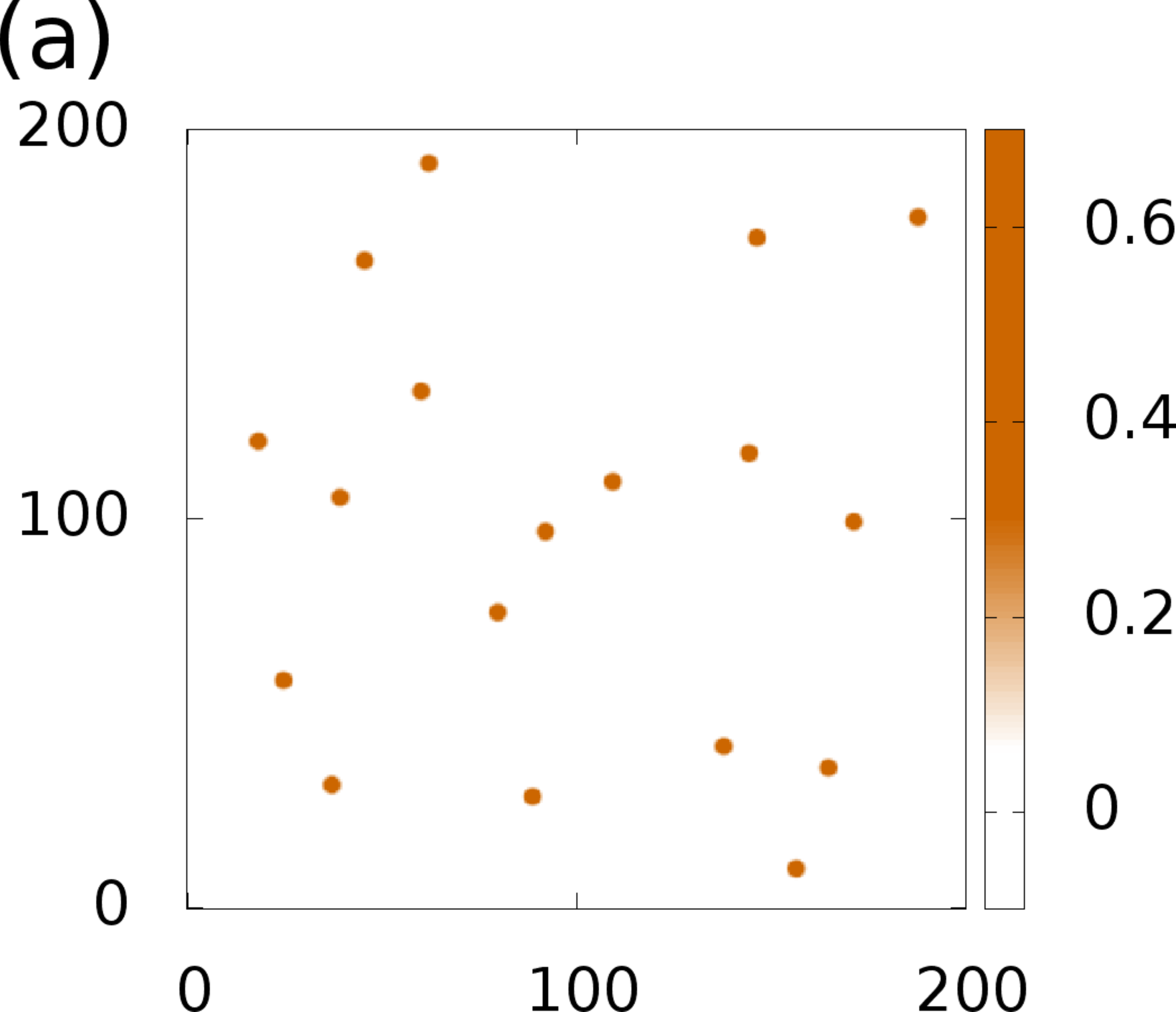} \hspace{1 mm}
	\includegraphics[width=0.45\linewidth]{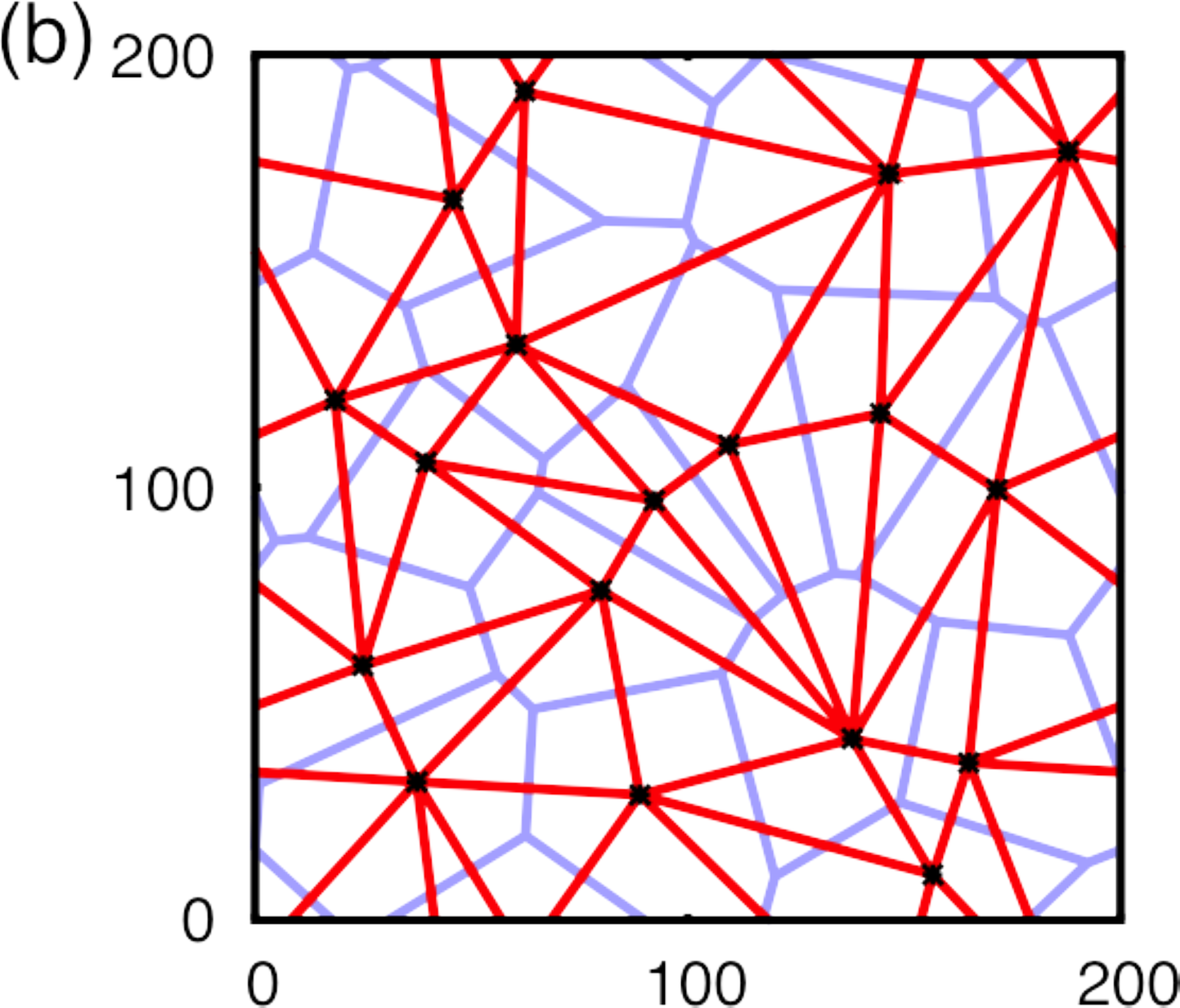} 
	\caption{An example Delaunay triangulation is shown for a simple one-component case.  
		      In (a) we display a typical order parameter profile for the one-component model where
		      we observe isolated peaks.  The coordinates of the maxima are calculated, these are
		      shown as black points in (b).  In (b) we show the Voronoi diagram (light blue polygon network) 
		      and the Delaunay triangulation (red triangles) for this particular set of coordinates.}
	\label{figTriangulation}
\end{figure}

The Delaunay triangulation is a triangulation of points (in our case the coordinates of the peaks of the bumps 
in both order parameter fields) which maximises the minimum angles of every triangle (i.e., avoids `skinny' 
triangles).  This triangulation can be calculated from the Voronoi diagram \cite{BKOS97,HjDa06} of any set of 
points on a 2d plane.  The Voronoi diagram is a set of polygons, where each polygon represents an area in 2d 
space which is closer to a particular point than to any of the other points (i.e.,~the locus of points contained in 
each polygon is closer to the bump inside the polygon than any other bump).  In Fig.~\ref{figTriangulation} we 
show an example of how we calculate the Delaunay triangulation for a given order parameter profile.  The 
example shows the triangulation for a one-component profile (as the pairing between bumps in the two-component 
model makes the triangulation harder to see) but the process is applied in the same manner to the 
two component model.  We take the coordinates of all the bumps to be our points on a 2d plane.  We then 
calculate the Voronoi diagram (shown as the light blue lines in Fig.~\ref{figTriangulation}(b)) which can be 
used to calculate the Delaunay triangulation (shown as the red lines in Fig.~\ref{figTriangulation}(b)).  This can 
be done using any of the algorithms outlined in Refs.~\cite{BKOS97,HjDa06}.  For an efficient method of 
calculating Voronoi diagrams and Delaunay triangulations see Ref.~\cite{Bowy81}.  (Note that Delaunay 
triangulation becomes degenerate when points appear in certain lines of symmetry.  However, the initial noise 
added to the order parameter fields prevents bumps from forming in perfect symmetry).  We use the statistics of 
the triangles in the Delaunay triangulation to characterise the structures produced by the bumps.

We extract three quantities from the triangulation: the area of the triangles, the length of the sides and
the angles in each of the triangles.  This information is gathered for five different realisations of the initial noise
profile for systems of size $200 \times 200$ and the information is sorted into bins.  From these bins we obtain 
the probability distribution function for each quantity.  Comparing the different distributions for various values 
of $c$ allows us to observe how the triangles in the triangulation change as we go from hexagonal to 
square ordering.  Here we concentrate on the probability distributions of the angles in the triangulation to 
characterise the crystalline structures. For results from the other measures see Ref.~\cite{Robb12}.


\begin{figure}
	\includegraphics[width=0.32\linewidth]{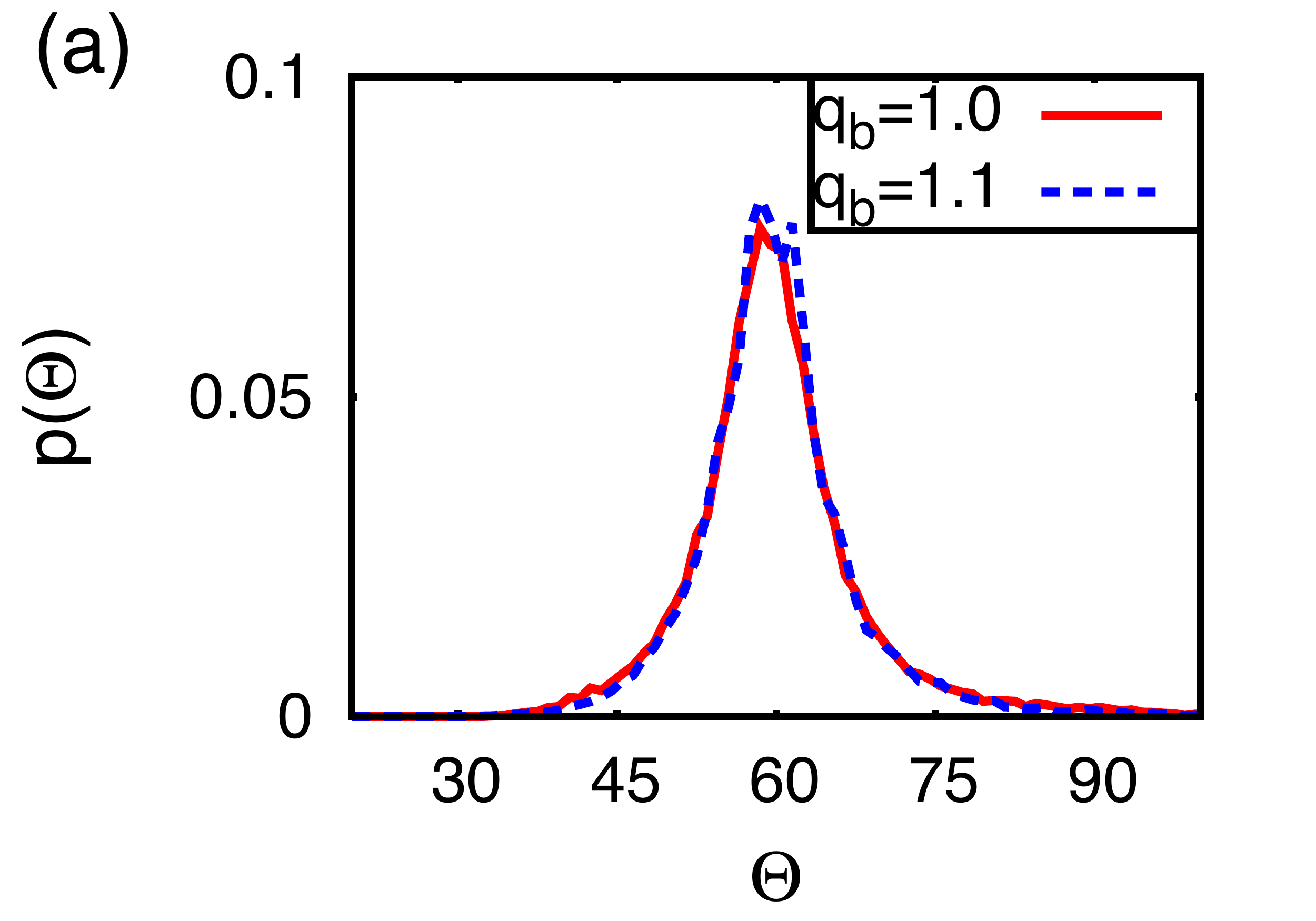}
	\includegraphics[width=0.32\linewidth]{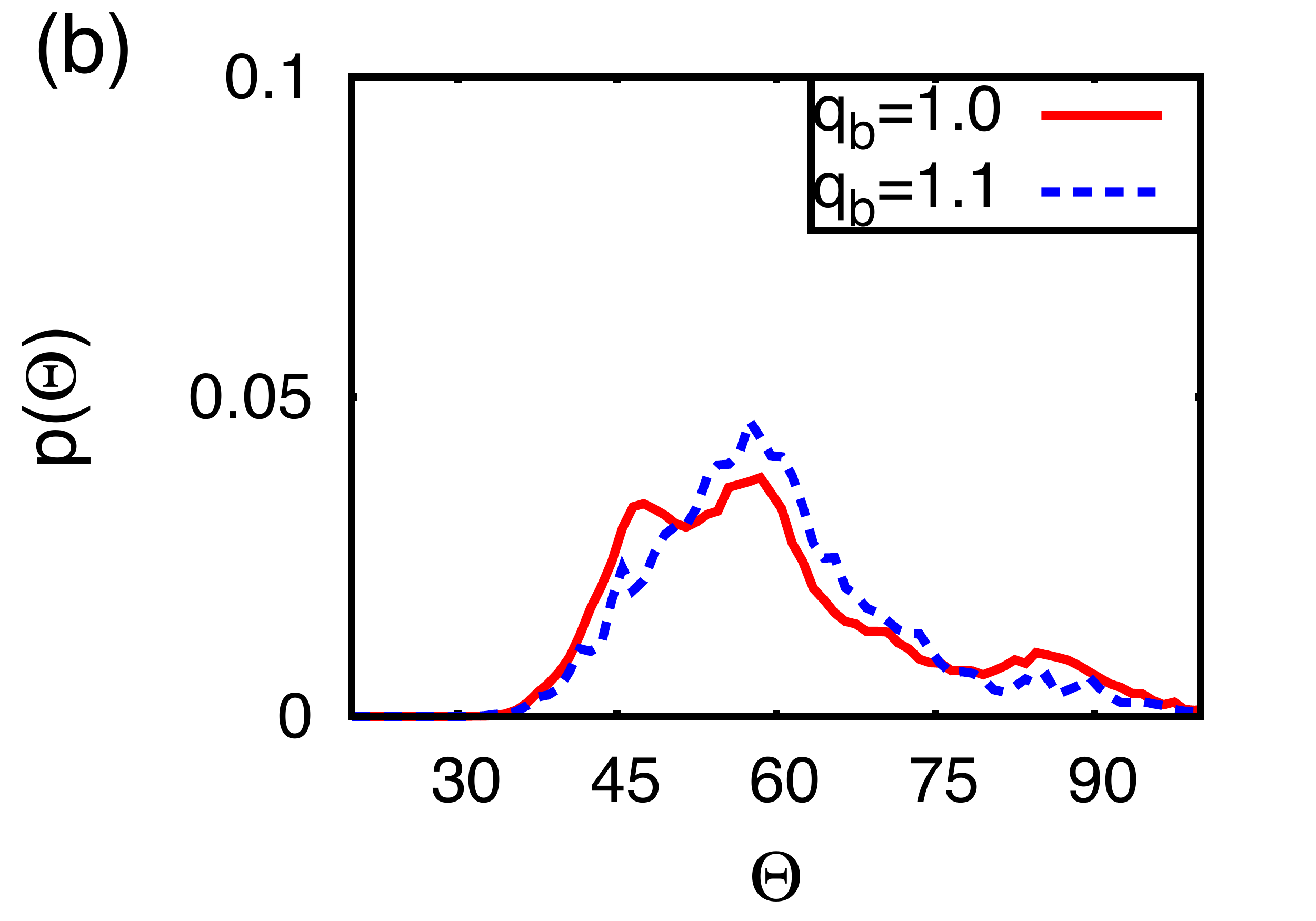}
	\includegraphics[width=0.32\linewidth]{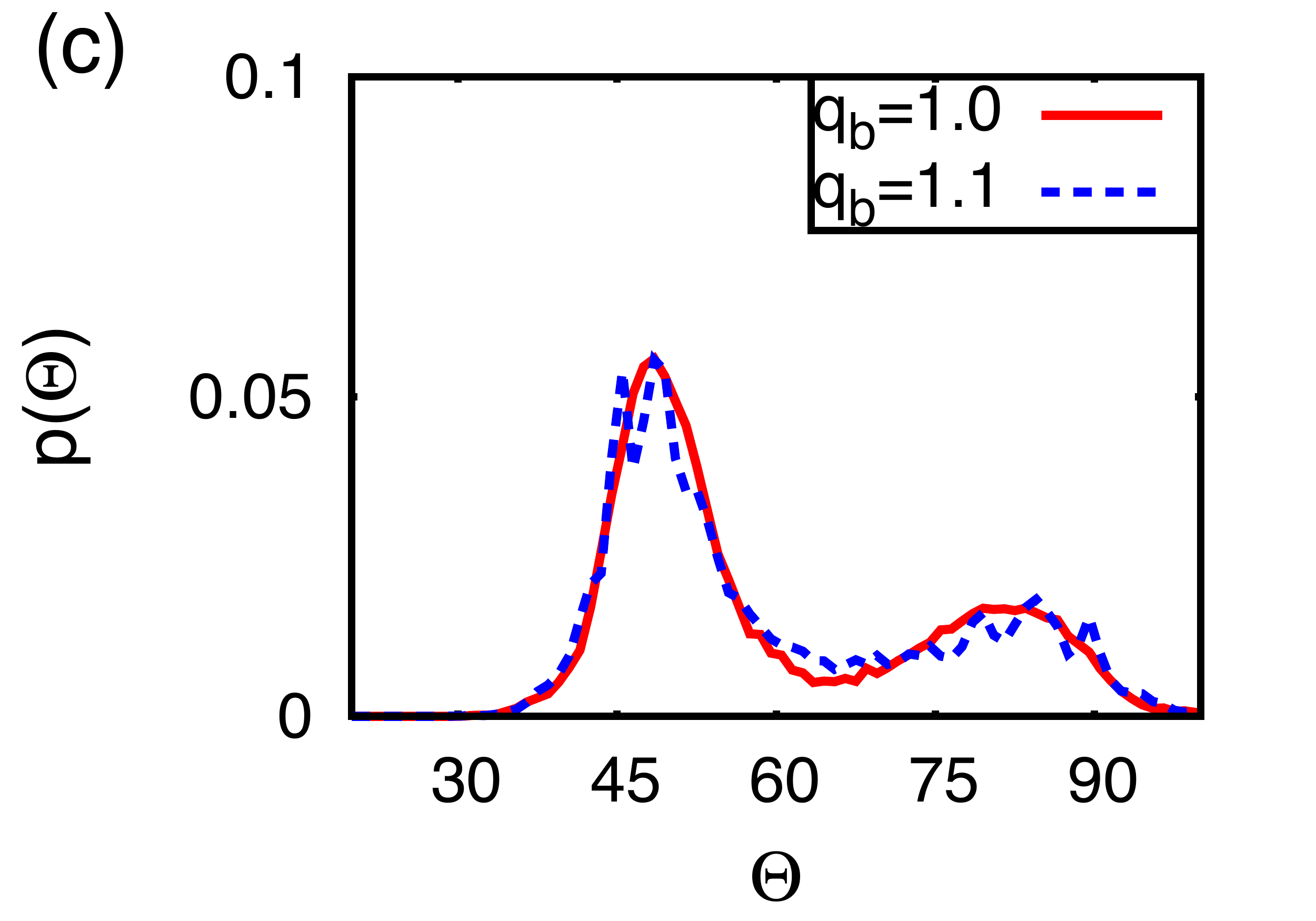}\\
	\includegraphics[width=0.32\linewidth]{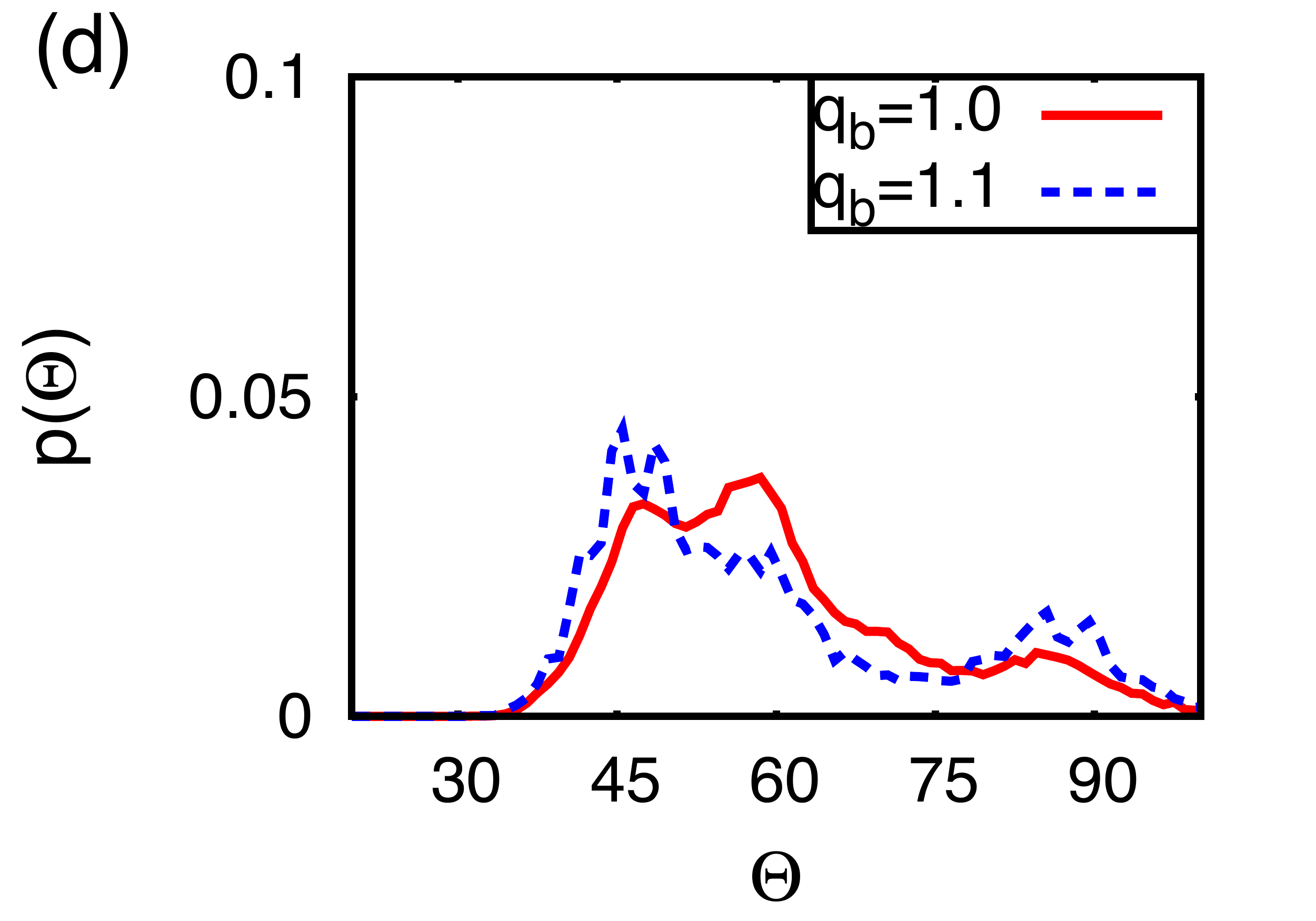}
	\includegraphics[width=0.32\linewidth]{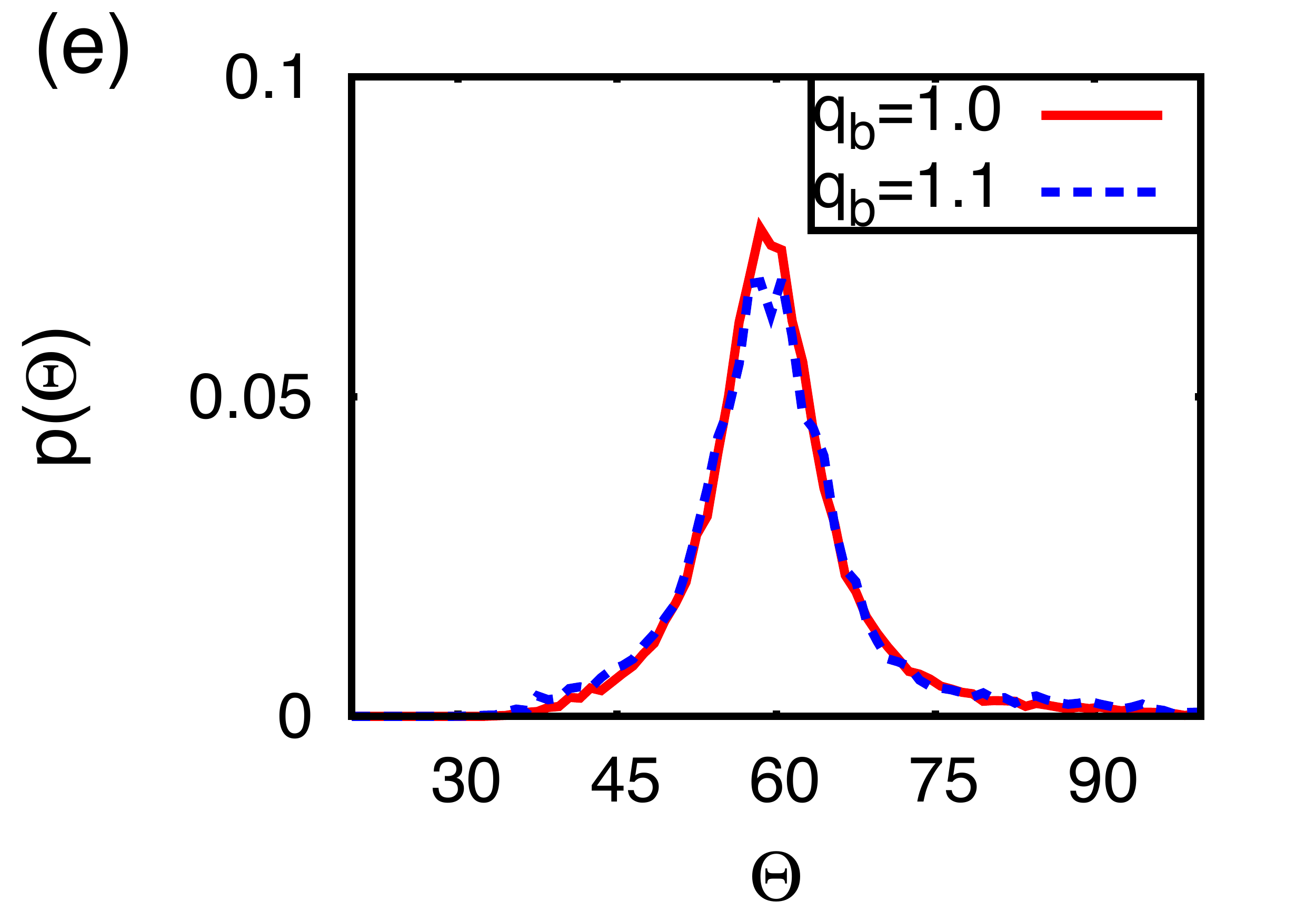}
	\caption{The bond angle distribution $p(\Theta)$ for a constant total order parameter value $\barA 
		       + \barB = 0.24$. The concentration of species $a$ is (a) $c = 0$, (b) $c = 0.25$, 
		       (c) $c = 0.5$, (d) $c = 0.75$, and (e) $c = 1$ (corresponding to the simulation snapshots 
		       shown in Fig.~\ref{figTransProf}).  Results for the symmetric system are shown as the red 
		       solid lines and the asymmetric system results are shown as blue dashed lines.  The 
		       parameter values are: $q_a=1$, $\eta = 4$ and $r = -0.9$.}
	\label{figAngleDist}
\end{figure}

In Fig.~\ref{figAngleDist} we display the probability distribution $p(\Theta)$ for the triangle corner angles, as 
the concentration $c$ is varied from $0$ to $1$.  We show results for the symmetric (solid red line) and the 
asymmetric (dashed blue line) systems.  The distribution of the angles of the triangles clearly shows the 
transition between hexagonal and square ordering.  When $c = 0$ we observe hexagonal ordering in both the
symmetric and the asymmetric systems, which results in the formation of roughly equilateral triangles in 
the Delaunay triangulation.  This produces angle distributions which have a single peak at $60^\circ$, as 
shown in Fig.~\ref{figAngleDist}(a).  As the value of $c$ increases the structure changes to square ordering, 
transforming the triangles into right-angled triangles.  Hence the angle distribution changes and we observe a 
peak slightly above the value $45^\circ$ and another peak (half the size) slightly below $90^\circ$, as shown 
in Fig.~\ref{figAngleDist}(c).  Increasing the concentration further to $c = 1$ restores the hexagonal ordering, 
hence the angle distribution returns to the single peak at $60^\circ$ [Fig.~\ref{figAngleDist}(e)].  In between the 
purely hexagonal and the purely square ordered structures we observe states where the distribution of bond 
angles is more evenly spread, with small peaks occurring just above $45^\circ$, at around $60^\circ$ and just 
below $90^\circ$ [Figs.~\ref{figAngleDist}(b) and \ref{figAngleDist}(d)].  These represent the somewhat 
amorphous structures which lack the long range ordering which is present in the hexagonally and square 
ordered structures.  The position of these peaks in the bond angle distributions $p(\Theta)$ does not depend 
on the quantity or size of the bumps and so the peaks occur in (almost) the same position for the symmetric 
and the asymmetric systems for all values of $c$ (this is not the case for the area or length distributions).  This 
makes the bond angle distributions ideal for comparing the structure of bump formations in different systems.  
On comparing the symmetrical and the asymmetrical cases we observe that $p(\Theta)$ appears smoother in 
the symmetrical case.  The distribution function $p(\Theta)$ has a more jagged appearance for the asymmetrical mixture, which we 
believe is due to the fact that there are different sized bumps in this mixture, making it more difficult for the 
bumps to organise themselves into regular structures.  In Figs.~\ref{figAngleDist}(a) and \ref{figAngleDist}(e) 
the distributions appear very similar for the symmetric and the asymmetrical cases, however, in the other 
distributions (in particular, Figs.~\ref{figAngleDist}(b) and \ref{figAngleDist}(d)) we observe a distinct difference 
in the height of the three peaks.  This suggests that the transition between the different ordered states occurs 
differently in the symmetric and the asymmetric systems. 

To examine more closely the transition from the hexagonal to the square ordered states we introduce an order 
parameter $\Phi$ which is calculated from the distribution of the angles from the Delaunay triangulation.
We integrate the angle distributions over three regions which cover the three different peaks (these 
regions are determined arbitrarily from close examination of the angle distributions in 
Fig.~\ref{figAngleDist}) and define the quantities:

\begin{eqnarray}
	R_0 &=& \int^{53}_{25} p(\Theta) \hspace{1mm} d\Theta,
	\nonumber \\
	R_1 &=& \int^{72}_{53} p(\Theta) \hspace{1mm} d\Theta,
	\nonumber \\
	R_2 &=& \int^{115}_{72} p(\Theta) \hspace{1mm} d\Theta.
\end{eqnarray}

We then define the order parameter $\Phi$ in the following way:

\begin{equation}
	\Phi = \frac{R_0 + R_2}{R_1}.
	\label{eqTrans}
\end{equation} 

When a structure consists of mainly hexagonal configurations of bumps the value of $\Phi$ is small (since
$\Phi \to 0$ as $R_0 \to 0$ and $R_2 \to 0$) and when a profile is dominated by square ordering the 
value of $\Phi$ is large (since $\Phi \to \infty$ as $R_1 \to 0$).  Calculating this quantity for the angle 
distributions for different values of $c$ gives us a measure for the hexagonal vs.~square ordering of the
bumps. 

\begin{figure}
	\includegraphics[width=0.6\linewidth]{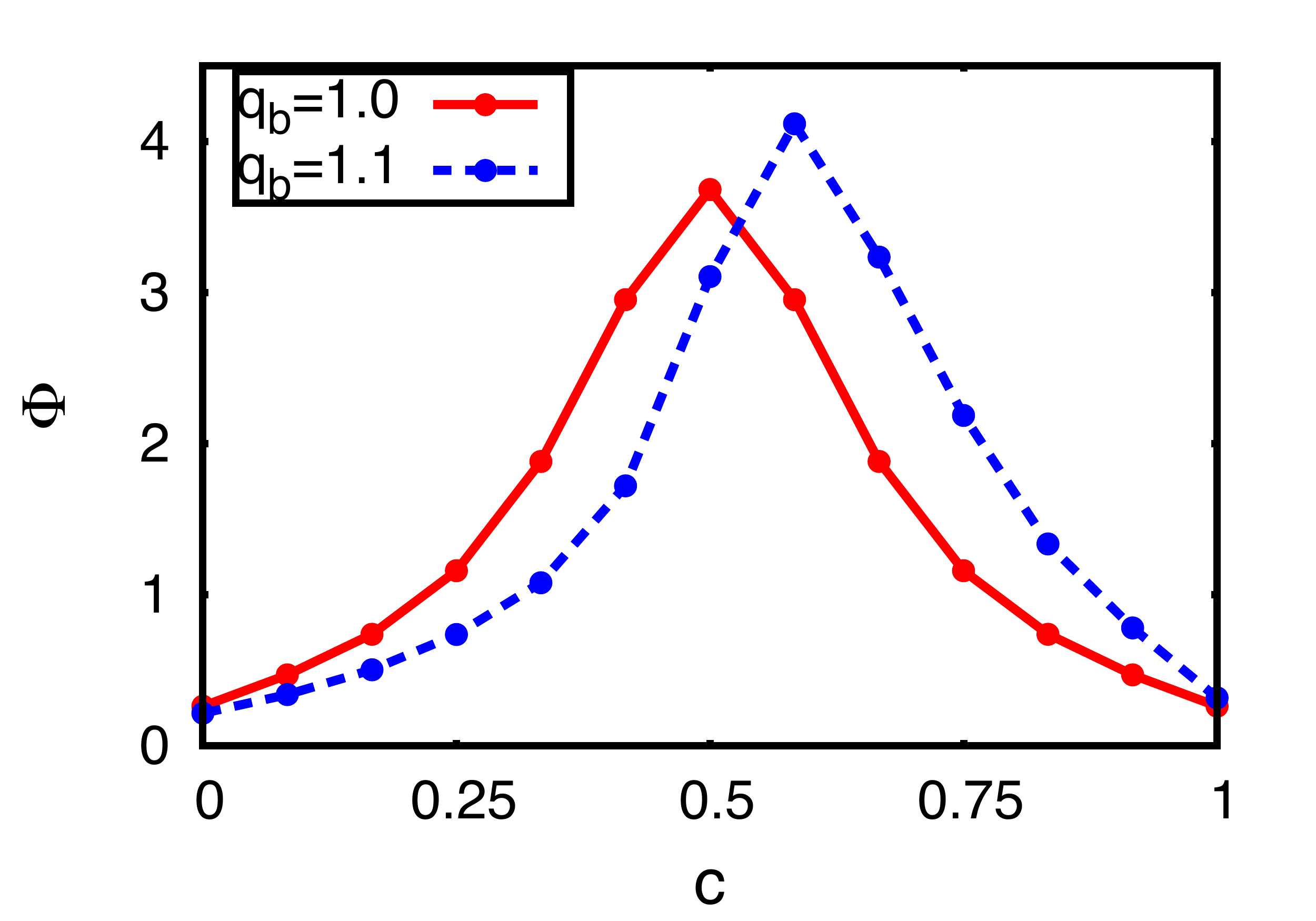}
	\caption{Plot showing the order parameter $\Phi$ (defined in Eq.~(\ref{eqTrans})) as a function of the concentration 
		      $c$ of species $a$, where $\barA + \barB = 0.24$.  The red solid line and points show the 
		      symmetric case and the blue dashed line and points show the asymmetrical case.  The 
		      parameter values are: $q_a=1$, $\eta = 4$ and $r = -0.9$.}
	\label{figOrderTran}
\end{figure}

In Fig.~\ref{figOrderTran} we show how the order parameter $\Phi$ changes with the concentration $c$
for the symmetric (solid red line) and the asymmetric (dashed blue line) mixtures.  Both curves show
a smooth continuous transition from hexagonal ordering to square ordering and back again.  The different
sized bumps in the asymmetric system break the symmetry around  $c = 0.5$ and we observe that 
the maximum (which corresponds to the strongest square ordering) occurs at around $c \approx 0.6$, and is 
actually higher than the peak in the symmetric case.  The transition to and from square ordering appears to be 
slightly sharper in the asymmetrical case.  Even though there is a difference in the transition between the 
different ordered states in the symmetric and the asymmetric mixtures, they appear to be qualitatively similar.  It 
may be the case that for a larger difference in the values of $q_a$ and $q_b$ a different type of transition from 
hexagonal to square ordering might occur, e.g.,~a discontinuous transition.  However, the effect of varying the 
ratio $q_a/q_b$ is not studied in detail here.  

\section{Conclusions}
\label{secConc}

In this paper we have investigated the VPFC model and its application to materials 
modelling.  We first considered the one-component model proposed by Chan et~al.~in 
Ref.~\cite{CGD09}.  We determined the linear stability of the homogeneous state and 
discussed the dispersion relation.  We examined the phase behaviour in one dimension 
and calculated the phase diagram, computing exactly the location of the tricritical point 
between the homogeneous and periodic states, and identified the region of phase space 
where localised structures occur.  Focusing on the latter region of the phase diagram, 
we investigated the localised steady state profiles and discussed the slanted homoclinic 
snaking which occurs in the bifurcation diagrams.  The one-component model was also 
studied in two dimensions and we determined the phase diagram, radial distribution 
functions and effective pair potentials from our simulation data.  Some of the behaviour
we have identified -- the presence of transitions resembling transitions from a solid phase
to a liquid phase and then to a gas-like phase -- replicates behaviour observed in 
nonconserved systems \cite{BCR09}. In section \ref{secTwoComp} of the paper, we extended 
the model to include two coupled order parameter fields.  We have considered how the 
coupling affects the linear stability of flat films and then briefly touched on the phase 
behaviour of this two component model.  We have focused on the bump structures which form, 
considering both a symmetrical mixture where the bumps are of equal size and an 
asymmetrical system where one of the bump species is slightly smaller than the other 
species.  The radial distribution functions and effective pair potentials for these systems
are somewhat similar to those in binary mixtures of oppositely charged colloidal particles.
We have investigated how varying the concentration $c$ of the mixture produces a crossover 
from hexagonal to square ordered crystalline structures and how the transition differs 
between the symmetrical and the asymmetrical systems. 

A key issue on which we should comment concerns the question of what precisely does the order parameter 
profile $\phi({\bf x},t)$ in the VPFC model represent? In the regular PFC model, the phase with the uniform flat 
profile is taken to represent the liquid phase, whilst the bump phase corresponds to the crystalline solid. This 
interpretation is underpinned by the fact that the regular PFC can be derived from density functional theory 
(DFT) \cite{ElGr04} and dynamical density functional theory (DDFT) \cite{TBVL09}, which is a theory for the 
dynamics of a system of interacting Brownian (colloidal) particles \cite{MaTa99,MaTa00,ArEv04,ArRa04}. DFT 
\cite{Evan79, Evan92, HaMc06} is a statistical mechanics theory for the one-body number density 
$\rho({\bf x})$ of a system of particles, where $\rho({\bf x})=\langle \hat{\rho}({\bf x})\rangle$ and where 
$\hat{\rho}({\bf x})=\sum_i\delta({\bf x-x}_i)$ is the density operator and $\langle \cdot \rangle$ denotes a 
statistical ensemble average \cite{Evan79}. The central quantity in DFT is the Helmholtz free energy functional 
$F[\rho]$ and the equilibrium fluid density profile $\rho({\bf x})$ is that which minimises the 
grand free energy $\Omega[\rho]=F[\rho]-\mu \int {\rm d}{\bf x} \rho({\bf x})$. The DDFT for Brownian particles 
\cite{MaTa99,MaTa00,ArEv04,ArRa04} takes as input this functional and so yields the correct equilibrium fluid 
density profile. Making a truncated gradient expansion approximation for $F[\rho]$, expanding the free energy 
around that of a reference liquid state with uniform density $\rho_0$, one can argue that the free energy is 
approximately given by Eqs.~\eqref{eqPFCFree} and \eqref{eqPFCOne}, where the order parameter 
$\phi({\bf x})\propto \rho({\bf x})-\rho_0$. Thus it is clear that in a bulk liquid, where $\rho({\bf x})$ is a constant, 
so too is $\phi({\bf x})$ a constant and in the solid phase, where $\rho({\bf x})$ consists of a periodic array of 
density peaks, then $\phi({\bf x})$ also contains periodic modulations. However, there are some problems 
extending this interpretation to the VPFC. Consider for example Fig.~\ref{figOneCompCorr}(a) where we see 
a few isolated localised peaks surrounded by a uniform background where $\phi({\bf x})\approx 0$. 
Maintaining the above PFC interpretation, this would correspond to a few individual `frozen' particles, 
surrounded by a fluid of mobile particles. One might be tempted to think of this as some sort of glass transition 
\cite{HaMc06, BeBi11, Cava09}, but the glass transition is a collective phenomenon: in a glass one does see 
`dynamical heterogeneity' i.e.,~regions where the particles are totally jammed and other regions which are 
more mobile, but to our knowledge one never sees a single particle that is jammed on its own surrounded by 
more mobile particles. Thus, it may be possible to assume this interpretation may be maintained for the VPFC, 
i.e.,~by considering the localised peaks surrounded by a uniform background to be a dynamically 
heterogeneous glassy system, but there are problems with this point of view.

An alternative interpretation for the order parameter profile in the VPFC model is that $\phi(x)$ is related to a 
coarse-grained density profile (rather than an ensemble average density profile) for the system 
$\tilde{\rho}({\bf x},t)$, i.e.,~$\phi({\bf x},t)\propto \tilde{\rho}({\bf x},t)$.  Following Ref.~\cite{ArRa04}, we may 
define the temporally coarse-grained density profile for a system of Brownian colloidal particles as 
$\tilde{\rho}({\bf x},t)=\int K(t-t')\hat{\rho}({\bf x},t){\rm d} t'$, where $K(t)$ is a normalised function of finite 
support which defines a time window over which the density is coarse-grained. One can then argue 
\cite{ArRa04} that the time evolution equations for $\tilde{\rho}({\bf x},t)$ must be very similar or even the same 
as the DDFT equations for the time evolution of the ensemble average density $\rho({\bf x},t)$, as long as the 
width in time $\tau$ of $K(t)$ is large enough. By choosing the time $\tau$ so that it is large compared to the 
time between the colloidal particles receiving Brownian `kicks' from the solvent, but is short compared to the 
diffusive time scale, corresponding to the typical time for a particle to diffuse a distance equal to its own 
diameter, then the coarse-grained density $\tilde{\rho}({\bf x},t)$ and the order parameter $\phi({\bf x},t)$ will be 
quantities which contain peaks, each of which correspond to an individual particle in the system. Thus, in a 
low density colloidal suspension one should see isolated peaks in the coarse-grained density, surrounded by 
regions where $\phi({\bf x},t)\approx 0$, corresponding to no particles being present in that region of the 
system. This is the justification for the interpretation made by Chan {\it et~al.}~in Ref.~\cite{CGD09}, that the 
peaks in the order parameter correspond to particles and the uniform background corresponds to a portion of 
solvent free of particles. In order to observe the long time Brownian motion of the particles in this description, 
one should add a stochastic noise term to the dynamical equations for the system \eqref{eqDynTwo}, that 
continuously drives the system (as opposed to the small amount of noise that is present in our initial order 
parameter profiles). However, in numerical simulations there can be problems with such an approach, 
because the particles can become pinned in place by the discrete grid on which they are defined, and so do 
not move. We did not make a detailed investigation of the of the VPFC model with additional noise.  Further 
issues arise as the noise renormalises the parameters of the continuum model.

There are state points in the PFC and VPFC phase diagram where all possible interpretations of $\phi$ break 
down: these are the state points where the equilibrium state is the stripe or the hole phase, such as those 
displayed in Fig.~\ref{figHighPhive}.  Systems of spherical particles do not have an ensemble average density 
profile $\rho$ nor a coarse-grained density profile $\tilde{\rho}$ with stripes/holes, unless the particles in the 
system interact via pair potentials containing competing attractive and repulsive parts 
\cite{APER07, Arch08, ArWi07}. We must conclude that for the parameter values corresponding to these state 
points, the gradient expansion that is implicit in the PFC and VPFC free energy functionals has broken 
down and that these order parameter profiles are unphysical.

The radial distribution functions for the one-component model displayed in Fig.~\ref{figOneCompCorr} (see 
also Fig.~5 of Ref.~\cite{CGD09}) are very similar to those in real fluids.  We observe static correlations which 
are very similar to what one observes in fluids.  Increasing the value of $\bar{\phi}$ increases the number of 
bumps and close packing causes long range (crystalline) ordering of the bumps.  Calculating the effective pair 
potential between isolated pairs of bumps, we find a pair potential having an attractive minimum at a pair 
separation distance which is slightly larger than the diameter of the bumps.  Thus, the interactions and 
correlations between bumps share certain features with some colloidal fluids \cite{BaHa03}.  We 
also extend the model to consider a two component mixture, with a simple repulsive coupling between the two 
order parameter profiles.  At low values of $\bar{\phi}$ the bumps commonly appear in pairs and at 
intermediate values they tend to form chains.  At higher values of $\bar{\phi}$ the system exhibits crystalline 
ordering.  The appearance of these structures is somewhat reminiscent of the arrangement of the particles in a 
binary mixture of oppositely charged colloidal particles - see e.g.,~Ref.~\cite{LCH05, BaCa05} and references 
therein.  The radial distribution functions and the effective pair potentials show there is a fairly strong attraction 
between bumps of the opposite species $a$ and $b$.  The minimum in the $ab$ effective pair potential is at a 
shorter pair separation distance than the minimum in the $aa$ and $bb$ pair potentials and so we observe 
square ordering when the concentration $c\approx 1/2$ and $\bar{\phi}$ is high enough for the bumps to pack 
into a crystalline structure. However, when $c\approx 0$ or $c\approx 1$, we observe hexagonal ordering and 
so we observe a transition from hexagonal to square ordering as the concentration $c$ is varied.  We find that
this transition occurs smoothly but can become skewed by changing the size of one of the species of bumps
($q_a \ne q_b$).

It would be interesting to further investigate the effect that varying the ratio $q_a/q_b$ has on the structures 
which form.  In particular, determining the range of values of $q_a$ and $q_b$ for which bump profiles 
form in the 2d system would allow one to determine the range of size ratios of particles (bumps) that can 
be modelled.  The transition between hexagonal and square structures could then be studied for systems with 
very different sized bumps and if the VPFC in this regime continues to be able to model mixtures of charged 
colloidal particles, then a wide range of different crystal structures should be observed \cite{LCH05}. 
  
Note also that the localised structures that we observe are not a unique property of the VPFC model but are in 
fact also present in the regular PFC model for a small range of parameter values outside the limit of linear 
stability.  This is something that we will focus on in future work. 

\begin{acknowledgements}
	This work was supported by the EU via the ITN MULTIFLOW
	(PITN-GA-2008-214919). MJR also gratefully acknowledges support
	from EPSRC and AJA thanks RCUK for support.
\end{acknowledgements}

	\bibliographystyle{prsty}
	\bibliography{references}

\end{document}